\documentclass[12pt,a4paper]{article}
\usepackage{jheppub}
\usepackage{amsmath,amssymb,amsfonts,pstricks,mathtools}
\usepackage{float}
\usepackage{scalefnt}
\usepackage{subfig}
\usepackage{ifsym}
\usepackage{wrapfig}
\usepackage{fancybox}
\usepackage{float}
\newcommand{\bea}{\begin{eqnarray}\displaystyle}
\newcommand{\eea}{\end{eqnarray}}
\newcommand{\nn}{\nonumber}
\newcommand{\fgel}{\mathfrak{f}}

\newcommand{\figref}[1]{Fig.~\protect\ref{#1}}
\newcommand{\nome}{\mathfrak{q}}
{\setlength{\fboxsep}{15pt}
\setlength{\mylength}{\linewidth}%
\addtolength{\mylength}{-2\fboxsep}%
\addtolength{\mylength}{-2\fboxrule}%
\Sbox
\minipage{\mylength}%
\setlength{\abovedisplayskip}{0pt}%
\setlength{\belowdisplayskip}{0pt}%
\equation}%
{\endequation\endminipage\endSbox
\[\fbox{\TheSbox}\]}

\begin{document}
\title{\begin{flushright} ${}$\\[-130pt] \small CERN-PH-TH/2013-237 \\[130pt]\end{flushright}M-strings, Elliptic Genera and $\mathcal{N}=4$ String Amplitudes
%.M-Strings and $\mathcal{N}=4$ Topological Amplitudes\\
%Variations on $\mathcal{N}=2$ partition functions\\
%Four approaches to $\mathcal{N}=2$ partition functions\\Mass deformed theories from M-strings and Topological strings
}
\author[\clubsuit]{Stefan Hohenegger,}
\author[\spadesuit]{Amer Iqbal}
\affiliation[\clubsuit]{Department of Physics, CERN - Theory Division, CH-1211 Geneva 23, Switzerland}
\affiliation[\spadesuit]{Department of Physics, LUMS School of Science \& Engineering, U-Block, D.H.A, Lahore, Pakistan.}
\affiliation[\spadesuit]{Department of Mathematics, LUMS School of Science \& Engineering, U-Block, D.H.A, Lahore, Pakistan.}

\emailAdd{Stefan.Hohenegger@cern.ch}
\emailAdd{amer@alum.mit.edu}

\abstract{We study mass-deformed $\mathcal{N}=2$ gauge theories from various points of view. Their partition functions can be computed via three dual approaches: firstly, $(p,q)$-brane webs in type II string theory using Nekrasov's instanton calculus, secondly, the (refined) topological string using the topological vertex formalism and thirdly, M theory via the elliptic genus of certain M-strings configurations. We argue for a large class of theories that these approaches yield the same gauge theory partition function which we study in detail. To make their modular properties more tangible, we consider a fourth approach by connecting the partition function to the equivariant elliptic genus of $\mathbb{C}^2$ through a (singular) theta-transform. This form appears naturally as a specific class of one-loop scattering amplitudes in type~II string theory on $T^2$, which we calculate explicitly.}
\maketitle

%\tableofcontents

\section{Introduction}
The world-volume theory of multiple coincident  M5 branes is among the most fascinating objects of current interest in high energy physics. It is a conformal field theory in six dimensions with $\mathcal{N}=(2,0)$ supersymmetry and a non-abelian gauge group of ADE type~\cite{Witten:1995zh}. While highly interesting both from a mathematical as well as a physics perspective, it remained largely mysterious so far, mostly due to the lack of a Lagrangian description.\footnote{More precisely, no non-abelian six-dimensional action is known. However, several promising proposals in lower dimensions for actions of the compactified theory have been put forward \cite{Bonetti:2012st}.} Therefore, in the recent past, new approaches have been considered, with the goal to extract at least some information about this theory, or its lower dimensional cousins.

In this paper we study various approaches to computing partition functions of certain mass-deformed gauge theories with eight supercharges related to the six-dimensional $\mathcal{N}=(2,0)$ theory. We will collectively refer to such theories as $\mathcal{N}=2^*$ gauge theories.\footnote{When compactified to four dimensions the simplest of these theories is the ${\cal N}=2^{*}$ $SU(N)$ gauge theory i.e., it has a massive hypermultiplet in the adjoint representation.} A pictorial overview of our approaches is given in  \figref{Fig:OverviewPartFct}, depicting the fact that we use three dual representations of the gauge theories, as well as a perturbative approach using particular one-loop amplitudes in string theory. Let us begin by reviewing the dual formulations of mass-deformed supersymmetric gauge theories:
\begin{figure}[h]
\begin{center}
\epsfig{file=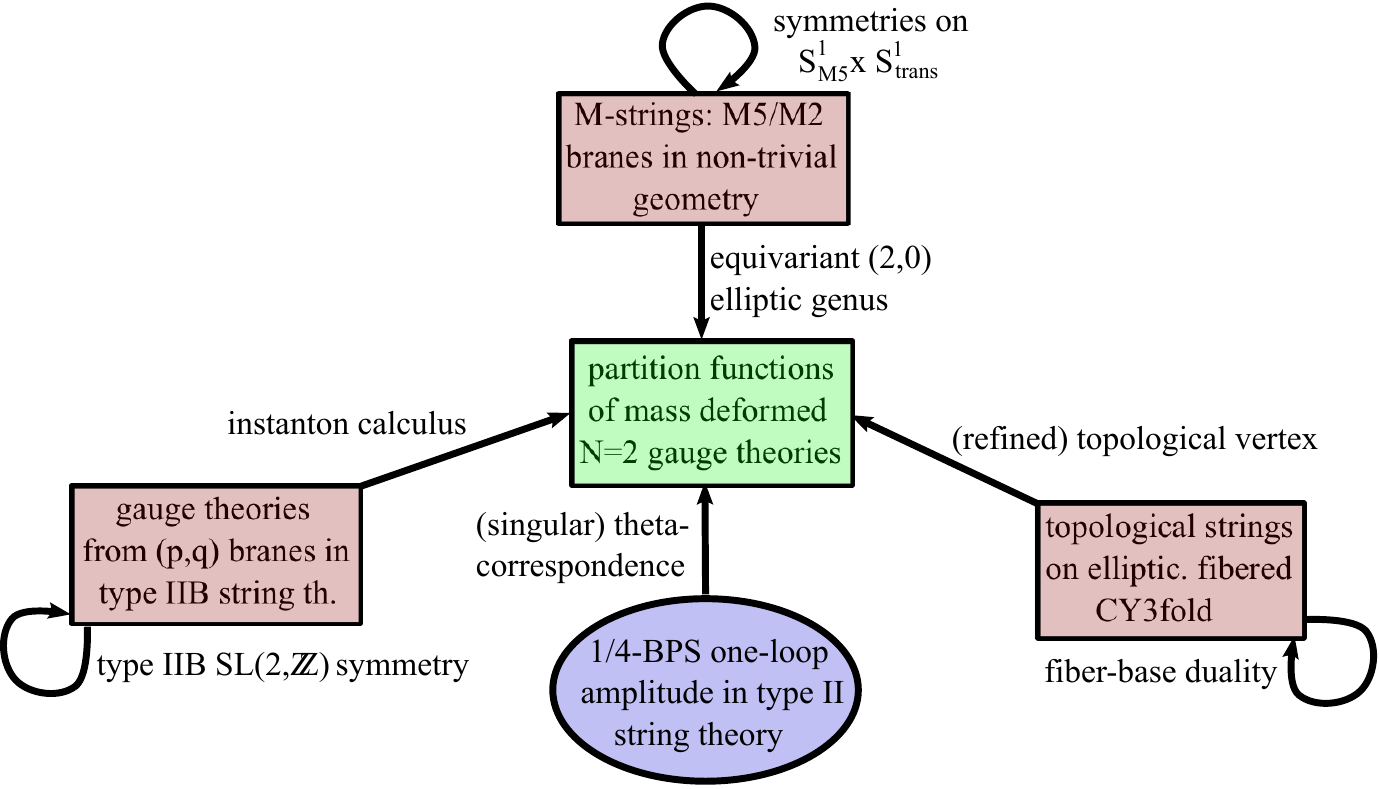,width=15cm}
\end{center}
\caption{\emph{Four-fold approach to $\mathcal{N}=2^*$ gauge theories:} The red squares represent three dual settings from which the partition function of certain mass deformed $\mathcal{N}=2$ gauge theories can be computed using various techniques. Each of these approaches has interesting dualities which can be exploited. The blue circle denotes a specific one-loop string amplitude which reproduces the partition function for a particular configuration of branes.}
\label{Fig:OverviewPartFct}
\end{figure}
\begin{enumerate}

\item \emph{Gauge theory on $(p,q)$  5-brane web in type IIB:}\\
 Five dimensional supersymmetric gauge theories can be realized on $(p,q)$ 5-brane webs in type IIB string theory \cite{Aharony:1997bh}. The theory we are interested in corresponds, before mass deformation, to a brane web compactified on a torus consisting of D5- and NS5-branes. Mass deformations of the gauge theory arise from the splitting of the D5-brane on the NS5-brane worldvolume as discussed in \cite{Hollowood:2003cv}. The gauge theory appears on the world-volume of the D5-branes and its partition function can be calculated using Nekrasov's instanton calculus \cite{Nekrasov:2002qd}. Symmetries of the underlying string theory (\emph{e.g.} $SL(2,\mathbb{Z})$) have an interesting impact also on the gauge theory partition function, which we will study in detail.

\item \emph{(Refined) topological string on elliptically fibered Calabi-Yau three folds:}\\
The gauge theory can also be obtained from the point-particle limit of M-theory on a specific elliptically fibered Calabi-Yau threefold (we will write CY3fold in the rest of the paper). The CY3fold which can be used to engineer the theory is the $\mathbb{Z}_{N}\times \mathbb{Z}_{M}$ orbifold of the CY3fold which is dual to the compactified brane web with a single D5- and NS5-brane. The orbifold action gives a CY3fold which is dual to the compactified brane web with $N$ D5-branes and $M$ NS5-branes. The CY3fold obtained from an $\mathbb{Z}_{N}\times \mathbb{Z}_{M}$ action and $\mathbb{Z}_{M}\times \mathbb{Z}_{N}$ action turn out to be the same hence the corresponding gauge theories are dual to each other. The case discussed in \cite{mstrings} corresponded to $M=k,N=1$ which was dual to $M=1,N=k$.
%Its partition function can be computed in various ways, reflecting several different realisations of the topological string discussed in the %literature.
The gauge theory partition function is given by the (refined) topological string partition function \cite{BCOV,Gopakumar:1998ii,Gopakumar:1998jq,Hollowood:2003cv} which can be calculated using the (refined) topological vertex \cite{Iqbal:2007ii}. The geometric engineering of the gauge theory from a CY3fold requires choosing a curve (or a chain of curves if there are more than one factor in the gauge group) whose area gives the gauge coupling (the worldsheet instantons wrapping this curve give space-time instantons) \cite{Katz:1996fh}. It may happen that there are more than one choice for this curve in which case the corresponding gauge theories are dual to each other, which is known as fiber-base duality \cite{Katz:1997eq}. The choice of different curves, corresponding to spacetime instantons, is reflected in the refined topological strings as the choice of preferred direction of the refined topological vertex. Different choices for the preferred directions lead to differently looking (but equivalent) expressions for the final partition functions each giving the instanton expansion for the corresponding gauge theory. In the case of the CY3fold obtained from the orbifold $\mathbb{Z}_{N}\times \mathbb{Z}_{M}$ we can choose the preferred direction corresponding to the curves coming from the resolution of $\mathbb{Z}_{N}$ action or the $\mathbb{Z}_{M}$ action leading to two different, yet equal, expressions for the partition functions.

%%%%
\item \emph{M-strings: M5/M2-branes with non-trivial geometry}\\
From an M-theoretic perspective, away from its conformal fixed point the six-dimensional $\mathcal{N}=(2,0)$ theory essentially describes interacting (almost) tensionless strings. With this picture in mind, in \cite{mstrings} it was proposed to consider a setup in which the M5 branes are slightly separated from one another and (a stack of) M2 branes are suspended in between them. From the M5 brane point of view, the latter appear as strings, which in \cite{mstrings} where dubbed \emph{M-strings} (see figure~\ref{Fig:Mstrings}).

\begin{figure}[h!]
\begin{center}
\epsfig{file=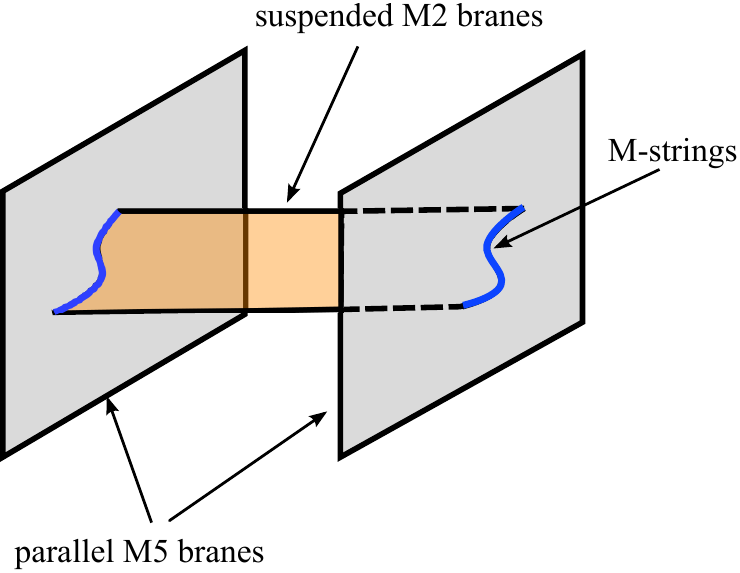,width=8cm}
\end{center}
\caption{M-strings appearing in a setup of (a stack of) M2 branes suspended between two parallel M5 branes.}
\label{Fig:Mstrings}
\end{figure}
Compactifying this setup to lower dimensions on $S^1_{\text{M5}}\times S^1_{\text{trans}}$, where the M5-branes wrap $S^1_{\text{M5}}$ while $S^1_{\text{trans}}$ is transverse, the only BPS states appearing are the Kaluza-Klein modes of the (now mass-deformed) $\mathcal{N}=(2,0)$ superconformal theory as well as the suspended M2 branes. The degeneracy of the BPS states (\emph{i.e.} the 5D gauge theory partition function) is captured by the supersymmetric partition function of the M-strings, \emph{i.e.} the (2,0) equivariant elliptic genus. Particularly the symmetries arising from  $S^1_{\text{M5}}\times S^1_{\text{trans}}$ lead to interesting dualities among different theories. Indeed, we can identify $S^1_{\text{M5}}\times S^1_{\text{trans}}$ with the torus on which the brane web was compactified. Thus it follows that the analog of the $\mathbb{Z}_{N}\times \mathbb{Z}_{M}$ orbifold action on the brane configuration with a single M5 brane on $S^{1}_{\text{M5}}$ and M2 branes wrapping $S^{1}_{\text{trans}}$ gives generalised M-string configuration corresponding to the mass deformed gauge theories we are considering. We will not discuss in detail these generalised M string configurations and refer the reader to \cite{BGKV} for details. However, we will show that the $(2,0)$ equivariant elliptic genus of the product of instanton moduli spaces gives the refined topological string partition function of the $\mathbb{Z}_{N}\times \mathbb{Z}_{M}$ orbifold CY3fold and hence the corresponding $(2,0)$ theory describes these generalised M-string configurations. To define the $(2,0)$ elliptic genus we need to specify a bundle on the target space \cite{Witten:1993yc}, which we show can directly be determined from the brane web picture.
\end{enumerate}
%%%%%%%%
Using these approaches, we write explicit expressions for the corresponding partition functions for generic values of $M$ and $N$. For simplicity, and since they are more tractable, we will discuss some low values of $(M,N)$ in more detail. One of these cases has generic $N$ but $M=1$, which in the M-theory picture corresponds to a single M5-brane with $N$ stacks (distinguished by their position in the transverse space) of M2-branes beginning and ending on it. The corresponding partition function exhibits very interesting modular properties, which we make more tangible by considering yet another computational approach. We link the gauge theory partition function to the equivariant elliptic genus $\chi_{\text{ell}}(\mathbb{C}^2)$ of $\mathbb{C}^2$ through a particular (singular) theta-transform, which is represented by the blue circle in figure~\figref{Fig:OverviewPartFct}. The equivariant elliptic genus is a generalisation of the usual elliptic genus (which would vanish in the case of $\mathbb{C}^2$) to non-compact manifolds, as introduced in \cite{Gritsenko:1999nm}. The theta transform of $\chi_{\text{ell}}(\mathbb{C}^2)$ can be written as a particular (torus-) integral with the kernel given by a Siegel-Narain theta function corresponding to a $(2,2)$ Narain lattice. Integrals of this type appear naturally in one-loop scattering amplitudes of string amplitudes. In fact, amplitudes of this type (also at higher loop orders) have been used in the past to obtain a direct world-sheet description of the topological string. This approach has mostly been used to compute higher genus partition functions or more general correlators for the unrefined topological string (with $\mathcal{N}=2$ \cite{Antoniadis:1993ze} as well as $\mathcal{N}=4$ \cite{Lerche:1999ju} supersymmetry). However, it has been proposed to study also the refined topological string along these lines \cite{Billo:2007va,Antoniadis:2013bja,Billo:2013fi,AHFNZ} (see also \cite{Antoniadis:2010iq}), which also extends beyond the perturbative level. Inspired by this idea, we show that the torus integral mentioned before can indeed be obtained from a very particular series of one-loop amplitudes in type II string theory compactified on $T^2$ with massive external legs. These amplitudes generalise a similar class of amplitudes studied in \cite{Hohenegger:2011us}, which are related to the (usual) elliptic genus of $K3$.

This paper is organised as follows: In section~\ref{Sect:N2Gauge}, we introduce the three dual settings which we use to compute gauge theory partition functions. In section~\ref{Sect:PartFct}, we explicitly compute partition functions for $M$ D5-branes and $N$ NS5 branes in type II string theory and extensively discuss their dualities. In section~\ref{Sect:TopString} we relate the gauge theory partition functions to the equivariant elliptic genus of $\mathbb{C}^2$, thereby making its modular properties more tangible. In section~\ref{Sect:EllGen} we discuss more properties of the equivariant elliptic genera and provide several interesting series expansions. In section~\ref{Sect:OneLoopStringAmp} we find a series of one-loop scattering amplitudes in type II string theory on $T^2$, which capture the equivariant elliptic genus of $K3$ and are related to the gauge theory partition functions. Finally section~\ref{Sec:Conclusions} contains our conclusions. There are several appendices, which contain additional mathematical definitions and conventions, as well as further useful material on some concepts used in the main body of this paper.\\[5pt]

\noindent
\textbf{Note added:} While finalising our manuscript, we learned of reference~\cite{BGKV} being in preparation and both papers appeared on the same day on the ArXiv. Ref.~\cite{BGKV} discusses generalised M-strings configurations corresponding to orbifolded geometries.

\section{${\cal N}=2^{*}$ theories: Brane webs, CY3folds and M-strings}\label{Sect:N2Gauge}

In this section we discuss various dual realisations of the five dimensional mass deformed gauge theories with eight supercharges. These theories arise from maximally supersymmetric (sixteen supercharges) gauge theories. As we mentioned above, there are three dual descriptions in terms of brane webs, elliptic CY3folds and M-strings. We now explain each of these pictures in detail.

\subsection{Brane webs}

\begin{figure}[h!]
  \centering
  % Requires \usepackage{graphicx}
  \includegraphics[width=10cm]{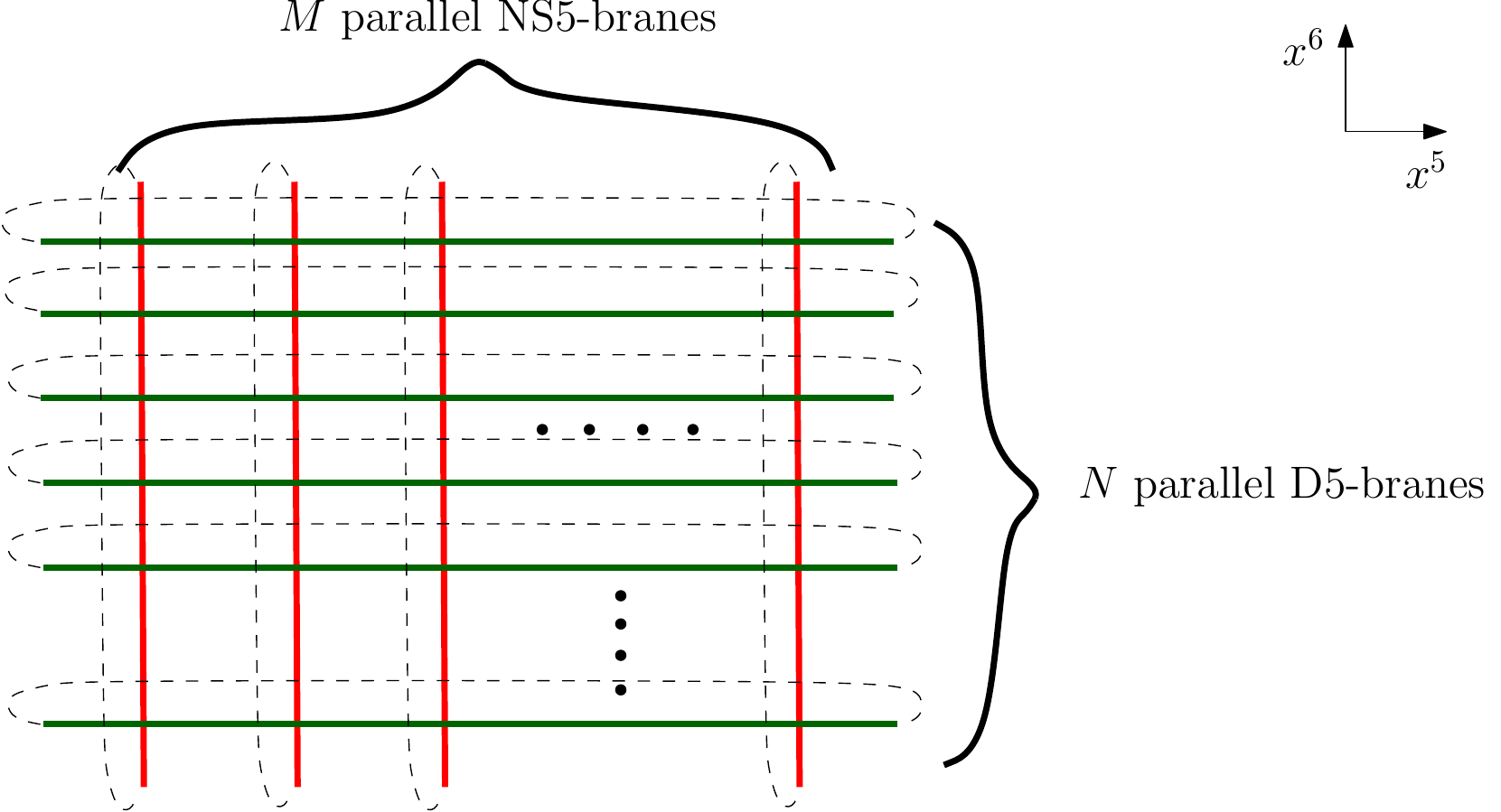}\\
  \caption{5-brane web with $N$ D5-branes (horizontal) and $M$ NS5-branes (vertical)}\label{NMweb}
\end{figure}

The theories we are interested in, have a simple description in terms of certain webs of $(p,q)$ 5-branes in type IIB string theory on $\mathbb{R}^{1,9}$ \cite{Aharony:1997bh}. Denoting the coordinates of the latter as $X^{A}$ for $A=0,1,\cdots, 9$, our setup is summarised in the following table:
\begin{center}
\parbox{14.5cm}{\begin{center}\begin{tabular}{|c||c|c|c|c|c||c|c||c|c|c|}\hline
\bf{brane} & $X^0$ & $X^1$ & $X^2$ & $X^3$ & $X^4$ & $X^5$ & $X^6$ & $X^7$ & $X^8$ & $X^9$  \\\hline\hline
D5-branes & $\bullet$ & $\bullet$ & $\bullet$ & $\bullet$ & $\bullet$ & $\bullet$ & --&&&\\\hline
NS5-branes & $\bullet$ & $\bullet$ & $\bullet$ & $\bullet$ & $\bullet$ & -- & $\bullet$&&&\\\hline
\end{tabular}\end{center}
${}$\\[-55pt]
\begin{align}
{}\hspace{2.6cm}\underbrace{\hspace{4.9cm}}_{\text{gauge theory}}\underbrace{\hspace{1.85cm}}_{(p,q)-\text{plane}}\underbrace{\hspace{3cm}}_{\text{transverse } \mathbb{R}^3}\nonumber
\end{align}}
\end{center}
The gauge theory lives on the common worldvolume of the 5-branes along  $X^{0,1,2,3,4}$. The $(p,q)$ brane web lives in the $X^{5,6}$ plane, along which the branes are oriented according to their $(p,q)$-charges, and encodes the details of the five-dimensional gauge theory~\cite{Aharony:1997bh}. In the following we consider theories on a $(p,q)$ web which consists of $N$ parallel D5-branes and $M$ parallel NS5-branes, as shown in \figref{NMweb}. This brane configuration allows us to compactify the $(p,q)$-plane ($X^{5,6}$ plane) to a $T^2$ where we denote the radius of the $X^5$ and $X^6$ circle by $R_{5}$ and $R_{6}$ respectively. Note that compactification of the $(p,q)$ plane is in general not possible for an arbitrary brane web.
The $SL(2,\mathbb{Z})$ symmetry of type IIB string theory maps the $M$ NS5-branes and $N$ D5-branes to $N$ NS5-branes and $M$ D5-branes. Thus the gauge theory corresponding to the brane configuration in \figref{NMweb} is dual to the gauge theory on the brane web with $N$ and $M$ interchanged.

\begin{figure}[h]
  \centering
  % Requires \usepackage{graphicx}
  \includegraphics[width=12cm]{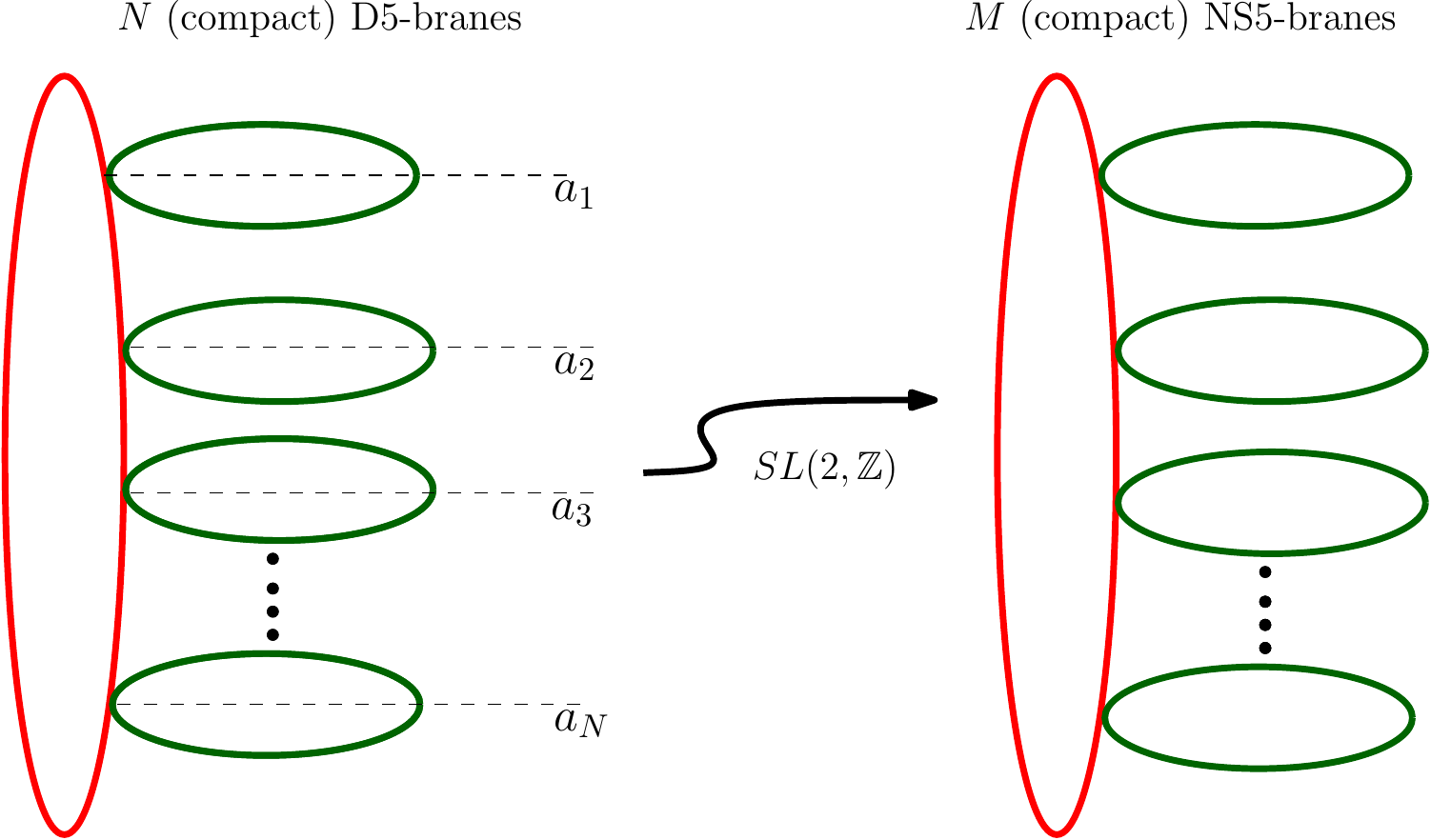}\\
  \caption{(a) Web giving rise to 5D maximally supersymmetric $U(N)$ theory on the D5-branes, (b) the web giving rise to $U(1)^{N}$ theory on the D5-branes.}\label{sl2z}
\end{figure}

The case $M=1$ (shown in \figref{sl2z}(a)) was discussed in \cite{mstrings} and the dual brane configuration obtained by $SL(2,\mathbb{Z})$ symmetry of the type IIB string theory is shown in \figref{sl2z}(b).

The $\{a_{i}\,|\,i=1,2,\cdots, N\}$ label the position of the D5-branes on the transverse circle. When the D5-branes are at the same point on the circle the gauge theory in $\mathbb{R}^{1,4}$, given by $X^{0,1,2,3,4}$, has $U(N)$ gauge group. Separating the branes on the circle breaks $U(N)$ to $U(1)^N$ with position of the branes $a_{i}$ being the Coulomb branch parameters. The dual gauge theory which arises on the D5-branes in the dual brane configuration shown in \figref{sl2z}(b) is also a $U(1)^{N}$ gauge theory such that,
\bea
a_{i}-a_{i+1}=\frac{1}{g_{i}^2}\,,
\eea
where $g_{i}$ the gauge coupling of the $i$-th $U(1)$ factor.

\begin{figure}[h]
  \centering
  % Requires \usepackage{graphicx}
  \includegraphics[width=12cm]{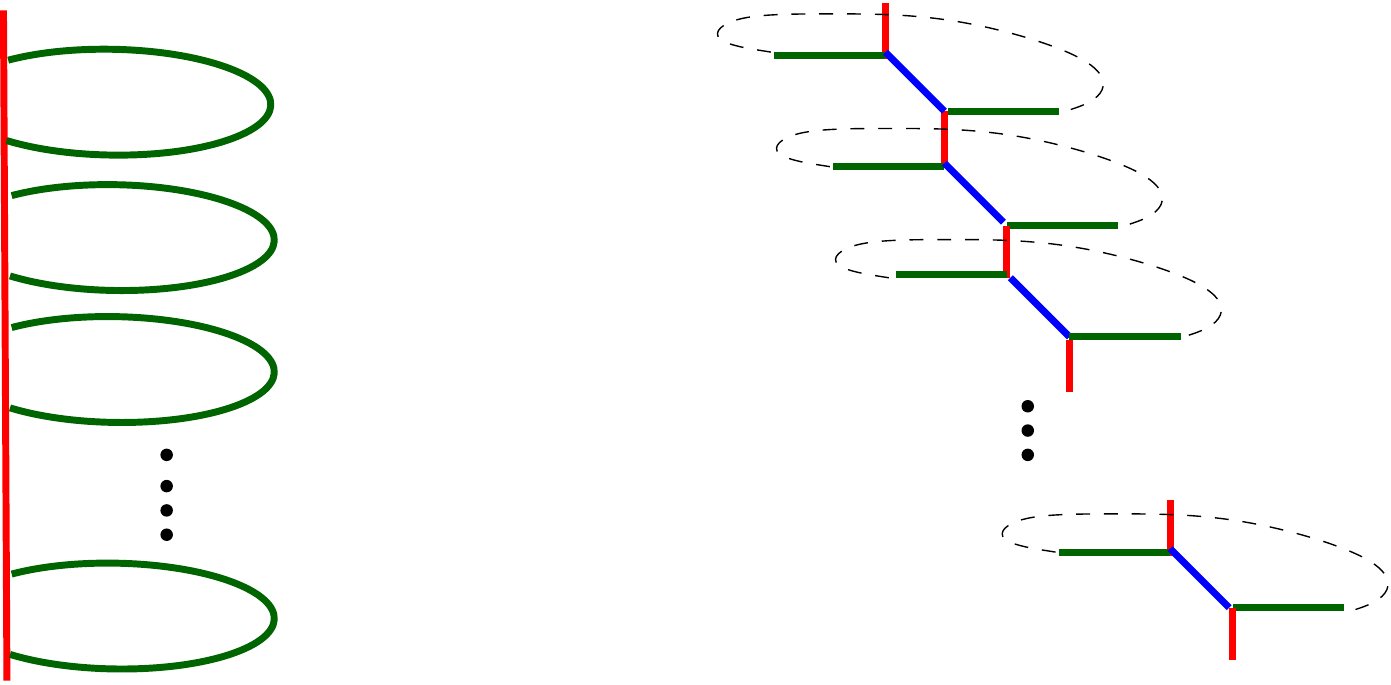}\\
  \caption{The mass to the adjoint hypermultiplet is given by a deformation of the brane web. The mass parameter $m$ corresponds to the length of the blue interval.}\label{adjointmass}
\end{figure}

The five dimensional theory on a generic $(p,q)$ brane web has ${\cal N}=1$ supersymmetry with eight supercharges. However, the brane webs shown in \figref{NMweb} have ${\cal N}=2$ supersymmetry, with sixteen supercharges, on the common worldvolume. The enhancement is essentially due to massless bifundamental hypermultiplets (in the ${\cal N}=1$ language) coming from strings stretched between the D5-branes separated by the NS5-branes. For the case $M=1$ these strings give a hypermultiplet in the adjoint representation since it is the same stack of D5-branes on the either side of the NS5-brane. This massless adjoint hypermultiplet can be given mass to break ${\cal N}=2$ supersymmetry to ${\cal N}=1$ supersymmetry, by splitting the stack of D5-branes on the NS5-branes. This deforms the 5-brane web and $(p,q)$ charge conservation generates $(1,1)$ 5-branes in the web as shown in \figref{adjointmass}.

%Another configuration we study in detail below, corresponds to having a single D5-brane ($N=1$) compactified on a circle and multiple NS5-branes ($M\geq 1$) %as shown in \figref{oned5} below. Notice the difference in geometry as with respect to~\figref{sl2z}.
%\begin{figure}[h]
%  \centering
  % Requires \usepackage{graphicx}
%  \includegraphics[width=7cm]{oned5.pdf}\\
%  \caption{A single D5-brane wrapped on a circle with $M$ NS5-branes transverse to the circle.}\label{oned5}
%\end{figure}

For the brane web with $N$ D5-branes and $M>1$ NS5-branes, with both D5-branes and NS5-branes compactified on two different circles, we consider deformations by giving masses to the bifundamental hypermultiplets. However, we limit ourselves to a special case where all the hupermultiplets get the same mass $m$. The gauge theory on the D5-branes has:
\begin{align}
&\mbox{Gauge group}: \,\,G= U(1)\times\underbrace{SU(N)\times SU(N)\times \cdots\times SU(N)}_{M\text{ factors}}\nn\\
&\mbox{Matter}: \mbox{hypermultiplets in}\,\, \oplus_{a=1}^{M}\Big[({\bf N}_{a},\overline{\bf N}_{a+1})\oplus (\overline{\bf N}_{a},{\bf N_{a+1}})\Big]\,,\label{pqTheory}
\end{align}
where ${\bf N}_{a}$ is the $N$-dimensional fundamental representation of $a$-th $SU(N)$ and $\overline{\bf N}_{a}$ is its complex conjugate representation.

Using $SL(2,\mathbb{Z})$ symmetry of the type IIB string theory we can map the above configuration of $N$ D5-branes and $M$ NS5-branes to a configuration of $M$ D5-branes and $N$ NS5-branes. In this case the theory on the D5-branes has:
\begin{align}
&\mbox{Gauge group}: \,\,G= U(1)\times\underbrace{SU(M)\times SU(M)\times \cdots \times SU(M)}_{N \text{ factors}}\nn\\
&\mbox{Matter}: \mbox{hypermultiplets in}\,\, \oplus_{a=1}^{N}\Big[({\bf M}_{a},\overline{\bf M}_{a+1})\oplus (\overline{\bf M}_{a},{\bf M_{a+1}})\Big]\,,\label{pqDual}
\end{align}
where ${\bf M}_{a}$ is the $M$-dimensional fundamental representation of $a$-th $SU(M)$ and $\overline{\bf M}_{a}$ is its complex conjugate representation.

Per construction, these two gauge theories (equ.(\ref{pqTheory}) and equ.(\ref{pqDual})), are dual to each other. The partition function of these two theories can be calculated using Nekrasov's instanton calculus and duality suggests that after appropriately identifying the parameters of the two gauge theories the partition functions should be equal.

We denote by $\{a_{i}\,|\,i=1,2,\cdots,N\}$ and $\{b_{j}\,|\,j=1,2,\cdots,M\}$ the positions of the $N$ D5-branes on the $X^{6}$ circle and $M$ NS5-branes on the $X^5$ circle respectively. Then the following table summarizes the relation between the parameters of the dual theories:

\vskip 0.6cm
{\begin{center}
\begin{tabular}{|c|c|}\hline
  $N$ D5-branes/$M$ NS5-branes                          & $M$ D5-branes/$N$ NS5-branes \\\hline\hline
  Coulomb parameters:                                   &  Gauge couplings:\\
  $a_{i}-a_{i+1}\,,i=1,2,\cdots,N$                      & $\frac{1}{g_{i}^2}\,,i=1,2,\cdots, N$ \\[4pt]\hline
  Gauge couplings:  & Coulomb parameters:\\
         $\frac{1}{g_{j}^2}\,,j=1,2,\cdots,M$                                               & $b_{j}-b_{j+1}\,,j=1,2,\cdots M$ \\[6pt]\hline
\end{tabular}
\end{center}}
\vskip 0.6cm

\subsection{Calabi-Yau Threefolds}\label{Sect:OverviewSectCY3}

In this section we will review the construction of CY3folds dual to the 5-brane webs discussed in the previous section. The five dimensional ${\cal N}=1$ supersymmetric theory on the 5-brane web has a dual description in terms of M-theory compactification on a CY3fold. The geometry of the CY3fold is encoded in the 5-brane web.

\begin{figure}[h]
  \centering
  % Requires \usepackage{graphicx}
  \includegraphics[width=7cm]{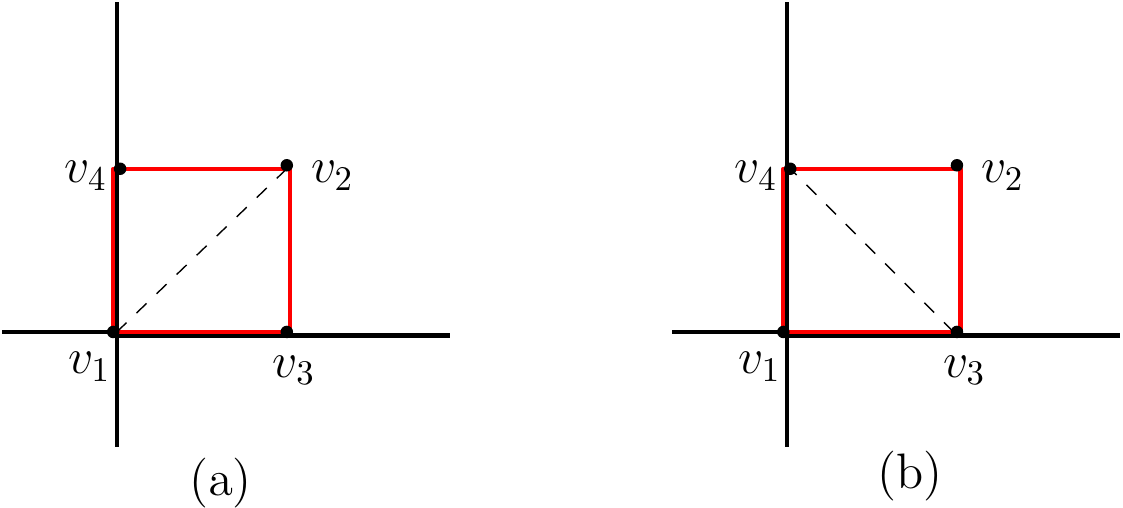}\\
  \caption{The Newton polygon of the conifold with triangulations shown.}\label{NewtonC}
\end{figure}

We will be concerned with toric non-compact CY3folds, which are constructed by gluing together $\mathbb{C}^3$ patches. The data of how to glue the patches is encoded in a geometric object called the Newton polygon. Consider the case of the conifold given by the equation,
\begin{align}
&x_{1}x_{2}-x_{3}x_{4}=0\,,&&x_{1,2,3,4}\in \mathbb{C}\,.
\end{align}
Solving the equation we can express $x_{4}=\frac{x_{1}x_{2}}{x_{3}}$ which implies that the conifold can be constructed from two $\mathbb{C}^3$ patches. Let us denote the coordinate ring of the two $\mathbb{C}^3$ patches by $U_{1}$ and $U_2$ then,
\begin{align}
&U_{1}=\mathbb{C}[x_{1},x_{2},x_{3}]\,,&&U_{2}=\mathbb{C}[x_{1},x_{2},x_{4}]=\mathbb{C}[x_{1},x_{2},\frac{x_{1}x_{2}}{x_{3}}]\,.
\label{CR1}
\end{align}
To each $x_{i}$ we associate a vector $v_{i}$ such that the relation $x_{1}x_{2}=x_{3}x_{4}$ is now encoded in the relation between the vectors $v_{1}+v_{2}-v_{3}-v_{4}=0$. Since the conifold is a three dimensional manifold, these vectors are taken to lie in the integer lattice $\mathbb{Z}^3 \subseteq \mathbb{R}^3$. The manifold being CY3fold imposes the extra condition that the points representing the vectors lie in a plane. We can take these vectors to be
\begin{align}
&v_{1}=(0,0,1)\,,&&v_{2}=(1,1,1)\,,&&v_{3}=(1,0,1)\,,&&v_{4}=(0,1,1)\,.
\end{align}
The above four vectors give a polygon in the $z=1$ plane shown in \figref{NewtonC}.

The triangulation of the Newton polygon given in \figref{NewtonC}(a) corresponds to choosing the coordinate rings of the two $\mathbb{C}^3$ patches as given in Eq.(\ref{CR1}). The coordinate rings of the two $\mathbb{C}^3$ patches for the triangulation given in \figref{NewtonC}(b) are,
\begin{align}
&\tilde{U}_{1}=\mathbb{C}[x_{1},x_{3},x_{4}]\,,&&\tilde{U}_{2}=\mathbb{C}[x_{2},x_{3},x_{4}]=\mathbb{C}[\frac{x_{3}x_{4}}{x_{1}},x_{3},x_{4}]\,.
\end{align}

\begin{figure}[h]
  \centering
  % Requires \usepackage{graphicx}
  \includegraphics[width=7cm]{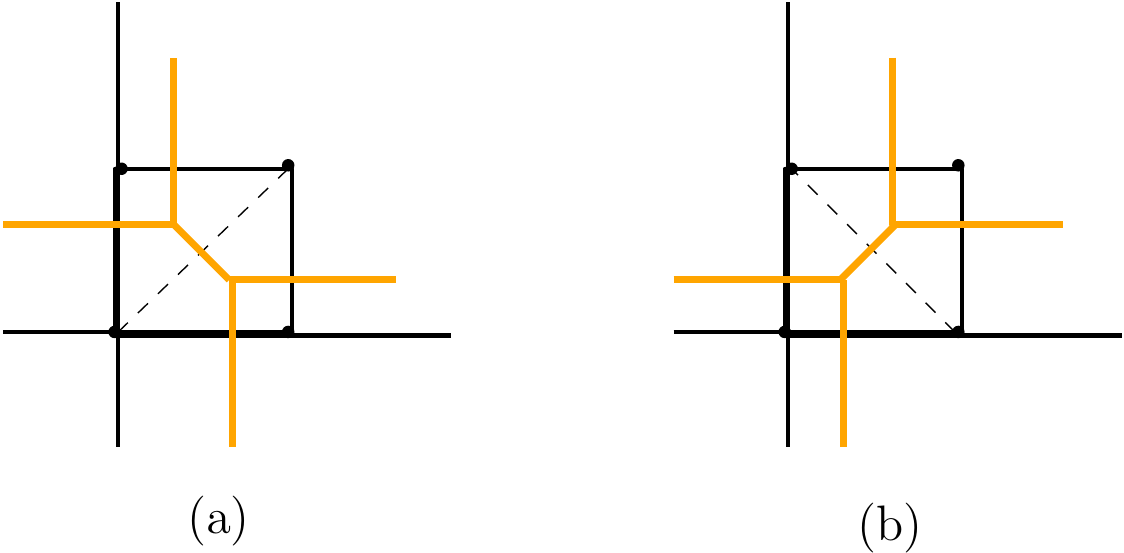}\\
  \caption{The Newton polygon and the brane web.}\label{NewtonC2}
\end{figure}

The Newton polygon and the triangulations are directly related to the brane webs. The brane web is just the dual of the Newton polygon i.e., the edges of the Newton polygon are orthogonal to the edges of the brane web and for each face in the Newton polygon we get a vertex in the brane web. The two brane webs corresponding to the two triangulations of the Newton polygon are shown in \figref{NewtonC2}.

\begin{figure}[h]
  \centering
  % Requires \usepackage{graphicx}
  \includegraphics[width=7cm]{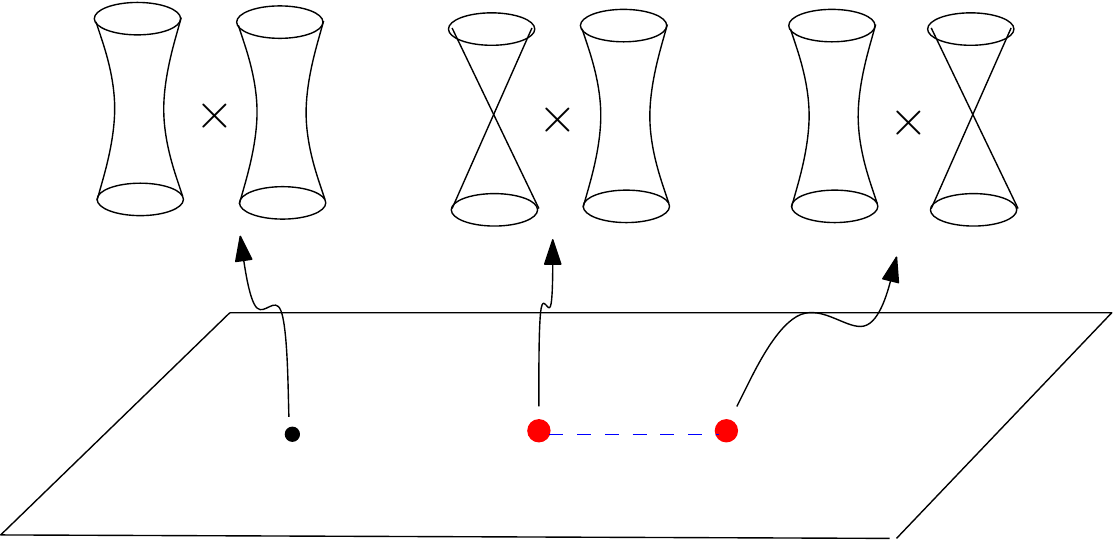}\\
  \caption{The geometry of deformed conifold. The dotted line indicates an $S^3$
formed by the line joining $z = 0$ with $z =\varepsilon$ together with the two cigars coming
from $\mathbb{C}^{\times}$ fibers as they go from $z = 0$ to $z =\varepsilon$.}\label{fibration}
\end{figure}

The conifold is a singular variety and there are two ways to remove the singularity, we can either deform the equation of the conifold (deformed conifold) or we can resolve the singularity (resolved conifold). The deformed conifold is given by the equation,
\bea
x_{1}x_{2}-x_{3}x_{4}=\varepsilon\,.
\eea
To understand the geometry of this deformation we write the above equation as
\begin{align}
&x_{1}x_{2}=z\,,&&x_{3}x_{4}=z-\varepsilon\,,&&z\in \mathbb{C}\,.
\end{align}
Over the z-plane these two equations define a fibration. The equation $x_{1}x_{2}=z$ gives a $\mathbb{C}^{\times}$ fiber for each $z\neq 0$, which degenerates to $\mathbb{C}$ over $z=0$. Similarly the equation $x_{3}x_{4}=z-\varepsilon$ gives a $\mathbb{C}^{\times}$ fiber for each $z\neq \varepsilon$ with $\mathbb{C}^{\times}$ degenerating to $\mathbb{C}$ over $z=\varepsilon$. The geometry of this fibration is shown in \figref{fibration}.

The deformation produces an $S^3$ in the geometry whose projection on the z-plane gives a path from $z=0$ to $z=\varepsilon$. The deformed conifold is the total space of the cotangent bundle on $S^3$. This can be seen by defining new coordinates $z_{1,2,3,4}$ such that $x_{1}=z_{1}+iz_{2},x_{2}=z_{1}-iz_{2},x_{3}=-(z_{3}+iz_{4}),x_{4}=z_{3}-iz_{4}$ and the equation of the deformed conifold is given by
\begin{align}
z_{1}^2+z_{2}^2+z_{3}^2+z_{4}^2=\varepsilon\,.\label{S3deform}
\end{align}
The real and imaginary part of (\ref{S3deform}) gives (taking $z_{i}=y_{i}+i\,u_{i}$, $y_{i},u_{i}\in \mathbb{R}$, and assuming $\varepsilon$ to be real otherwise we can absorb its phase by redefining $z_{i}$),
\begin{align}
&\sum_{i=1}^{4}y_{i}^2=\varepsilon+\sum_{i=1}^{4}u_{i}^2\,,&&\sum_{i=1}^{4}y_{i}\,u_{i}=0
\end{align}
$u_{i}=0$ give the zero section of the cotangent bundle which is the $S^3$ with area proportional to $\varepsilon$. For each point on $S^3$ the second equation gives the cotangent plane hence the total space is that of the cotangent bundle.

The resolution of the singularity is obtained by blowing up the $\mathbb{C}^4$ in which the conifold is the hypersurface $x_{1}x_{2}-x_{3}x_{4}=0$. The exceptional locus of the blowup of $\mathbb{C}^4$ is a $\mathbb{P}^3$. The intersection of the conifold and the latter is a quadric $\mathbb{P}^1\times \mathbb{P}^1$ in $\mathbb{P}^3$ given by $\sum_{i=1}^{4}z_{i}^2=0$ with $[z_{1},z_{2},z_{3},z_{4}]\in \mathbb{P}^3$. The exceptional divisor of the conifold after the resolution of the singularity is a $\mathbb{P}^1$ which is one of the $\mathbb{P}^1$ of the $\mathbb{P}^{1}\times \mathbb{P}^1$. The two choices for the exceptional $\mathbb{P}^1$ corresponds to the two triangulations of the Newton polygon and are related by flop transition. The exceptional $\mathbb{P}^1$ is dual to the $(1,1)$ or the $(-1,1)$ brane in the brane web, depending on the triangulation, with the size of the $\mathbb{P}^1$ related to the length of the $(1,1)$ (or $(-1,1)$) brane in the $(p,q)$-plane \cite{Aharony:1997bh}. The resolved conifold is the total space of the bundle ${\cal O}(-1)\mapsto {\cal O}(-1)\mapsto \mathbb{P}^{1}$. $(x_{1},x_{3})$ and $(x_{2},x_{4})$ gives the two line bundles on $\mathbb{P}^1$ such that
\begin{align}
&x_{1}=\xi\,x_{3}\,,&&x_{2}=\xi^{-1}\,x_{4}\,,&&\xi\in \mathbb{P}^1\,.
\end{align}
Choosing instead $(x_{1},x_{4})$ and $(x_{2},x_{3})$ related by
\begin{align}
&x_{1}=\tilde{\xi}\,x_{4}\,,&&x_{2}=\tilde{\xi}^{-1}\,x_{3}\,,&&\tilde{\xi}\in \mathbb{P}^1\,,
\end{align}
gives the flopped version of the resolved conifold.

\begin{figure}[h]
  \centering
  % Requires \usepackage{graphicx}
  \includegraphics[width=7cm]{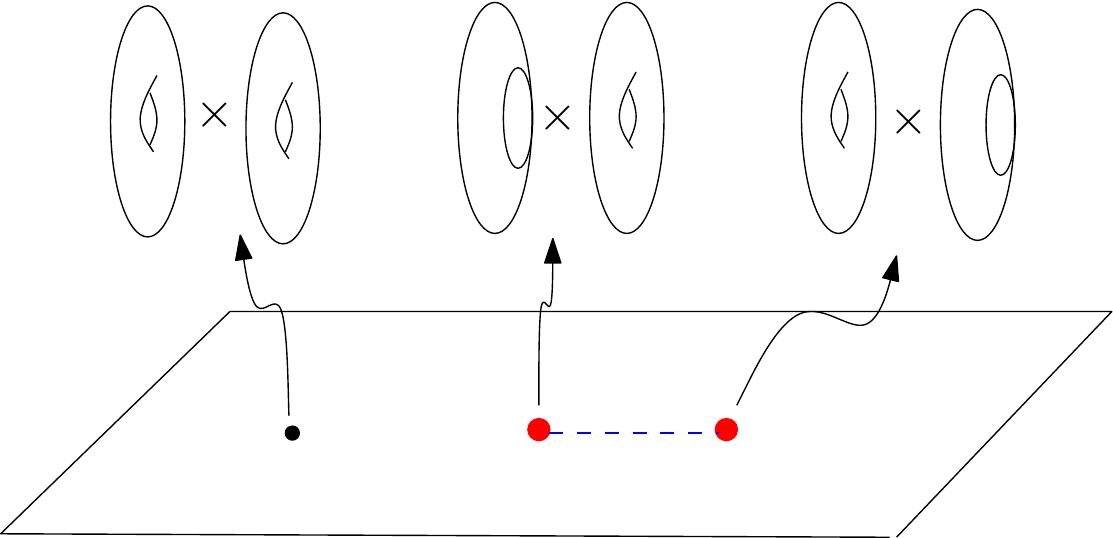}\\
  \caption{The geometry of partially compactified conifold as $T^2\times T^2$ fibration over the $z$-plane.}\label{compConifold}
\end{figure}

Now that we have the web dual to the conifold we can try to compactify the web by putting it on a $T^2$ and try to determine the geometry corresponding to it. Notice that the conifold is given by a $\mathbb{C}^{\times}\times \mathbb{C}^{\times}$ fibration over the $z$-plane. The geometry which corresponds to the compactified web is obtained by compactifying each of the two $\mathbb{C}^{\times}$ fibers to a $T^2$ \cite{Hollowood:2003cv} as shown in \figref{compConifold}. We will denote this partially compactified conifold by $X_{1,1}$ since the dual web has one D5-brane and one NS5-brane and in anticipation of generalizing to the case of multiple D5/NS5-branes.

The K\"ahler parameters of the two elliptic curves are related to the radius of the circles on which the D5-brane and the NS5-branes are wrapped,
\begin{align}
T=i\,R_{5}\,,&&U=i\,R_6\,,%(\mbox{$X^5$ is the D5-brane circle}) \,\,\,\rho=i\,R_{6}\,,(\mbox{$X^6$ is the NS5-brane circle})\,.
\end{align}
where $R_5$ is the circle in the $X^5$ direction (on which the D5-branes are wrapped) while $R_6$ is the radius of the circle in the $X^6$ direction (on which the NS5-branes are wrapped). The elliptic CY3fold $X_{1,1}$ has three K\"ahler parameters $\tau,\rho$ and $T$. $\tau$ and $\rho$ correspond to the elliptic fibers and $T$ is corresponds to the exceptional $\mathbb{P}^1$ of the resolved conifold. The gauge theory engineered from this CY3fold via M-theory the parameter corresponds to the mass $m$ of the hypermultiplets.

\begin{figure}[h]
  \centering
  % Requires \usepackage{graphicx}
  \includegraphics[width=3.5in]{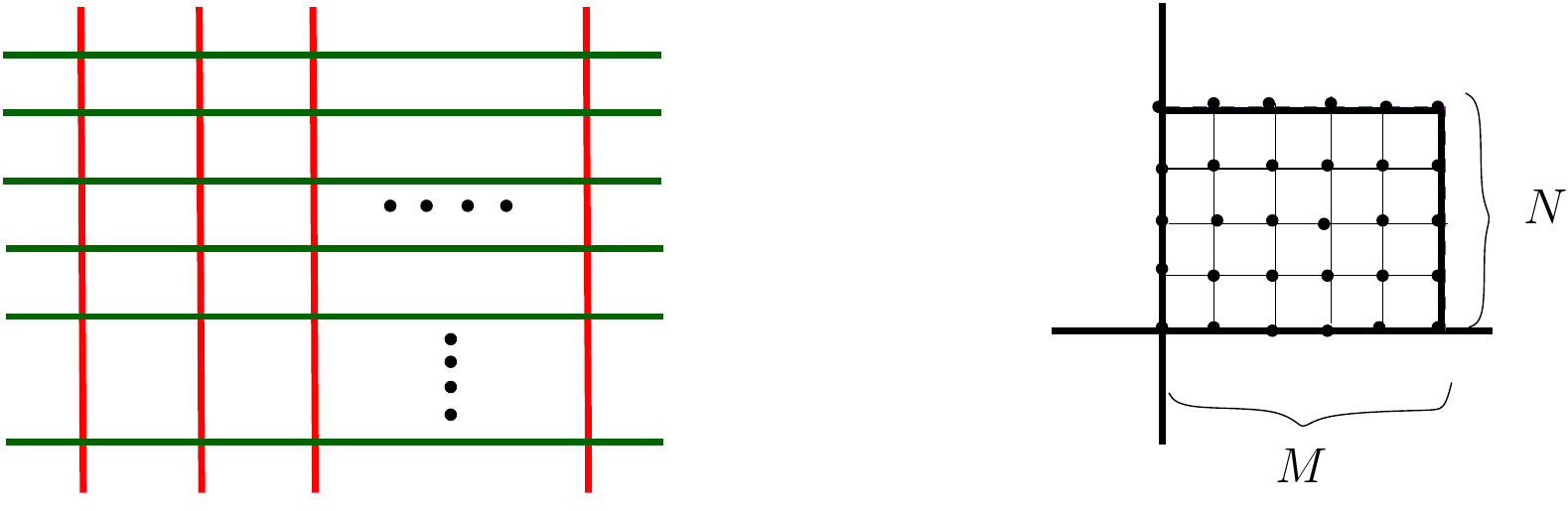}\\
  \caption{The $(N,M)$ brane web and the corresponding Newton polygon.}\label{newton}
\end{figure}

We will denote by $X_{N,M}$ the elliptic CY3fold dual to the brane web with $N$ D5-branes and $M$ NS5-branes wrapped on the $X^5$ and $X^6$ circles respectively. The brane web can be used to construct the Newton polygon of the dual toric geometry which is shown in \figref{newton}.
$X_{N,M}$ is a toric non-compact local CY3fold which is an $\mathbb{Z}_{N}\times \mathbb{Z}_{M}$ orbifold of $X_{1,1}$. Before partial compactification it is the $\mathbb{Z}_{N}\times \mathbb{Z}_{M}$ orbifold of the conifold given by \cite{Aganagic:1999fe}:
\begin{align}
&x_{1}\,x_{2}=z^{N}\,,&&x_{3}\,x_{4}=z^{M}\,.
\end{align}
From the above we obtain $X_{N,M}$ by partially compactifying the two $\mathbb{C}^{\times}$ fibers to elliptic curves as before. Thus the brane configuration of multiple D5-branes and NS5-branes on $S^1\times S^1$ along $X^{5,6}$ respectively is dual to an elliptic CY3fold which is the orbifold of $X_{1,1}$. If we take $\rho\mapsto i \infty$ then $X_{1,1}$ is $A_{0}$ fibration over an elliptic curve which was discussed in \cite{mstrings}.

From the geometry it is clear that $X_{N,M}$ is topologically the same as $X_{M,N}$. Thus the topological string partition function associated with $X_{N,M}$ should be the same as the one associated with $X_{M,N}$ if we interchange $\tau$ and $\rho$. We will see in section~\ref{Sect:Vertex} that this is indeed the case.

\subsection{M-strings}

The brane web theories we described in the last couple of sections also have a description in terms of M5/M2 brane configurations which are a generalisation of the configuration studied in \cite{mstrings}.

\begin{figure}[h]
  \centering
  % Requires \usepackage{graphicx}
  \includegraphics[width=14cm]{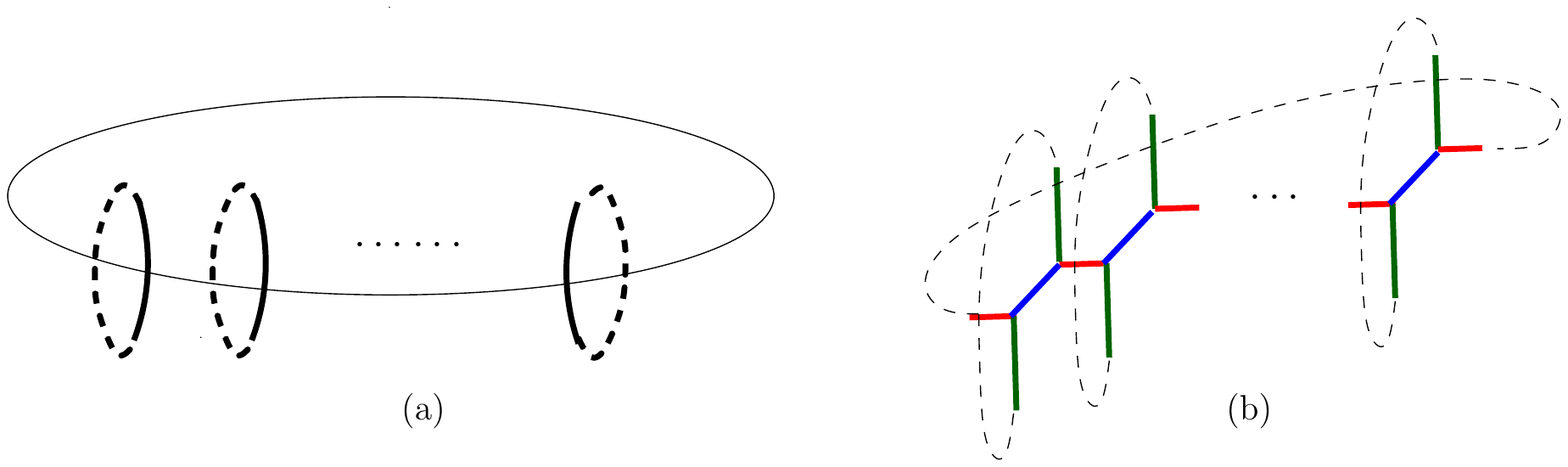}\\
  \caption{(a) Multiple M5-branes with single stack of M2-branes in between different M5-branes and (b) corresponding brane web. }\label{mstrings1}
\end{figure}

Consider $M$ parallel M5-branes, whose worldvolume theory is the six dimensional $(2,0)$ $A_{M-1}$ theory \cite{Witten:1995zh}. We denote the coordinates of the $\mathbb{R}^{1,10}$ spactime of M-theory by $Y^{I},I=0,1,\cdots,10$ such that the M5-branes fill $\mathbb{R}^{1,5}$ parameterized by $Y^{0,1,2,3,4,5}$ and wrap the circle parameterized by $Y^1$. All M5-branes are at the origin of $\mathbb{R}^{5}_{\perp}$ which has coordinates $Y^{6,7,8,9,10}$. We can separate these coincident M5-branes along the $Y^6$ direction so that their position in $Y^6$ is given by $b_{j}, j=1,2,\cdots,M$. For multiple M5-branes we can now have M2-branes stretched between the $i$-th and the $j$-th M5-branes such that the worldvolume of the M2-branes is given by $Y^0,Y^1$ and $b_{i}\leq Y^6\leq b_j$. This is similar to the configuration studied in \cite{mstrings}, where the $Y^6$ direction was not compactified to a circle. For the dual brane web with $N=1$ see \figref{mstrings1}(b).

The brane web with $N>1$ seems must be dual to configuration of M5-branes and M2-branes in which we have $N$ stacks of M2-branes at different points on the M5-branes. However, inspecting the M5/M2-brane worldvolume coordinates shown in the table below there is no possibility to separate the M2-branes to make different stacks, apart from separating them in the transverse $\mathbb{R}^{4}_{\perp}$ where they are all located at the origin.
\begin{center}
\parbox{14.5cm}{\begin{center}\begin{tabular}{|c||c|c|c|c|c||c|c||c|c|c|c|}\hline
\bf{brane} & $Y^0$ & $Y^1$ & $Y^2$ & $Y^3$ & $Y^4$ & $Y^5$ & $Y^6$ & $Y^7$ & $Y^8$ & $Y^9$ & $Y^{10}$ \\\hline\hline
M5-branes & $\bullet$ & $\bullet$ & $\bullet$ & $\bullet$ & $\bullet$ & $\bullet$ & &&&&\\\hline
M2-branes & $\bullet$ & $\bullet$ &           &           &           & & $\bullet$&&&&\\\hline
\end{tabular}\end{center}
${}$\\[-55pt]
\begin{align}
{}\hspace{2.2cm}\underbrace{\hspace{1.8cm}}_{\text{M-string}}\underbrace{\hspace{3.7cm}}_{\mathbb{R}^{4}_{\|}}\underbrace{\hspace{0.9cm}}_{S^{1}}\,\underbrace{\hspace{3.8cm}}_{\text{transverse } \mathbb{R}^4}\nonumber
\end{align}}
\end{center}
We refer the reader to \cite{BGKV} for a detailed discussion of this possibility which preserves  supersymmetry and the mass deformation of the theory, by orbifolding the transverse space. The fact that the transverse space has non-trivial geometry also follows if one T-dualizes the NS5-braces to convert them to pure geometry which is Taub-NUT.

\begin{figure}[h]
  \centering
  % Requires \usepackage{graphicx}
  \includegraphics[width=14cm]{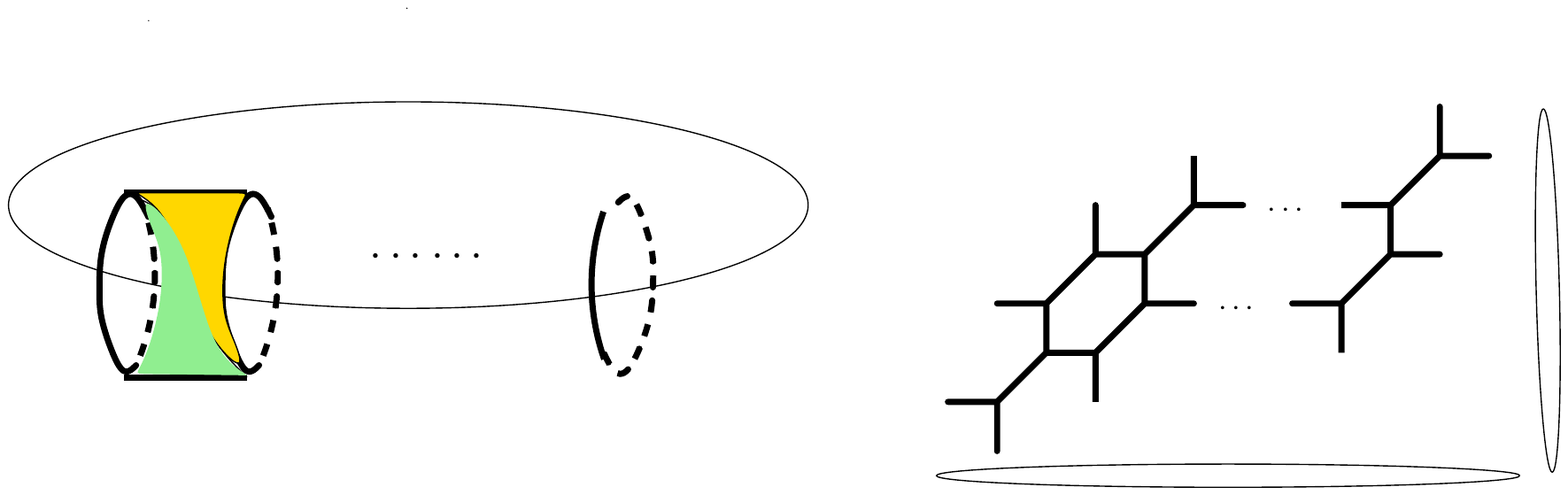}\\
  \caption{(a) Two stacks of M2-branes (shown in different colors) between the M5-branes and (b) the corresponding brane web. }\label{mstrings2}
\end{figure}

As a consequence of this modification of the transverse geometry we can have stacks of M2-branes separated from each other in the transverse space. The case of $2$ stacks of M2-branes along with the dual brane web is shown in \figref{mstrings2}.

This separation of the stacks of M2-branes in the dual brane web picture is given by the distance between the D5-branes. These stacks of M2-branes ending on the M5-branes give rise to M-strings in the worldvolume of the M5-branes which are points in $\mathbb{R}^{4}_{||}$ inside the M5-branes. Since we have different stacks of M2-branes these points can be thought of as colored points. For the case of $k$ stacks of M2-branes, with $n_i$ number of M2-branes in the $i$-th stack, it would seem that the moduli space should be,
\bea
\mbox{Hilb}^{n_{1}}[\mathbb{C}^2]\times \mbox{Hilb}^{n_{2}}[\mathbb{C}^2]\times \cdots \mbox{Hilb}^{n_{k}}[\mathbb{C}^2]\,.
\eea
However, it turns out to be $M(k,n)$, the $SU(k)$ instanton moduli space of charge $n=\sum_{i=1}^{k}n_{i}$. We show this explicitly in section 3 when we calculate the $(2,0)$ elliptic genus of the M-strings. The distance between the stack of M2-branes get related to the Coulomb branch parameters breaking $SU(k) \mapsto U(1)^{k-1}$ or equivalently the equivariant parameters of the $U(1)^{k-1}$ action on the $M(k,n)$.

%%%%%%%%%%%%%%%%%%%%%%%%%%
\section{Partition functions}\label{Sect:PartFct}
In this section we calculate the partition functions associated with various dual descriptions of the brane web theory. We give expressions for the partition functions for generic $(M,N)$ and discuss the cases for low values as examples in more detail. More specifically, we begin by briefly outlining the notation and some of the conventions we use in section~\ref{Sect:Conventions}. We discuss the topological vertex computation in section~\ref{Sect:Vertex}, the elliptic genus of the M-strings setup in section~\ref{Sect:Mstrings} and the instanton calculus in section~\ref{Sect:Instanton}. In the latter case, we will see that the $(2,0)$ theory on the M-string worldsheet can be determined from the brane web i.e., both the target space and the bundle on the target space required for the $(2,0)$ theory can be determined by following some simple rules from the 5-brane web. Also we will see that the supersymmetry on the M-string worldsheet is generically $\mathcal{N}=(2,0)$, however, in the case of a single NS5-brane an enhancement of supersymmetry takes place to $\mathcal{N}=(2,2)$.
%%%%%%%%%%%%%%%%%%%%%%%%%%
\subsection{Notation and Conventions}\label{Sect:Conventions}
In this section we briefly review our notation and the conventions we use in the following. We follow closely the notation of \cite{macdonald} which is also the source of most of the identities involving Schur functions that we will use.  Many of the expressions we write are given as sums over partitions which we will denote with Greek letters. Indeed, we call a set of positive integers $\mu_1\geq \mu_2 \geq\ldots\geq \mu_{\ell(\mu)}>0$ a partition $\mu$ of $m\in\mathbb{N}$ if $\sum_{i=1}^{\ell}\mu_i=m$. We denote by $\ell(\mu)$ the length of $\mu$ and define
\begin{align}
&|\mu|=\sum_{i=1}^{\ell(\mu)}\mu_i\,,&&||\mu||^2=\sum_{i=1}^{\ell(\mu)}\mu_i^2\,.\label{Norms}
\end{align}
Partitions can naturally be represented in terms of \emph{diagrams}, \emph{i.e.} arrays of left-aligned boxes with $\mu_i$ boxes in the $i$th row. We will denote by $\prod_{(i,j)\in \mu}$ the product over all boxes in the diagram of $\mu$ i.e.,
\bea
\prod_{(i,j)\in \mu}f(i,j)= \prod_{i=1}^{\ell(\mu)}\,\prod_{j=1}^{\mu_{i}}f(i,j)\,.
\eea
The conjugate partition $\mu^t$ corresponds to the transposed diagram such that $\mu^{t}_{i}=\#\{j\,|\,\mu_{j}\geq i\}$. Similar to (\ref{Norms}) we define
\begin{align}
||\mu^t||^2=\sum_{i=1}^{\ell(\mu^t)}(\mu_i^t)^2\,.
\end{align}
The empty partition will be denoted $\emptyset$.

Given partitions $\mu$ of $m$ and $\nu$ of $n$ with $l(\mu)\geq l(\nu)$, $m\geq n$ and $\mu_i\geq \nu_i$, we define a \emph{skew partition} $\mu/\nu$ through the diagram consisting of all boxes which are in $\mu$, but not in $\nu$. Given a set of variables ${\bf x}=(x_1,x_2,\ldots)$ we can define the \emph{skew Schur function} labelled by the skew partition $\mu/\nu$ as
\begin{align}
s_{\mu/\nu}({\bf x})=\sum_{\lambda}N^{\mu}_{\nu\lambda}\,s_{\lambda}({\bf x})\,,\label{Schur}
\end{align}
where $N^{\mu}_{\nu\lambda}$ are the Littlewood-Richardson coefficients and the sum is over all partitions.

Furthermore, for explicit computations our notation is as follows: The refined topological vertex which we use in section~\ref{Sect:Vertex} is given by
\bea\nn
C_{\lambda\mu\nu}(t,q)&=&q^{\frac{||\mu||^2}{2}}t^{-\frac{||\mu^t||^2}{2}}\,q^{\frac{||\nu||^2}{2}}\widetilde{Z}_{\nu}(t,q)\sum_{\eta}\Big(\frac{q}{t}\Big)^{\frac{|\eta|+|\lambda|-|\mu|}{2}}
s_{\lambda^t/\eta}(t^{-\rho}q^{-\nu})s_{\mu/\eta}(q^{-\rho}t^{-\nu^t})\,,\\\nn
\widetilde{Z}_{\nu}(t,q)&=&\prod_{(i,j)\in \nu}\Big(1-t^{\nu^{t}_{j}-i+1}\,q^{\nu_{i}-j}\Big)^{-1}\,,\label{RTV}
\eea
where the argument of the skew Schur functions $q^{-\rho}t^{-\nu}$ denotes the infinite set of variables $\{q^{i-\frac{1}{2}}\,t^{-\nu_{i}}\,|\,i=1,2,\cdots\}$. Furthermore, concerning our variables, we introduce the shorthand notation
\begin{align}
&q=e^{2\pi i\epsilon_1}\,,&&t=e^{-2\pi i\epsilon_2}\,,&&Q_m=e^{2\pi im}\,,&& Q_{T}=e^{2\pi i\,T}\,,&& Q_{U}=e^{2\pi i \,U}\,,\label{qtDef}
\end{align}
reflecting the equivariant $U(1)\times U(1)$ action on $\mathbb{C}^2$ as well as the mass deformation and the K\"ahler parameters of the underlying geometry, as we shall review in detail below.

%%%%%%%%%%%%%%%%%%%%%%%%%%
\subsection{Topological strings on elliptic CY3folds: Refined vertex computation}\label{Sect:Vertex}
In this section we calculate the (refined) topological string partition function for the elliptic CY3folds $X_{N,M}$ dual to the $(N,M)$ 5-brane web. We use the (refined) topological vertex formalism to compute the partition function.
%%%
\subsubsection{Generic Configuration}\label{Sect:Buildingblock}
\begin{figure}[h]
  \centering
  % Requires \usepackage{graphicx}
  \includegraphics[width=4in]{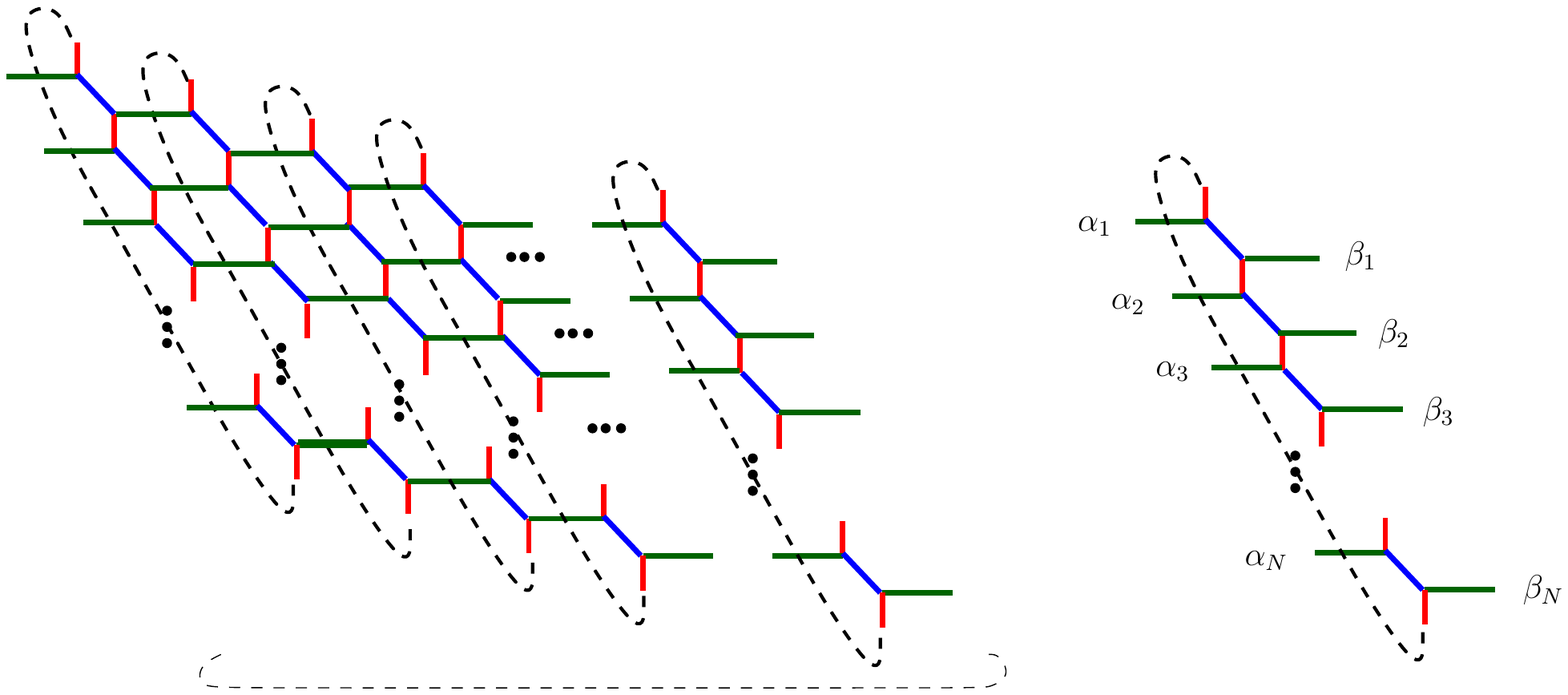}\\
  \caption{(a) The $(N,M)$ 5-brane web and (b) its building block.}\label{webdef}
\end{figure}

The refined topological string partition function of the CY3fold $X_{N,M}$, resolved orbifold of the elliptic conifold, can be calculated using the refined topological vertex and the web diagram corresponding to $X_{N,M}$. As we mentioned earlier the mass deformation of the gauge theory corresponds to deforming the brane web. After deformation the brane web is shown in \figref{webdef}(a). The brane web shown in \figref{webdef}(b) is what we will call the building block since we can construct the $(N,M)$ 5-brane web by gluing multiple building blocks together.

Recall that the distance between the $i$-th and the $j$-th D5-brane is given by $a_{i}-a_{j}$ and we denote by $Q_{i}$,
\begin{align}
&Q_{i}=e^{i(a_{i}-a_{i+1})}\,,&&i=1,2,\cdots, N\,,
\end{align}
where $N+1\equiv 1$ since the branes are on a circle of radius $U$,
\bea
Q_{N}=Q_{U}\,Q_{1}^{-1}Q_{2}^{-1}\cdots Q_{N-1}^{-1}\,.
\eea
The open string partition function associated with the building block (\figref{webdef}(b)) will be denoted by $W^{\alpha_{1}\alpha_{2}\cdots \alpha_{N}}_{\beta_{1}\beta_{2}\cdots \beta_{N}}(Q_{m},{\bf Q},\epsilon_{1},\epsilon_{2})$, where we denote by ${\bf Q}$ collectively all the $Q_{i}$. The topological string partition function of the CY3fold $X_{N,M}$ is given by
\bea
Z(N,M)=\sum_{\{\alpha\}}\prod_{a=1}^{M} \Big[Q_{b_{a}}^{|\alpha^{(a)}|}\,W^{\alpha^{(a)}}_{\alpha^{(a+1)}}\Big]\,,\label{NMPF}
\eea
where $\alpha^{(a)}$ is a set of $N$ partitions $\{\alpha^{(a)}_{1},\cdots \alpha^{(a)}_{N}\}$, $W^{\alpha^{(a)}}_{\alpha^{(a+1)}}=W^{\alpha^{(a)}_{1}\cdots \alpha^{(a)}_{N}}_{\alpha^{(a+1)}_{1}\cdots \alpha^{(a+1)}_{N}}$ and we define $|\alpha^{(a)}|=\sum_{b=1}^{N}|\alpha^{(a)}_{b}|$.

To express $W^{\alpha_{1}\cdots \alpha_{N}}_{\beta_{1}\cdots \beta_{N}}(Q_{m},{\bf Q},\epsilon_{1},\epsilon_{2})$ in terms of the refined topological vertex we have to choose a preferred direction. In this case we chose the latter to be horizontal so that the closed string partition functions will be equal to the partition function of the gauge theory on the branes which are horizontal i.e., gauge theory on the D5-branes. With this choice of the preferred direction we get,
\bea\nn
W^{\alpha_{1}\cdots \alpha_{N}}_{\beta_{1}\cdots\beta_{N}}(Q_{m},{\bf Q},\epsilon_{1},\epsilon_{2})&=&\sum_{\mu_{a}\,\nu_{a}} (-Q_{m})^{\sum_{b=1}^{N}|\mu_{b}|-|\nu_{b}|} \Big(\prod_{b=1}^{N}(-Q_{b})^{|\nu_{b+1}|}\Big) \Big(\prod_{a=1}^{N}C_{\nu_{a}\,\mu_{a}\alpha_{a}}(t,q)\Big)\\\nn
&&\times\Big(\prod_{b=1}^{N}C_{\nu_{b+1}^{t}\mu_{b}^{t}\beta^{t}_{b}}(q,t)\Big)\,,
\eea
where $N+1\equiv 1$ so that $\nu_{N+1}=\nu_{1}$. Using the definition of the refined topological vertex in terms of skew Schur functions, Eq.(\ref{RTV}), we get,
\begin{align}
&W^{\alpha_{1}\cdots \alpha_{N}}_{\beta_{1}\cdots\beta_{N}}=\Big(\prod_{a=1}^{N}q^{\frac{||\alpha_{a}||^2}{2}}t^{\frac{||\beta^{t}_{a}||^2}{2}}\widetilde{Z}_{\alpha_{a}}(t,q)
\widetilde{Z}_{\beta^{t}_{a}}(q,t)\Big)\sum_{\mu_{a},\nu_{a},\eta_{a},\sigma_{a}}(-Q_{m})^{\sum_{b=1}^{N}|\mu_{b}|-|\nu_{b}|}
\Big(\prod_{b=1}^{N}(-Q_{b})^{|\nu_{b+1}|}\Big)\nonumber\\
&\times\Big(\prod_{a=1}^{N}s_{\nu^{t}_{a}/\eta_{a}}(\sqrt{\frac{t}{q}}t^{-\rho}q^{-\alpha_{a}})s_{\mu_{a}/\eta_{a}}(q^{-\rho}t^{-\alpha^{t}_{a}})\Big)
\Big(\prod_{b=1}^{N}s_{\nu_{b+1}/\sigma_{b}}(\sqrt{\frac{q}{t}}q^{-\rho}t^{-\beta^{t}_{b}})s_{\mu^{t}_{b}/\sigma_{b}}(t^{-\rho}q^{-\beta_{b}})\Big)\nonumber
\end{align}
The sum over Schur functions in the above expression can be written as,
\bea
G({\bf x},{\bf y},\widetilde{Q})=\sum_{\mu,\eta}\,(-\widetilde{Q}_{1})^{|\mu_{1}|}\cdots (-\widetilde{Q}_{2N})^{|\mu_{2N}|}\,
\prod_{a=1}^{2N}s_{\mu_{a}/\eta_{a}}(x_{a})s_{\mu^{t}_{a}/\eta_{a+1}}(y_{a})\,,\label{schursum}
\eea
%(-Q_{m})^{|\mu_{odd}|-|\mu_{even}|}\Big(\prod_{b=1}^{N}(-Q_{b})^{|\nu_{b+1}|}\Big)
where
\begin{align}\label{kahler}
\widetilde{Q}_{a}&=Q_{m}\,,&&\text{for }a=1,3,\cdots,2N-1\nonumber\\
\widetilde{Q}_{a}&=Q_{\frac{a}{2}}Q_{m}^{-1}\,,&&\text{for } a=2,4,\cdots,2N\,\nonumber\\
(x_{a},y_{a})&=(q^{-\rho}t^{-\alpha^{t}_{\frac{a+1}{2}}},t^{-\rho}q^{-\beta_{\frac{a+1}{2}}})\,,&&\text{for }a=1,3,\cdots, 2N-1\nonumber\\
(x_{a},y_{a})&=(\sqrt{\frac{q}{t}}q^{-\rho}t^{-\beta^{t}_{\frac{a}{2}}},\sqrt{\frac{t}{q}}t^{-\rho}q^{-\alpha_{\frac{a}{2}+1}})\,,&&\text{for } a=2,4,\cdots, 2N\,.
\end{align}
Using standard identities of the Schur functions it follows that (the proof is given in appendix~\ref{Sec:Identity})\footnote{The computation is similar to the one carried out in \cite{mstrings} for the case $N=1$. Another way to prove (\ref{id}) is to use the free fermion representation to express this as $\mbox{Tr}\Big({\cal O}(x_{1},y_{1}){\cal O}(x_{2},y_{2})\cdots {\cal O}(x_{2N},y_{2N})\Big)$ where ${\cal O}(x,y)=\Gamma_{-}(x)\omega \Gamma_{+}(y)\omega$.}
\begin{align}
G({\bf x},{\bf y},\widetilde{Q})&=\frac{1}{\widehat{\eta}(\widetilde{Q}_{\bullet})}\prod_{i,j,k}\prod_{a,\ell=1}^{2N}(1-\,\widetilde{Q}_{\bullet}^{k-1}\widetilde{Q}_{a}\widetilde{Q}_{a+1}
\cdots \widetilde{Q}_{a+\ell-1}x_{a,i}y_{a+\ell-1,j})^{(-1)^{\ell-1}}\nonumber\\
&=\frac{1}{\widehat{\eta}(\widetilde{Q}_{\bullet})}\prod_{i,j,k}\prod_{a=1}^{2N}\prod_{\ell=1}^{N}\frac{(1-\,\widetilde{Q}_{\bullet}^{k-1}\widetilde{Q}_{a}\widetilde{Q}_{a+1}
\cdots \widetilde{Q}_{a+2\ell-2}x_{a,i}y_{a+2\ell-2,j})}{(1-\widetilde{Q}_{\bullet}^{k-1}\widetilde{Q}_{a}\widetilde{Q}_{a+1}
\cdots \widetilde{Q}_{a+2\ell-1}x_{a,i}y_{a+2\ell-1,j})}\,,\label{id}
\end{align}
where $\widehat{\eta}(\widetilde{Q}_{\bullet})=\prod_{k=1}^{\infty}(1-\widetilde{Q}_{\bullet}^{k})$ and $\widetilde{Q}_{\bullet}=\widetilde{Q}_{1}\widetilde{Q}_{2}\cdots\widetilde{Q}_{2N}= Q_{U}$. Using Eq.(\ref{kahler}) and Eq.(\ref{id}) the open string amplitude $W^{\alpha_{1}\cdots \alpha_{N}}_{\beta_{1}\cdots \beta_{N}}$ becomes
\bea
W^{\alpha_{1}\cdots \alpha_{N}}_{\beta_{1}\cdots \beta_{N}}(Q_{m},{\bf Q},\epsilon_{1},\epsilon_{2})&=&\frac{1}{\widehat{\eta}(Q_{U})}\Big(\prod_{a=1}^{N}q^{\frac{||\alpha_{a}||^2}{2}}t^{\frac{||\beta^{t}_{a}||^2}{2}}\widetilde{Z}_{\alpha_{a}}(t,q)
\widetilde{Z}_{\beta^{t}_{a}}(q,t)\Big)\\\nn
&&\times\prod_{i,j,k}\prod_{r,\ell=1}^{N}
\Big[\frac{1-Q_{U}^{k-1}Q_{r,r+\ell}Q_{m}^{-1}t^{-\beta^{t}_{r,j}+i-\frac{1}{2}}\,q^{-\alpha_{r+\ell,i}+j-\frac{1}{2}}}
{1-Q_{U}^{k-1}Q_{r,r+\ell}t^{-\beta^{t}_{r,j}+i-1}\,q^{-\beta_{r+\ell,i}+j}}\\\nn
&&\times\frac{1-Q_{U}^{k-1}Q_{r,r+\ell-1}Q_{m}\,t^{-\alpha^{t}_{r,j}+i-\frac{1}{2}}\,q^{-\beta_{r+\ell-1,i}+j-\frac{1}{2}}}
{1-Q_{U}^{k-1}Q_{r,r+\ell}t^{-\alpha^{t}_{r,j}+i}\,q^{-\alpha_{r+\ell,i}+j-1}}\Big]\,.
\eea
In the above expression $Q_{a,b}$ is the such that
\bea
Q_{ab}=\left\{\begin{array}{lcl}Q_{a}Q_{a+1}\cdots Q_{b-1} & \text{for} & a<b\,,\\
Q_{U}\,Q_{ba}^{-1} & \text{for} & a>b\,,\\
1 & \text{for} & a=b\,.\end{array}\right.
\eea
Define furthermore
\begin{align}
J_{\mu\nu}(x;t,q)&:=\prod_{(i,j)\in \mu}(1-x\,t^{\nu^{t}_{j}-i+\frac{1}{2}}\,q^{\mu_{i}-j+\frac{1}{2}})\prod_{(i,j)\in \nu}(1-x\,t^{-\mu^{t}_{j}+i-\frac{1}{2}}\,q^{-\nu_{i}+j-\frac{1}{2}})\nn\\
{\cal J}_{\mu\nu}(x;t;q)&=\prod_{k=1}^{\infty}J_{\mu\nu}(Q_{U}^{k-1}\,x;t,q)
\end{align}
and using additionally the identity
\bea\nn
\prod_{i,j=1}^{\infty}\frac{1-Q\,t^{-\mu^{t}_{j}+i-\frac{1}{2}}\,q^{-\nu_{i}+j-\frac{1}{2}}}{1-Q\,t^{i-\frac{1}{2}}\,q^{j-\frac{1}{2}}}=J_{\mu\nu}(Q;t,q)=J_{\nu\mu}(Q;t^{-1},q^{-1})\eea
we get the relation
\begin{align}
&\frac{W^{\alpha_{1}\cdots \alpha_{N}}_{\beta_{1}\cdots \beta_{N}}(Q_{m},{\bf Q},\epsilon_{1},\epsilon_{2})}{W^{\emptyset\cdots \emptyset}_{\emptyset\cdots \emptyset}(Q_{m},{\bf Q},\epsilon_{1},\epsilon_{2})}=\Big(\prod_{a=1}^{N}q^{\frac{||\alpha_{a}||^2}{2}}t^{\frac{||\beta^{t}_{a}||^2}{2}}\widetilde{Z}_{\alpha_{a}}(t,q)
\widetilde{Z}_{\beta^{t}_{a}}(q,t)\Big)\nn\\
&\hspace{1.5cm}\times\prod_{r,\ell=1}^{N}\frac{{\cal J}_{\beta_{r}\alpha_{r+\ell}}(Q_{r,r+\ell}\,Q_{m}^{-1};t,q)\,
{\cal J}_{\alpha_{r}\beta_{r+\ell-1}}(Q_{r,r+\ell-1}\,Q_{m};t,q)}{{\cal J}_{\beta_{r}\beta_{r+\ell}}(Q_{r,r+\ell}\sqrt{\frac{q}{t}};t,q)
\,{\cal J}_{\alpha_{r}\alpha_{r+\ell}}(Q_{r,r+\ell}\sqrt{\frac{t}{q}};t,q)}\,,\label{BB}
\end{align}
where we have introduced
\bea
W^{\emptyset\cdots \emptyset}_{\emptyset\cdots \emptyset}(Q_{m},{\bf Q},\epsilon_{1},\epsilon_{2})&=&\frac{1}{\widehat{\eta}(Q_{U})}\,\prod_{i,j,k}\prod_{r,\ell=1}^{N}
\Big[\frac{1-Q_{U}^{k-1}Q_{r,r+\ell}Q_{m}^{-1}t^{i-\frac{1}{2}}\,q^{j-\frac{1}{2}}}
{1-Q_{U}^{k-1}Q_{r,r+\ell}t^{i-1}\,q^{j}}\\\nn
&&\times\frac{1-Q_{U}^{k-1}Q_{r,r+\ell-1}Q_{m}\,t^{i-\frac{1}{2}}\,q^{j-\frac{1}{2}}}
{1-Q_{U}^{k-1}Q_{r,r+\ell}t^{i}\,q^{j-1}}\Big]\,.
\eea
The function ${\cal J}_{\mu\nu}(x;t;q)$ satisfies the following two important identities relating it to the theta function:
\bea
{\cal J}_{\mu\nu}(x;t,q){\cal J}_{\nu\mu}(Q_{U}x^{-1};t,q)&=&x^{\frac{|\mu|+|\nu|}{2}}\,t^{\frac{||\nu^t||^2-||\mu^t||^2}{4}}\,q^{\frac{||\mu||^2-||\nu||^2}{4}}\,
\vartheta_{\mu\nu}(x;U)\,,\\\nn
\frac{1}{\vartheta_{\alpha\alpha}(\sqrt{\frac{q}{t}};U)}=\frac{1}{\vartheta_{\alpha\alpha}(\sqrt{\frac{t}{q}};U)}&=&
\frac{(-1)^{|\alpha|}q^{\frac{||\alpha||^2}{2}}\,t^{\frac{||\alpha^t||^2}{2}}\,\widetilde{Z}_{\alpha}(t,q)\widetilde{Z}_{\alpha^t}(q,t)}
{{\cal J}_{\alpha\alpha}(Q_{U}\sqrt{\frac{t}{q}}){\cal J}_{\alpha\alpha}(Q_{U}\sqrt{\frac{q}{t}})}\,,\label{identities}
\eea
where for $x=e^{2\pi iz}$, we used the following $\vartheta$-functions
\begin{align}
\vartheta_{\mu\nu}(x;U)&:=\prod_{(i,j)\in \mu}\vartheta(x^{-1}t^{-\nu^{t}_{j}+i-\frac{1}{2}}q^{-\mu_{i}+j-\frac{1}{2}};U)\prod_{(i,j)\in \nu}\vartheta(x^{-1}t^{\mu^{t}_{j}-i+\frac{1}{2}}q^{\nu_{i}-j+\frac{1}{2}};U)\label{vartheta}\\
\vartheta(x;U)&=(x^{\frac{1}{2}}-x^{-\frac{1}{2}})\prod_{k=1}^{\infty}(1-x\,e^{2\pi i\,k\, U})(1-x^{-1}e^{2\pi i\,k\,U})=\frac{ie^{-\frac{i\pi\,U}{4}}\theta_{1}(U;z)}{\prod_{k=1}^{\infty}(1-e^{2\pi i\,k\,U})}\,.\nonumber
\end{align}
In the last expression, $\theta_{1}(U,z)$ is the Jacobi theta function, whose properties are reviewed in appendix~\ref{App:Defs}.

The partition function of the $(M,N)$ brane configurations Eq.(\ref{NMPF}) is then
\begin{align}
Z(M,N)=&W_{N}(\emptyset)^{M}\sum_{\alpha^{(i)}_{a}}Q_{T}^{|\alpha^{(M)}|}\,\Big(\prod_{i=1}^{M}Q_{B_{i}}^{|\alpha^{(i)}|-|\alpha^{(M)}|}\Big)
\prod_{i=1}^{M}\prod_{a=1}^{N}
\frac{\vartheta_{\alpha^{(i+1)}_{a}\alpha^{(i)}_{a}}(Q_{m})}{\vartheta_{\alpha^{(i)}_{a}\alpha^{(i)}_{a}}(\sqrt{t/q})}\nonumber\\
&\times \prod_{1\leq a<b\leq N}\prod_{i=1}^{M}\frac{\vartheta_{\alpha^{(i)}_{a}\alpha^{(i+1)}_{b}}(Q_{ab}Q_{m}^{-1})
\vartheta_{\alpha^{(i+1)}_{a}\alpha^{(i)}_{b}}(Q_{ab}Q_{m})}{\vartheta_{\alpha^{(i)}_{a}\alpha^{(i)}_{b}}(Q_{ab}\sqrt{t/q})
\vartheta_{\alpha^{(i)}_{a}\alpha^{(i)}_{b}}(Q_{ab}\sqrt{q/t})}\,.\label{GenTopVertex}
\end{align}
In the above partition function,
\bea
W_{N}(\emptyset)&:=&W^{\emptyset\cdots \emptyset}_{\emptyset\cdots \emptyset}(Q_{m},{\bf Q},\epsilon_{1},\epsilon_{2})\,,\,\,\mbox{where there are $N$ empty partitions}\,,\\\nn
Q_{B_{i}}&=&e^{i(b_{i}-b_{i+1})}\,,\,\,i=1,2,\cdots,M\,,\\\nn
Q_{T}&=&Q_{B_{1}}Q_{B_{2}}\cdots Q_{B_{M}}\,,
\eea
such that $M+1\equiv 1$ since the branes are on a circle.

An interesting question concerns the modular properties of the partition function (\ref{GenTopVertex}). Let us define
\bea\nn
\widehat{Z}(M,N)&:=&\frac{Z(M,N)}{W_{N}(\emptyset)^{M}}
\eea
which is explicitly given as
\bea
\widehat{Z}(M,N)&=&\sum_{\alpha^{(i)}_{a}}Q_{T}^{|\alpha^{(M)}|}\,\Big(\prod_{i=1}^{M}Q_{B_{i}}^{|\alpha^{(i)}|-|\alpha^{(M)}|}\Big)
\prod_{i=1}^{M}\prod_{a=1}^{N}
\frac{\vartheta_{\alpha^{(i+1)}_{a}\alpha^{(i)}_{a}}(Q_{m})}{\vartheta_{\alpha^{(i)}_{a}\alpha^{(i)}_{a}}(\sqrt{t/q})}\nonumber\\\label{zmnpf}
&&\times \prod_{1\leq a<b\leq N}\prod_{i=1}^{M}\frac{\vartheta_{\alpha^{(i)}_{a}\alpha^{(i+1)}_{b}}(Q_{ab}Q_{m}^{-1})
\vartheta_{\alpha^{(i+1)}_{a}\alpha^{(i)}_{b}}(Q_{ab}Q_{m})}{\vartheta_{\alpha^{(i)}_{a}\alpha^{(i)}_{b}}(Q_{ab}\sqrt{t/q})
\vartheta_{\alpha^{(i)}_{a}\alpha^{(i)}_{b}}(Q_{ab}\sqrt{q/t})}\,
\eea
Under the modular transformation,
\bea\label{MT}
(U,m,a_{i})\mapsto (-\frac{1}{U},\frac{m}{U},\frac{a_i}{U})\,,\label{ModularProperties}
\eea
the partition function $\widehat{Z}(M,N)$ is not invariant. However, it can be made modular invariant at the expense of making it non-holomorphic in $U$. Recall that the theta function can written in terms of the Eisenstein series as,
\bea\nn
\theta_{1}(U;z)=\eta^{3}(\tau)(2\pi iz)\,\mbox{exp}\Big(\sum_{k\geq 1}\frac{B_{2k}}{(2k)(2k)!}E_{2k}(U)(2\pi i z)^{2k}\Big)\,.
\eea
If we replace the $E_{2}(U)$ in the above expression of the theta function with the non-holomorphic $\widehat{E}_{2}(U,\overline{U})=E_{2}(U)-\frac{3}{\pi\,U_{2}}$ ($U_{2}=Im\,U$) then the partition function $\widehat{Z}(M,N)$ is modular invariant but no longer holomorphic and satisfies a holomorphic anomaly equation,
\bea
\frac{\partial \widehat{Z}(M,N)}{\partial \widehat{E}_{2}(U,\overline{U})}=\frac{1}{24}D^{(2)}_{b_{i},m}(\epsilon_{1},\epsilon_{2})\widehat{Z}(M,N)\,,
\eea
where $D^{(2)}_{b_{i}}(m,\epsilon_{1},\epsilon_{2})$ is a second order differential operator with respect to $b_{i}$ and depends of $m,\epsilon_{1}$ and $\epsilon_{2}$. This operator can be determined from following equation satisfied by $\vartheta_{\lambda\mu}(x;U)$,
\bea\nn
\frac{\partial \mbox{ln}\vartheta_{\mu\nu}(x;U)}{\partial \widehat{E}_{2}(U,\overline{U})}&=&\frac{1}{24}\sum_{(i,j)\in \mu}\Big(\mbox{ln}(x)+\epsilon_{1}(\mu_{i}-j+\frac{1}{2})-\epsilon_{2}(\nu^{t}_{j}-i+\frac{1}{2})\Big)\\\nn
&&+\frac{1}{24}\sum_{(i,j)\in \nu}\Big(\mbox{ln}(x)-\epsilon_{1}(\nu_{i}-j+\frac{1}{2})+\epsilon_{2}(\mu^{t}_{j}-i+\frac{1}{2})\Big)\,.
\eea

\subsubsection{Specific Examples}\label{Sect:ExamplesTopString}
%%%%%%%%%%%%%%%%%%%%%%%%%%%%%%
To illustrate the general expression (\ref{GenTopVertex}) we will consider a number of explicit examples in the following for small numbers of $M$ and $N$.
%%%%
\begin{itemize}
\item $M=1$ and $N=k$:\\
We begin with the case of $M=1$ and $N$ being generic, which we will denote by $k$ to avoid confusion. From the discussion in section 2 it follows that we have two dual M-string descriptions of this brane configuration. This case corresponds to $k$ parallel M5-branes compactified on a circle with one transverse direction compactified on a circle. The dual picture is that of a single M5-brane compactified on a circle with transverse space a charge $k$ Taub-NUT. Following the notation introduced in the previous section, we will consider the directions $(Y^{1},Y^{6})$ compactified on $S^1_{\text{M5}}\times S^1_{\text{trans}}$.\footnote{The case of the transverse direction $Y^6$ non-compact was discussed in \cite{mstrings}.} As already mentioned, the M5 branes can be separated along $X^6$ and the corresponding web diagram is shown in \figref{km5branes}(b).

\begin{figure}[h]
  \centering
  % Requires \usepackage{graphicx}
  \includegraphics[width=12cm]{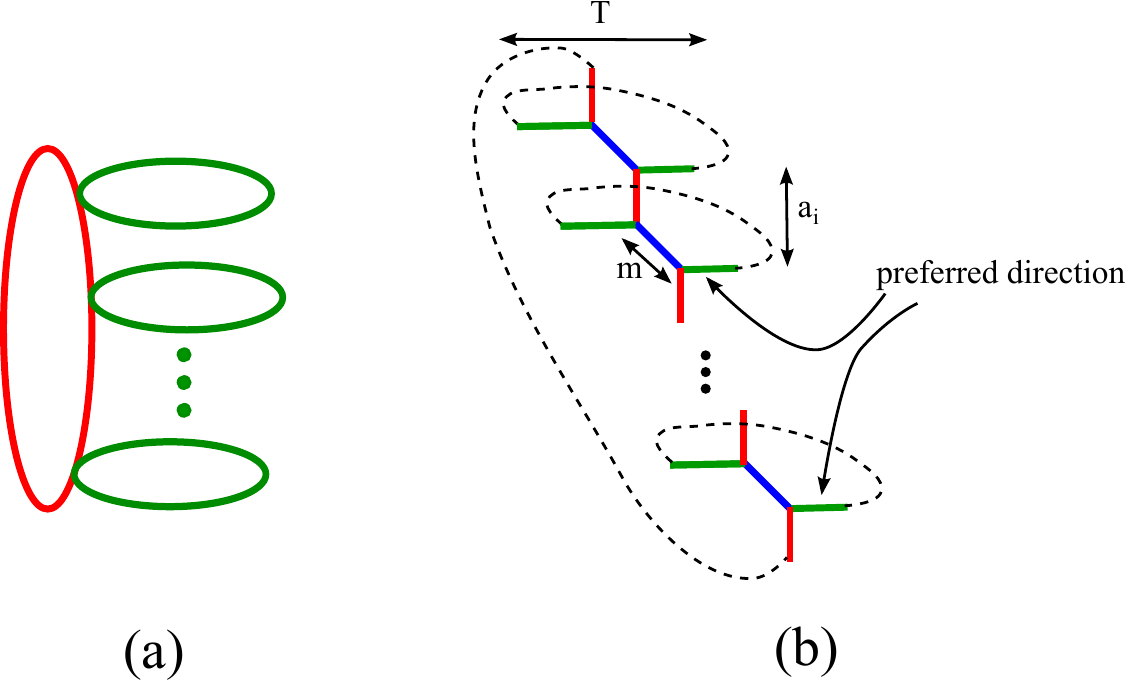}\\
  \caption{\emph{Multiple M5 and M2 branes on $S^1_{\text{M5}}\times S^1_{\text{trans}}$: }(a) $k$ M5-brane compactified on a circle with one transverse direction compactified; (b) the corresponding web diagram with preferred direction horizontal. We have also denoted the parameters $U$ and $t_{f_a}$ as well as the mass deformation parameter $m$. }\label{km5branes}
\end{figure}

The partition function can be determined using the formalism developed in section~\ref{Sect:Buildingblock} or by simply taking $M=1, N=k$ in Eq.(\ref{NMPF}) and is given by,
\bea\nn
Z(1,k;T,U,a_{i})&:=&\sum_{\alpha_{1,2,\cdots,k}}\,Q_{T}^{|\alpha_{1}|+|\alpha_{2}+\cdots+|\alpha_{k}|}\,
W^{\alpha_{1}\alpha_{2}\cdots\alpha_{k}}_{\alpha_{1}\alpha_{2}\cdots\alpha_{k}}\,,\\\nn
&=&W_{k}(\emptyset)\sum_{\alpha_{1,2,\cdots,k}}\,Q_{T}^{|\alpha_{1}|+|\alpha_{2}+\cdots+|\alpha_{k}|}\prod_{i=1}^{k}
\frac{\vartheta_{\alpha_{i}\alpha_{i}}(Q_{m})}{\vartheta_{\alpha_{i}\alpha_{i}}(\sqrt{\frac{t}{q}})}\\\nn
&& \prod_{1\leq i<j\leq k}\frac{\vartheta_{\alpha_{i}\alpha_{j}}(Q_{ij}Q_{m}^{-1})\vartheta_{\alpha_{i}\alpha_{j}}(Q_{ij}Q_{m})}
{\vartheta_{\alpha_{i}\alpha_{j}}(Q_{ij}\sqrt{\frac{t}{q}})\vartheta_{\alpha_{i}\alpha_{j}}(Q_{ij}\sqrt{\frac{q}{t}})}\,,\label{PartFct20}
\eea
where $Q_{ij}=Q_{i}Q_{i+1}\cdots Q_{j-1}$ and $Q_{i}=e^{i(a_{i}-a_{j})}$. $a_{i}-a_{j}$ is the distance between the M5 branes along $S^1_{\text{trans}}$ (see figure~\ref{km5branes}). The above partition function should be equal to the partition function $Z(k,1)$ after $T\mapsto U$ and $a_{i}\mapsto b_{i}$,
\bea
Z(k,1;T,U,a_{i})=Z(1,k;U,T,b_{i})\,.
\eea
This can checked by evaluating the partition function $Z(k,1)$ using Eq.(\ref{NMPF}),
\bea
Z(k,1)&=&W_{1}(\emptyset)^{k}\sum_{\nu_{1},\mathellipsis,\nu_{k}}Q_{T}^{|\nu_{k}|}\left(\prod_{i=1}^{k}(Q_{B_{i}})^{|\nu_{i}|-|\nu_{k}|}\right)
\prod_{i=1}^{k}\frac{\vartheta_{\nu_{i+1}\nu_{i}}(Q_{m})}{\vartheta_{\nu_{i}\nu_{i}}(\sqrt{\frac{t}{q}})}
\eea
In the above expression
\begin{align}
W_{1}(\emptyset)=\prod_{n=1}^{\infty}(1-Q_{U}^{n})^{-1}\,\prod_{i,j=1}^{\infty}\frac{(1-Q_{U}^{n-1}\,Q_{m}\,q^{i-\frac{1}{2}}\,t^{j-\frac{1}{2}})
(1-Q_{U}^{n}Q_{m}^{-1}\,q^{i-\frac{1}{2}}t^{j-\frac{1}{2}})}{
(1-Q_{U}^{n}q^{i}t^{j-1})(1-Q_{U}^{n}q^{i-1}t^{j})}\,.\label{u1pf}
\end{align}
It is the topological string partition function of the elliptic CY3fold $X_{1,1}$ in the limit $T\mapsto i\infty$, it also captures the massless and the KK modes of the $(2,0)$ $A_{0}$ theory compactified on the circle \cite{Lockhart:2012vp, mstrings} and it can also be written in terms of chi-y genus of the Hilbert scheme of points \cite{Iqbal:2008ra}. We will study the properties of this partition function and the associated partition function $Z(1,1)$ in detail in section \ref{Sect:TopString}.

\item $M=2$ and $N=1,2,k$:\\
Explicitly, for the values $M=2$ and $N=1,2$ the partition functions can be evaluated to be
\bea
Z(2,1)&:=&(W^{\emptyset}_{\emptyset})^2\,\sum_{\alpha,\beta}(-Q_{1})^{|\alpha|}(-Q_{2})^{|\beta|}\,W^{\alpha}_{\beta}\,W^{\beta}_{\alpha}\\\nn
&=&(W^{\emptyset}_{\emptyset})^2\,\sum_{\alpha,\beta}\,Q_{B_{1}}^{|\alpha|-|\beta|}\,Q_{T}^{|\beta|}\,
\frac{\vartheta_{\alpha\beta}(Q_{m})\vartheta_{\beta\alpha}(Q_{m})}
{\vartheta_{\alpha\alpha}(\sqrt{\frac{t}{q}})\vartheta_{\beta\beta}(\sqrt{\frac{t}{q}})}
\eea
as well as
\bea
Z(2,2)&:=&(W^{\emptyset\,\emptyset}_{\emptyset\,\emptyset})^2\,\sum_{\alpha_{1,2},\beta_{1,2}}
(-Q_{1})^{|\alpha_{1}|+|\alpha_{2}|}(-Q_{2})^{|\beta_{1}|+|\beta_{2}|}\,W^{\alpha_{1}\alpha_{2}}_{\beta_{1}\beta_{2}}
\,W^{\beta_{1}\beta_{2}}_{\alpha_{1}\alpha_{2}}\\\nn
&=&(W^{\emptyset\,\emptyset}_{\emptyset\,\emptyset})^2\,\sum_{\alpha_{1,2},\beta_{1,2}}
Q_{B_1}^{|\alpha_{1}|+|\alpha_{2}|-|\beta_{1}|-|\beta_{2}|}\,Q_{T}^{|\beta_{1}|+|\beta_{2}|}\,
\prod_{a=1}^{2}\frac{\vartheta_{\alpha_{a}\beta_{a}}(Q_{m})
\vartheta_{\beta_{a}\alpha_{a}}(Q_{m})}{\vartheta_{\alpha_{a}\alpha_{a}}(\sqrt{\frac{t}{q}})\vartheta_{\beta_{a}\beta_{a}}(\sqrt{\frac{t}{q}})}\\\nn
&&\times
\prod_{1\leq a<b\leq 2}\frac{
\vartheta_{\alpha_{a}\beta_{b}}(Q_{ab}Q_{m})\vartheta_{\alpha_{a}\beta_{b}}(Q_{ab}Q_{m}^{-1})\vartheta_{\beta_{a}\alpha_{b}}(Q_{ab}Q_{m})
\vartheta_{\beta_{a}\alpha_{b}}(Q_{ab}Q_{m}^{-1})}
{\vartheta_{\alpha_{a}\alpha_{b}}(Q_{ab}\sqrt{\frac{t}{q}})\vartheta_{\alpha_{a}\alpha_{b}}(Q_{ab}\sqrt{\frac{q}{t}})
\vartheta_{\beta_{a}\beta_{b}}(Q_{ab}\sqrt{\frac{t}{q}})
\vartheta_{\beta_{a}\beta_{b}}(Q_{ab}\sqrt{\frac{q}{t}})}
\eea
The general expression for $M=2$ and $N=k$ generic takes the form
\bea\nn
Z(2,k)&=&(W^{\emptyset\cdots\emptyset}_{\emptyset\cdots\emptyset})^2\,\sum_{\alpha_{1,\cdots,k},\beta_{1,\cdots,k}}
Q_{f}^{\sum_{a=1}^{k}(|\alpha_{a}|-|\beta_{b})|}\,Q_{\rho}^{\sum_{a=1}^{k}|\beta_{a}|}\,
\prod_{a=1}^{k}\frac{\vartheta_{\alpha_{a}\beta_{a}}(Q_{m})
\vartheta_{\beta_{a}\alpha_{a}}(Q_{m})}{\vartheta_{\alpha_{a}\alpha_{a}}(\sqrt{\frac{t}{q}})\theta_{\beta_{a}\beta_{a}}(\sqrt{\frac{t}{q}})}\\\nn
&&\times
\prod_{1\leq a<b\leq k}\frac{
\vartheta_{\alpha_{a}\beta_{b}}(Q_{ab}Q_{m})\vartheta_{\alpha_{a}\beta_{b}}(Q_{ab}Q_{m}^{-1})\vartheta_{\beta_{a}\alpha_{b}}(Q_{ab}Q_{m})
\vartheta_{\beta_{a}\alpha_{b}}(Q_{ab}Q_{m}^{-1})}
{\vartheta_{\alpha_{a}\alpha_{b}}(Q_{ab}\sqrt{\frac{t}{q}})\vartheta_{\alpha_{a}\alpha_{b}}(Q_{ab}\sqrt{\frac{q}{t}})
\vartheta_{\beta_{a}\beta_{b}}(Q_{ab}\sqrt{\frac{t}{q}})
\vartheta_{\beta_{a}\beta_{b}}(Q_{ab}\sqrt{\frac{q}{t}})}
\eea
\end{itemize}
%%%%%%%%%%%%%%%%%%%%%%%%%%%%%%%

\subsection{M-string Partition Function}\label{Sect:Mstrings}
In this section we will calculate the equivariant elliptic genus of the $(2,0)$ two-dimensional theory on the M-string worldsheet. We will see that the target space and the vector bundle to which the right moving fermions couple can be determined from the brane web diagram. Moreover, the partition function of the theory on the $(M,N)$ 5-brane web is related to the equivariant $(2,0)$ elliptic genus with target space $M(N,\vec{k}):=M(N,k_1)\times M(N,k_2)\times \mathellipsis M(N,k_M)$,
\bea
Z(M,N)=\sum_{\vec{k}}\,\vec{\varphi}^{|\vec{k}|}\,\chi_{\text{ell}}\Big(M(N,\vec{k}),V_{\vec{k}}\Big)\,,
\eea
where we have denoted by $\chi_{\text{ell}}(M,V)$ the equivariant $(2,0)$ elliptic genus of $M$ with $V$ as the bundle to which the right moving fermions couple and $\vec{\varphi}^{|\vec{k}|}=\prod_{i=1}^{M}\varphi_{i}^{k_{i}}$.

\subsubsection{$(2,0)$ Elliptic Genus}
Recall that the ${\cal N}=(2,2)$ sigma model with K\"ahler target space $M$ has a Lagrangian of the form
\bea
{\cal L}=\int d^{4}\theta\,K(\Phi_i,\overline{\Phi_{j}})\,,
\eea
where $K(\phi^{i},\phi^{\bar{j}})$ is the K\"ahler potential of $M$ and $\Phi_i$ are the chiral superfields. In terms of component fields the Lagrangian is given by\footnote{We consider a Lorentzian worldsheet so that $(\theta^{+}, \bar{\theta}^{+})$ are positive chirality spinors and $(\theta^{-},\bar{\theta}^{-})$ are negative chirality spinors. The fermions $\psi^{i}_{+}$ and $\psi^{j}_{-}$ are of negative and positive chirality respectively. Besides we introduce $\partial_{\pm}=\frac{\partial_{0}\pm\partial_{1}}{2}$.}
\bea\nn
{\cal L}=g_{i\bar{j}}\partial_{+}\phi^{i}\partial_{-}\phi^{\bar{j}}+g_{i\bar{j}}\partial_{-}\phi^{i}\partial_{+}
\phi^{\bar{j}}-2ig_{i\bar{j}}\psi_{+}^{i}D_{-}\psi^{\bar{j}}_{+}
-2ig_{i\bar{j}}\psi^{i}_{-}D_{+}\psi^{\bar{j}}_{-}+R_{\bar{k}i\bar{l}j}\psi^{i}_{+}\psi^{j}_{-}\psi^{\bar{k}}_{-}\psi^{\bar{l}}_{+}\,,
\eea
where the covariant derivatives are given by
\begin{align}
&D_{-}\psi^{\bar{j}}_{+}=(\partial_{-}\delta^{\bar{j}}_{\bar{l}}+\Gamma^{\bar{j}}_{\bar{l}\bar{k}}\partial_{-}\phi^{\bar{k}})\psi^{\bar{l}}_{+}\,,&&
D_{+}\psi^{\bar{j}}_{-}=(\partial_{+}\delta^{\bar{j}}_{\bar{l}}+\Gamma^{\bar{j}}_{\bar{l}\bar{k}}\partial_{+}\phi^{\bar{k}})\psi^{\bar{l}}_{-}\,.
\end{align}
The bosonic fields $\phi^{i}$ are local coordinates on the target space $M$ and since they are functions of the worldsheet coordinates $z,\overline{z}$ they describe locally the map $\Phi:\Sigma\mapsto M$. Denoting by $K$ and $\overline{K}$ the canonical and the anti-canonical bundle on $\Sigma$, the spinor bundles of opposite chirality are given by $K^{\frac{1}{2}}$ and $\overline{K}^{\frac{1}{2}}$. Since $K$ is the bundle of $(1,0)$ and $\overline{K}$ is the bundle of $(0,1)$ forms, the spinor bundles are just bundles of holomorphic and anti-holomorphic half-differentials. The fermions $\psi^{i}_{\pm}$ and $\psi^{\bar{i}}_{\pm}$ are spinors on $\Sigma$ and vectors on $M$ and therefore
\begin{align}
&\psi^{i}_{-}=K^{\frac{1}{2}}\otimes \Phi^{*}TM\,,&&\psi^{\bar{i}}_{-}=K^{\frac{1}{2}}\otimes \Phi^{*}\overline{TM}\\\nn
&\psi^{i}_{+}=\overline{K}^{\frac{1}{2}}\otimes \Phi^{*}TM\,,&&\psi^{\bar{i}}_{+}=\overline{K}^{\frac{1}{2}}\otimes \Phi^{*}\overline{TM}\,.
\end{align}
In the case of $(2,0)$ supersymmetry the right moving fermions, which we will denote by $\eta^{a}$, are instead taken to be \cite{Witten:1993yc}
\begin{align}
&\eta^{a}=\overline{K}^{\frac{1}{2}}\otimes \Phi^{*}V\,,&&\eta_{a}=\overline{K}^{\frac{1}{2}}\otimes \Phi^{*}V^{*}
\end{align}
where $V$ is a holomorphic bundle on $M$. The Lagrangian of the $(2,0)$ theory in component form is given by
\bea\nn
{\cal L}=g_{i\bar{j}}\partial_{+}\phi^{i}\partial_{-}\phi^{\bar{j}}
+g_{i\bar{j}}\partial_{-}\phi^{i}\partial_{+}
\phi^{\bar{j}}-2ig_{i\bar{j}}\psi^{i}_{-}D_{+}\psi^{\bar{j}}_{-}
-2i\eta_{a}D_{-}\eta^{a}
+F^{a}_{bi\bar{j}}\eta_{a}\eta^{b}\psi^{i}_{-}\psi^{\bar{j}}_{-}\,,
\eea
where by $F$ we denote the curvature 2-form of $V$ with connection $A$ and
\begin{align}
&D_{+}\psi^{\bar{j}}_{-}=(\partial_{+}\delta^{\bar{j}}_{\bar{l}}+\Gamma^{\bar{j}}_{\bar{l}\bar{k}}\partial_{+}\phi^{\bar{k}})\psi^{\bar{l}}_{-}\,,&&
D_{-}\eta^{a}=(\partial_{-}\delta^{a}_{b}+A^{a}_{b\,i}\partial_{-}\phi^{i})
\eta^{b}\,.
\end{align}
The supersymmetry transformations are given by
\begin{align}
&\delta\,\phi^{i}=\epsilon_{-}\psi^{i}_{-}\,,&&\delta\,\phi^{\bar{i}}=\epsilon_{-}\,A^{a}_{b\,i}\eta^{b}\psi^{i}_{-}+\overline{\epsilon}_{-}\psi^{\bar{i}}\,,\nonumber\\
&\delta\,\psi^{i}_{-}=-\overline{\epsilon}_{-}\partial_{\bar{z}}\phi^{i}\,,&&\delta\,\psi^{\bar{i}}=-\epsilon_{-}\partial_{\bar{z}}\phi^{\bar{i}}\,,\nonumber\\
&\delta\eta^{a}=\epsilon_{-}\,A^{a}_{b\,i}\eta^{b}\psi^{i}_{-}\,,&&\delta\eta_{a}=0\,.
\end{align}
The $(2,0)$ elliptic genus of the target space $M$ can be computed as the following trace over the sigma-model spectrum \cite{Kawai:1994np}
\bea
Z(\tau,z)=\mbox{Tr}(-1)^{F}y^{J_{L}}q^{H_{L}}\,\bar{q}^{H_{R}}
\eea
where $J_{L}$ is the zero mode of the left moving $U(1)$ current and $(-1)^{F}=e^{i\pi (J_{L}-J_{R})}$. Assuming that $\text{rank}(V)=\text{rank}(TM)$ we parameterize the Chern classes in terms of the Chern roots
\begin{align}
&c(TM)=\prod_{i}(1+x_{i})\,,&&\mbox{and}\,&&c(V)=\prod_{i}(1+\tilde{x}_{i})\,.
\end{align}
The $(2,0)$ elliptic genus can then be written in the form
\bea\label{20EG}
Z(\tau,z)=\int_{M}\,\prod_{i}\frac{x_{i}\,\theta(\tau,\tilde{x}_{i}+z)}{\theta(\tau,x_{i})}\,.
\eea
Thus, the elliptic genus is essentially determined by the Chern roots $x_i$ and $\tilde{x}_i$ of the bundles $TM$ and $V$ respectively.

\subsubsection{Fixed Point Calculus of Instanton Moduli Spaces}\label{Sect:hilbert}

We will denote by $M(r,k)$ the moduli space of $U(r)$ instantons of charge $k$ in $\mathbb{C}^2$. $M(r,k)$ is a hyper-K\"ahler manifold of real dimension $4rk$. It is isomorphic to the hyper-K\"ahler quotient (the ADHM construction),
\bea
&&M(r,k)=\{ (B_{1},B_{2},i,j)\,|\, [B_{1},B_{2}]+i\,j=0\,,\,\,[B_{1},B_{1}^{\dag}]+[B_{2},B_{2}^{\dag}]-i\,i^{\dag}-j\,j^{\dag}=\zeta\, \text{Id}\}\,,\nn
\eea
where we have introduced
\begin{align}
&B_{1},B_{2}\in \mbox{End}(\mathbb{C}^{k})\,,&&i\in \mbox{Hom}(\mathbb{C}^{r},\mathbb{C}^k)\,,&&j\in \mbox{Hom}(\mathbb{C}^k,\mathbb{C}^{r})\,.\nonumber
\end{align}
There is a $U(1)^{r}\times U(1)\times U(1)$ action on $M(r,k)$ with parameters $(e_{1},\mathellipsis,e_{r},t_{1},t_{2})$ given by
\begin{align}
&(B_1,B_2,i,j)\mapsto (t_{1}B_{1},t_{2}B_{2},i\,e^{-1},t_{1}t_{2}e\,j)\,,&&e=\mbox{diag}(e_1,e_2,\mathellipsis,e_r)\,.
\end{align}
The fixed points of $M(r,k)$ are labelled by a set of $r$ partitions $(\nu_{1},\nu_{2},\mathellipsis,\nu_{r})$
\bea
|\nu_{1}|+|\nu_{2}|+\mathellipsis +|\nu_{r}|=k\,.
\eea
A fixed point labelled by $(\nu_{1},\nu_{2},\mathellipsis,\nu_{r})$ is a direct sum $I_{\nu_1}\oplus I_{\nu_2}\oplus \mathellipsis\oplus I_{\nu_r}$ where $I_{\nu_a}$ is an ideal sheaf supported at the origin and generated by monomials $\{x^{i-1}\,y^{j-1}\,|\,(i,j)\in \nu_{a}\}$.

For $r=1$, $M(1,k)=\mbox{Hilb}^{k}[\mathbb{C}^2]$ is the Hilbert scheme of $k$ points in $\mathbb{C}^2$. It is the resolution of singularities of the $k$-th symmetric product of $\mathbb{C}^2$ and is given by
\bea
\mbox{Hilb}^{k}[\mathbb{C}^2]=\{I\subset \mathbb{C}[x,y]| \mbox{dim}(\mathbb{C}[x,y]/I)=k\}
\eea
and parameterizes the codimension $k$ ideals in $\mathbb{C}[x,y]$. The tangent space at $I\in \mbox{Hilb}^{k}[\mathbb{C}^{2}]$ is such that
\bea
T_{I}(\text{Hilb}^{k}[\mathbb{C}^2])\simeq \mbox{Hom}(I,\mathbb{C}[x,y]/I)\,.
\eea
The torus $T=U(1)\times U(1)$ action on $\mathbb{C}^2$
\bea
(x,y)\mapsto (e^{2\pi i \varepsilon_{1}}\,x,e^{2\pi i\varepsilon_{2}}\,y)\,,
\eea
induces an action on $\mbox{Hilb}^{k}[\mathbb{C}^2]$ with finite number of isolated fixed points labelled by the partitions of $k$. The fixed point corresponding to the partition $\lambda$ is
\bea
I_{\lambda}=\oplus_{(i,j)\ni \lambda}\mathbb{C}\,z_{1}^{i-1}z_{2}^{j-1}
\eea
The equivariant weights of the tangent bundle under the $U(1)\times U(1)$ action at a fixed point labelled by $\lambda$ are given by
\bea
\{t^{\lambda^{t}_{j}-i}q^{\lambda_{i}-j+1},t^{-\lambda^{t}_{j}+i-1}
q^{-\lambda_{i}+j}\,|\,(i,j)\in \lambda\}\,,\,\,\mbox{where} \,\,(q,t)=(t_{1},t_{2}^{-1})
\eea

For higher rank $r>1$ the equivariant weights of the tangent bundle are
\bea
\sum_{weights}e^{w}=\sum_{a,b=1}^{r}e_{a}e_{b}^{-1}\Big(\sum_{(i,j)\in \nu_{a}}\,t^{\nu^{t}_{b,j}-i}\,q^{\nu_{a,i}-j+1}+\sum_{(i,j)\in \nu_{b}}\,t^{-\nu^{t}_{a,j}+i-1}\,q^{-\nu_{b,i}+j}\Big)\,.\label{Tbundleweights}
\eea

\subsubsection{M-strings Dual to $(N,M)$ 5-Brane Web}

\begin{figure}[h]
  \centering
  % Requires \usepackage{graphicx}
  \includegraphics[width=10cm]{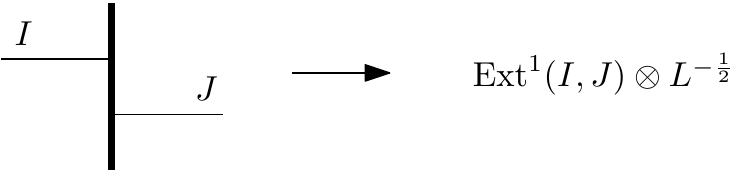}\\
\caption{Each pair of M2-brane stacks lead to a factor in the fiber wise description of the vector bundle. }  \label{bundle}
\end{figure}

\begin{figure}[h]
  \centering
  % Requires \usepackage{graphicx}
  \includegraphics[width=3cm]{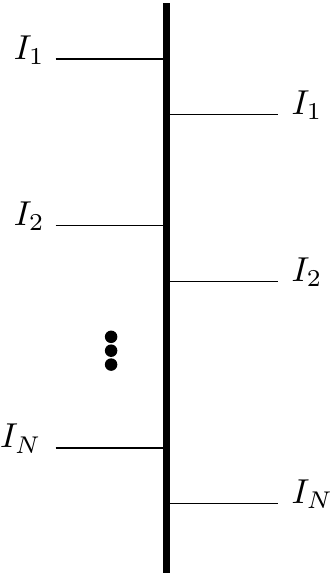}\\
  \caption{Pictorial representation of stacks of M2-branes ending on the M5-branes for the purpose of calculating the bundle.}\label{stacks}
\end{figure}

Before discussing the most general configuration let us consider the case of $M=1,N=k$ generic. In this case we have a single M5-brane wrapped on $S^1_{\text{M5}}$ with one transverse direction compactified to $S^1_{\text{trans}}$ and $k$ stacks of M2-branes ending on the M5-brane, after wrapping the circle, separated in the transverse space. The latter give $k$ colored points in the $\mathbb{R}^{4}_{\|}$ which is transverse to the M-string worldsheet inside the M5-brane worldvolume. Thus for the configuration in which there are $n_i$ M2-branes in the $i$-th stack one expects the moduli space of M-strings to be
\bea
H_{\vec{n}}:=\mbox{Hilb}^{n_1}[\mathbb{C}^2]\times \mbox{Hilb}^{n_2}[\mathbb{C}^2]\times \mathellipsis \times \mbox{Hilb}^{n_k}[\mathbb{C}^2]\,.
\eea
The vector bundle on $H_{\vec{n}}$ relevant for the $(2,0)$ theory can be determined from a simple rule, illustrated in \figref{bundle}, which gives a fiber wise description of the bundle \cite{mstrings}.
A stack of $n_{r}$ M2-branes ending on the M5-brane from the left and a stack of $n_{s}$ of M2-branes ending of the M5-branes from the left, labelled by $(r,s)$, gives rise to a factor $\mbox{Ext}^{1}(I,J)\otimes L^{-\frac{1}{2}}$ in the fiber of bundle over the point $(I,J)\in H_{n_1,n_2}$. The equivariant weights of the bundle at the fixed point of $H_{\vec{n}}$ labelled by the two partitions $\lambda$ and $\mu$ (with $|\lambda|=n_{r}$ and $|\mu|=n_{s}$) are given by \cite{Smirnov},
\bea
\{e^{id_{rs}}\,t^{\mu^{t}_{j}-i+\frac{1}{2}}q^{\lambda_{i}-j+\frac{1}{2}}\,|\,(i,j)\in \lambda\}\,\cup \{e^{id_{rs}}\,t^{-\lambda^{t}_{j}+i-\frac{1}{2}}\,q^{-\mu_{i}+j-\frac{1}{2}}\,|\,(i,j)\in \mu\}\,,\label{weights1}
\eea
where
\bea
e^{i\,d_{rs}}=e^{i(a_{r}-a_{s})}Q_{m}\,.
\eea
For $M=1,N=k$ we can draw the picture representing the $k$ stacks of M2-branes on the M5-brane as given in \figref{stacks}. We have depicted the stacks to be separated along the M5-brane, however, we mean by this their separation only in the transverse space, in order to clearly calculate the vector bundle of right moving fermions.

From \figref{stacks} it then follows that the fiber of the bundle over the point ${\bf I}=(I_{1},I_{2},\mathellipsis,I_{N})\in H_{\vec{n}}$ is given by
\bea
V|_{\bf I}=\oplus_{r,s=1}^{N}\mbox{Ext}^{1}(I_{r},I_{s})\otimes L^{-\frac{1}{2}}\,,
\eea
with weights given by
\bea\nonumber
\sum_{\text{weights}}e^{w}&=&\sum_{r,s=1}^{N}e^{id_{rs}}\Big(\sum_{(i,j)\in \nu_{r}}\,t^{\nu^{t}_{s,j}-i+\frac{1}{2}}\,q^{\nu_{r,i}-j+\frac{1}{2}}+\sum_{(i,j)\in \nu_{s}}\,t^{-\nu^{t}_{r,j}+i-\frac{1}{2}}\,q^{-\nu_{s,i}+j-\frac{1}{2}}\Big)\,,\\\label{weights2}
\eea
which, for $Q_{m}=\Big(\frac{q}{t}\Big)^{\pm \frac{1}{2}}$, are precisely the weights of the tangent bundle of the rank $N$ instanton moduli space at the fixed point labelled by ${\bf I}$ as discussed in section~\ref{Sect:hilbert}.

Thus the moduli space of the $m$ M-strings dual to the $(M,N)$ 5-brane web is the rank $N$ instanton moduli space of charge $m$ with vector bundle just the tangent bundle. The bundle has an extra $U(1)_{m}$ action with equivariant parameter $m$.

The $(2,0)$ elliptic genus for the configuration with $N$ stacks and total of $k$ M2-branes is then given by
\bea
{\cal Z}_{k}=\int_{M(N,k)}\prod_{i}\frac{x_{i}\theta(\tilde{x}+z;U)}{\theta(x;U)}=\int_{M(N,k)}\prod_{i}\frac{x_{i}\vartheta(\tilde{x}+z;U)}{\vartheta(x;U)}\,,
\eea
where $x_i$ and $\tilde{x}_{i}$ are the Chern roots of the tangent bundle and the bundle $V$ respectively (in this case $V$ is just the tangent bundle). Since $M(N,k)$ has isolated fixed points labelled by partitions as discussed in the last section we get using Eq.(\ref{weights2}),
\bea
{\cal Z}_{k}&=&\sum_{|\nu_{1}|+\mathellipsis+|\nu_{N}|=k}\prod_{r,s=1}^{N}\Big[\prod_{(i,j)\in \nu_{r}}\frac{\vartheta(e^{id_{rs}}t^{\nu^{t}_{s,j}-i+\frac{1}{2}}q^{\nu_{r,i}-j+\frac{1}{2}};U)}{\vartheta(e^{i(a_{r}-a_{s})}
t^{\nu^{t}_{s,j}-i}q^{\nu_{r,i}-j+1};U)}\\\nonumber
&&\times \prod_{(i,j)\in \nu_{s}}\frac{\vartheta(e^{id_{rs}}t^{-\nu^{t}_{r,j}+i-\frac{1}{2}}q^{-\nu_{s,i}+j-\frac{1}{2}};U)}{\vartheta(e^{i(a_{r}-a_{s})}
t^{-\nu^{t}_{r,j}+i-1}q^{-\nu_{s,i}+j};U)}\Big]
\eea
It turns out that the weights given above are precisely what were used to define $\vartheta(x,U)$ in Eq.(\ref{vartheta}). From Eq.(\ref{vartheta}) and Eq.(\ref{weights1}) we see that
\bea
\prod_{\text{\small weights of}\atop  {{\text{Ext}^{1}(I,J)\otimes L^{-\frac{1}{2}}} } }\vartheta(w;U)=(-1)^{|\lambda|+|\mu|}\,\vartheta_{\lambda\mu}(e^{i\,d_{rs}};U)
\eea
and this allows us to write
\bea
{\cal Z}_{k}&=&\sum_{|\nu_{1}|+\mathellipsis+|\nu_{N}|=k}\prod_{r,s=1}^{N}\frac{\vartheta_{\nu_{r}\nu_{s}}(e^{id_{rs}};U)}{\vartheta_{\nu_{r}\nu_{s}}(e^{i(a_{r}-a_{s})}\sqrt{\frac{q}{t}};U)}\\\nonumber
&=&\sum_{|\nu_{1}|+\mathellipsis+|\nu_{N}|=k}\prod_{r=1}^{N}\frac{\vartheta_{\nu_{r}\nu_{r}}(Q_{m};U)}{\vartheta_{\nu_{r}\nu_{r}}(\sqrt{\frac{q}{t}})}
\prod_{1\leq r<s\leq N}\frac{\vartheta_{\nu_{r}\nu_{s}}(e^{i(a_{r}-a_{s})}Q_{m};U)\vartheta_{\nu_{r}\nu_{s}}(e^{i(a_{r}-a_{s})}Q_{m}^{-1};U)}{
\vartheta_{\nu_{r}\nu_{s}}(e^{i(a_{r}-a_{s})}\sqrt{\frac{q}{t}};U)\vartheta_{\nu_{r}\nu_{s}}(e^{i(a_{r}-a_{s})}\sqrt{\frac{t}{q}};U)}\,,
\eea
where we have used the property $\vartheta_{\lambda\mu}(x)=\vartheta_{\mu\lambda}(x^{-1})(-1)^{|\lambda|+|\mu|}$. The full M-string partition function is given by
\bea
{\cal Z}(1,N)=\sum_{k}\varphi^{k}\,{\cal Z}_{k}\,.\label{MstringPF1}
\eea
If we identify $\varphi=Q_{T}$ then
\bea
{\cal Z}(1,N)=\frac{Z(1,N)}{W_{N}(\emptyset)}\,
\eea
where $Z(1,N)$ is the topological string partition function of the $(1,N)$ 5-brane web given in Eq.(\ref{PartFct20}).

In general we will have
\bea\nn
(M,N)\,\,\mbox{5-brane web} \mapsto \mbox{Moduli space of M-strings}&=&M(N,k_{1})\times M(N,k_{M})\\\nn
&& \mbox{with vector bundle $V(N,M)$}
\eea
Let $\vec{I}^{(i)}\in M(N,k_{i}),i=1,2,\mathellipsis,M$ be a fixed point. Then the fiber of the bundle $V(N,M)$ over the fixed point $(\vec{I}^{(1)},\vec{I}^{(2)},\mathellipsis,\vec{I}^{(M)})\in M(N,k_{1})\times \mathellipsis \times M(N,k_{M})$ is given by
\bea
V(N,M)|_{(\vec{I}^{(1)},\mathellipsis,\vec{I}^{(M)})}=\oplus_{i=1}^{M}\oplus_{r,s=1}^{N}\mbox{Ext}^{1}(I^{(i)}_{r},I^{(i+1)}_{s})\otimes L^{-\frac{1}{2}}\,.
\eea
The weights of this bundle at the fixed point $(\vec{I}^{(1)},\vec{I}^{(2)},\mathellipsis,\vec{I}^{(M)})$ labelled by partitions $(\nu^{(1)}_{1},\cdots,\nu^{(1)}_{N};\mathellipsis;\nu^{(1)}_{1},\cdots,\nu^{(1)}_{N})$ are given by
\bea\nonumber
\sum_{weights}e^{w}=\sum_{p=1}^{M}\sum_{r,s=1}^{N}e^{id_{rs}}\Big(\sum_{(i,j)\in \nu^{(p)}_{r}}\,t^{\nu^{t,(p+1)}_{s,j}-i+\frac{1}{2}}\,q^{\nu^{(p)}_{r,i}-j+\frac{1}{2}}+\sum_{(i,j)\in \nu^{(p+1)}_{s}}\,t^{-\nu^{t,(p)}_{r,j}+i-\frac{1}{2}}\,q^{-\nu^{(p+1)}_{s,i}+j-\frac{1}{2}}\Big)\,,\\\label{NMweights}
\eea
Using Eq.(\ref{Tbundleweights} we can determine the weights of the tangent bundle to the product of instanton moduli spaces to be,
\bea\nn
\sum_{weights}e^{w}=\sum_{p=1}^{M}\sum_{r,s=1}^{N}e^{i(a_{r}-a_{s})}\Big(\sum_{(i,j)\in \nu^{(p)}_{r}}\,t^{\nu^{t,(p)}_{s,j}-i}\,q^{\nu^{(p)}_{r,i}-j+1}+\sum_{(i,j)\in \nu^{(p)}_{s}}\,t^{-\nu^{t,(p)}_{r,j}+i-1}\,q^{-\nu^{(p)}_{s,i}+j}\Big)\,.\\\label{NMbundleweights}
\eea
The $(2,0)$ elliptic genus of $M(N,k_{1})\times \mathellipsis \times M(N,k_{M})$ with bundle $V(N,M)$ is
\begin{align}
{\cal Z}_{k_{1},\mathellipsis,k_{M}}&=\sum_{|\nu^{(p)}|=k_{p}}\prod_{p=1}^{M}\prod_{r,s=1}^{N}\frac{\prod_{(i,j)\in \nu^{(p)}_{r}}\vartheta(e^{id_{rs}}t^{\nu^{t,(p+1)}_{s,j}-i+\frac{1}{2}}q^{\nu^{(p)}_{r,i}-j+\frac{1}{2}};U)}
{\prod_{(i,j)\in \nu^{(p)}_{r}}\vartheta(e^{i(a_{r}-a_{s})}t^{\nu^{t,(p)}_{s,j}-i}q^{\nu^{(p)}_{r,i}-j+1};U)}\nonumber\\
&\times
\frac{
\prod_{(i,j)\in \nu^{(p+1)}_{s}}\vartheta(e^{id_{rs}}t^{-\nu^{t,(p)}_{r,j}+i-\frac{1}{2}}q^{-\nu^{(p+1)}_{s,i}+j-\frac{1}{2}};U)}
{\prod_{(i,j)\in \nu^{(p)}_{s}}\vartheta(e^{i(a_{r}-a_{s})}t^{-\nu^{t,(p)}_{r,j}+i-1}q^{-\nu^{(p)}_{s,i}+j};U)}\,.
\end{align}
This can furthermore be written in the form
{\allowdisplaybreaks\begin{align}
{\cal Z}_{k_{1},\mathellipsis,k_{M}}&=\sum_{|\nu^{(p)}|=k_{p}}\prod_{p=1}^{M}\prod_{r,s=1}^{N}\frac{\vartheta_{\nu^{(p)}_{r}\nu^{(p+1)}_{s}}(e^{id_{rs}};U)}{\vartheta_{\nu^{(p)}_{r}\nu^{(p)}_{s}}(e^{i(a_{r}-a_{s})}\sqrt{\frac{q}{t}};U)}\\\nn
&=\sum_{|\nu^{(p)}|=k_{p}}\prod_{p=1}^{M}\Big(\prod_{r=1}^{N}\frac{\vartheta_{\nu^{(p)}_{r}\nu^{(p+1)}_{r}}(Q_{m};U)}
{\vartheta_{\nu^{(p)}_{r}\nu^{(p)}_{r}}(\sqrt{\frac{q}{t}};U)}\Big)\\
&\hspace{0.3cm}\times \prod_{1\leq r<s\leq N}\frac{\vartheta_{\nu^{(p)}_{r}\nu^{(p+1)}_{s}}(e^{i(a_{r}-a_{s})}Q_{m};U)\vartheta_{\nu^{(p+1)}_{r}\nu^{(p)}_{s}}(e^{i(a_{r}-a_{s})}Q_{m}^{-1};U)}
{\vartheta_{\nu^{(p)}_{r}\nu^{(p)}_{s}}(e^{i(a_{r}-a_{s})}\sqrt{\frac{q}{t}};U)
\vartheta_{\nu^{(p)}_{r}\nu^{(p)}_{s}}(e^{i(a_{r}-a_{s})}\sqrt{\frac{t}{q}};U)}\,\label{MstringsNM1}
\end{align}}
If we form the full partition function
\bea
{\cal Z}(N,M)=\sum_{k_{1},\mathellipsis,k_{M}}\varphi_{1}^{k_{1}}\mathellipsis \varphi_{M}^{k_{M}}{\cal Z}_{k_{1}\mathellipsis k_{M}}
\eea
then from Eq.(\ref{MstringsNM1}) and Eq.(\ref{GenTopVertex}) we see that upon identifying $\varphi_{p}=Q_{B_{p}},p=1,2,\mathellipsis,M$ we have
\bea
{\cal Z}(N,M)=\frac{Z(N,M)}{W_{N}(\emptyset)^{M}}\,.
\eea
Thus the $(2,0)$ elliptic genus on the product of instanton moduli spaces, with bundle $V(N,M)$, gives the topological string partition function of the $(N,M)$ 5-brane configuration which is the same as the Nekrasov partition function of the corresponding gauge theory on the web.

\subsection{Nekrasov's instanton partition function}\label{Sect:Instanton}
In the previous sections we saw that the partition function of the $(N,M)$ web of branes can be calculated in two different ways using the topological vertex and $(2,0$ elliptic genus. The $(N,M)$ brane web gives rise to a quiver gauge theory and one can use Nekrasov's instanton calculus to determine the gauge theory instanton partition function.

In the limit $Q_{U}, Q_{T}\mapsto 0$ the circle transverse to the D5-branes and the circle on which the D5-branes are wrapped is decompactified and the corresponding quiver gauge theory was studied in \cite{Bao:2011rc}. The partition function is given by \cite{Bao:2011rc}\,,
\bea\nn
{\cal Z}_{\text {gauge}}=\sum_{\alpha^{(i)}_{a}}\Big(\varphi_{i}^{|\alpha^{(i)}|}\Big)\Big(\prod_{i=1}^{M}z_{\text {vec}}(\alpha^{(i)})\Big)
\Big(\prod_{i=1}^{M}z_{\text {bifun}}(\alpha^{(i-1)},\alpha^{(i)})z_{\text {bifun}}(\alpha^{(i)},\alpha^{(i+1)})\Big)\,.
\eea
where $\vec{\alpha}^{(0)}$ and $\vec{\alpha}^{(M)}$ are all trivial partitions. The factor $z_{\text{vec}}$ and $z_{\text{bifun}}$ which correspond to vector multiplet contribution and bifundamental hypermultiplet contribution are defined using the function $R_{\mu\nu}$,
\bea\label{Rfactor}
R_{\mu\nu}(Q,\epsilon)&=&\prod_{(i,j)\in \mu}(Q^{-\frac{1}{2}}\,q^{-\frac{\nu^{t}_{j}+\mu_{i}-i-j+1}{2}}-Q^{\frac{1}{2}}\,q^{\frac{\nu^{t}_{j}+\mu_{i}-i-j+1}{2}})\,\\\nn
&\times&\prod_{(i,j)\in \nu}(Q^{-\frac{1}{2}}\,q^{\frac{\mu^{t}_{j}+\nu_{i}-i-j+1}{2}}-Q^{\frac{1}{2}}\,q^{-\frac{\mu^{t}_{j}+\nu_{i}-i-j+1}{2}})\,.
\eea
and are given by
\bea\label{ZVH}
Z_{\text{vec}}(\vec{\alpha}^{(i)})&=&\prod_{a,b=1}^{N}Z_{\alpha^{(i)}_{a}\alpha^{(i)}_{b}}(Q_{ab},\epsilon)\,,\,\,\,
Z_{\alpha^{(i)}_{a}\alpha^{(i)}_{b}}(Q_{ab},\epsilon)=\frac{1}{R_{\alpha_{a}\alpha_{b}}(Q_{ab},\epsilon)}\,,\\\nn
Z_{\text{bifun}}(\vec{\alpha}^{(i)},\vec{\alpha}^{(j)})&=&
\prod_{a,b=1}^{N}Z_{\alpha^{(i)}_{a}\alpha^{(j)}_{b}}(Q_{ab}Q_{m},\epsilon)\,,\,\,\,\,
Z_{\alpha^{(i)}_{a}\alpha^{(j)}_{b}}(Q_{ab}Q_{m},\epsilon)=R_{\vec{\alpha}^{i},\vec{\alpha}^{j}}(Q_{ab}Q_{m},\epsilon)\,.
\eea
Using Eq.(\ref{Rfactor}) and Eq.(\ref{ZVH}) we see that ${\cal Z}_{\text{gauge}}$ is precisely $Q_{U},Q_{T}\mapsto 0$ limit of $\widehat{Z}(M,N)$ given in Eq.(\ref{zmnpf}),\footnote{We have restricted ourselves to discussing the unrefined case $\epsilon_{1}=-\epsilon_{2}=\epsilon$.}
\bea\nn
\lim_{Q_{U},Q_{T}\mapsto 0}\widehat{Z}(M,N)&=&\sum_{\alpha^{(i)}_{a}}\Big(\prod_{i=1}^{M}Q_{B_{i}}^{|\alpha^{(i)}|}\Big)
\prod_{i=1}^{M}\prod_{a=1}^{N}
\frac{R_{\alpha^{(i+1)}_{a}\alpha^{(i)}_{a}}(Q_{m},\epsilon)}{R_{\alpha^{(i)}_{a}\alpha^{(i)}_{a}}(1,\epsilon)}\\\nn
&\times & \prod_{1\leq a<b\leq N}\prod_{i=1}^{M}\frac{R_{\alpha^{(i)}_{a}\alpha^{(i+1)}_{b}}(Q_{ab}Q_{m}^{-1},\epsilon)
R_{\alpha^{(i+1)}_{a}\alpha^{(i)}_{b}}(Q_{ab}Q_{m},\epsilon)}{R_{\alpha^{(i)}_{a}\alpha^{(i)}_{b}}(Q_{ab},\epsilon)
R_{\alpha^{(i)}_{a}\alpha^{(i)}_{b}}(Q_{ab},\epsilon)}\,,
\eea
where in the above equation $\alpha^{(M)}$ is the set of trivial partitions.

\begin{figure}[h]
  \centering
  % Requires \usepackage{graphicx}
  \includegraphics[width=3in]{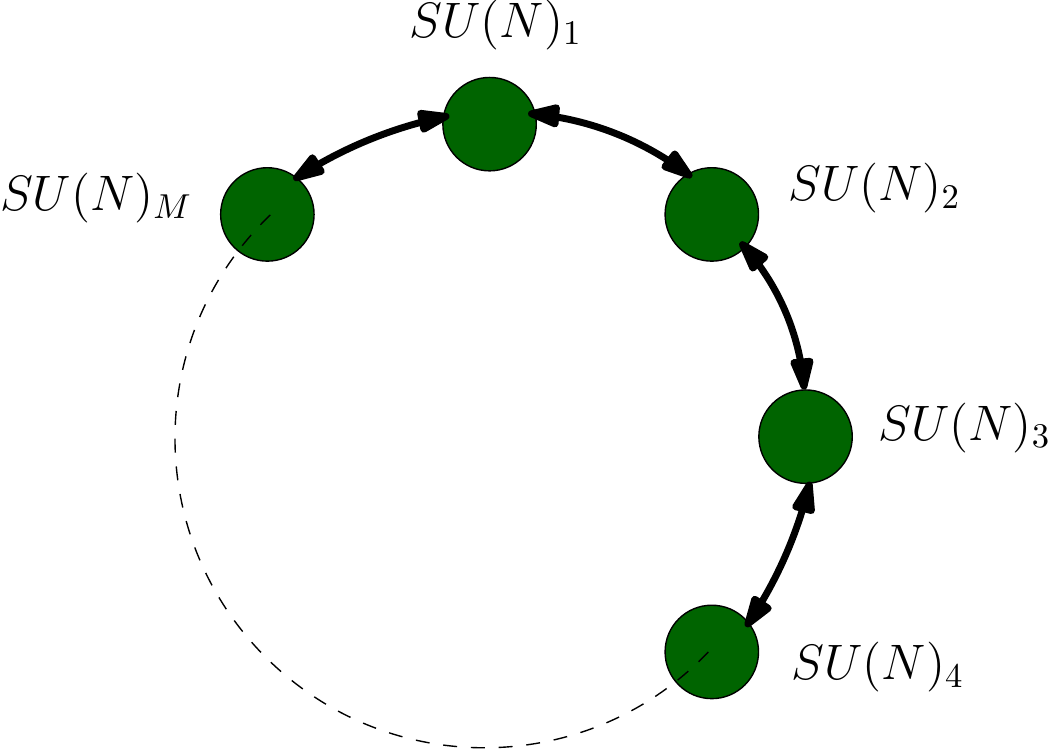}\\
  \caption{The quiver of the gauge theory corresponding to an $(N,M)$ web.}\label{quiver}
\end{figure}
The quiver corresponding to the $(N,M)$ brane web is shown in \figref{quiver}. In this case, because of the compactification of the direction transverse to the D5-branes, strings can wind around the circle before ending on the D5-branes. The factor $Z_{\alpha^{(i)}_{a}\alpha^{(i)}_{b}}(Q_{ab},\epsilon)$ is the contribution coming from strings stretched between the $a$-th and the $b$-th brane in the $i$-th stack. If the NS-branes are compactified on a circle then we have strings stretched between the $a$-th and the $b$-th brane which wind around the circle. Thus the contribution of the vector multiplet becomes,
\bea
Z_{\alpha^{(i)}_{a}\alpha^{(i)}_{b}}(Q_{ab},\epsilon)\mapsto
\prod_{k\in \mathbb{Z}}^{\infty}Z_{\alpha^{(i)}_{a}\alpha^{(i)}_{b}}(Q_{ab}Q_{U}^{k},\epsilon)=\frac{1}{\vartheta_{\alpha_{a}\alpha_{b}}(Q_{ab})}\,.
\eea
Similarly the bifundamental hypermultiplet contribution $Z_{\alpha^{(i)}_{a}\alpha^{(j)}_{b}}(Q_{ab}Q_{m},\epsilon)$ comes from strings stretched between the $a$-th and the $b$-th D5-brane of the $i$-th and the $(i+1)$-th stack respectively. Taking into account the winding of the strings we get
\bea
Z_{\alpha^{(i)}_{a}\alpha^{(j)}_{b}}(Q_{ab}Q_{m},\epsilon)\mapsto \prod_{k\in \mathbb{Z}}Z_{\alpha^{(i)}_{a}\alpha^{(j)}_{b}}(Q_{ab}Q_{m}Q_{U}^{k},\epsilon)=\vartheta_{\alpha^{(i+1)}_{a}\,\alpha^{(i)}_{b}}(Q_{ab}Q_{m})\,.
\eea
Thus the vector multiplet and bifundamental hypermultiplet contributions are
\bea
Z_{\text{vec}}=\prod_{i=1}^{M}\prod_{a,b=1}^{N}\frac{1}{\vartheta_{\alpha^{(i)}_{a}\alpha^{(i)}_{b}}(Q_{ab})}\,,\,\,
Z_{\text{bifun}}=\prod_{i=1}^{M}\prod_{a,b=1}^{N}\vartheta_{\alpha^{(i+1)}_{a}\alpha^{(i)}_{b}}(Q_{ab}Q_{m})\,.
\eea
Thus the full instanton partition function of the gauge theory corresponding to the $(N,M)$ brane web is given by
\bea\nn
&&\sum_{\alpha^{i}_{a}}\varphi_{i}^{|\alpha^{(i)}|}\,Z_{vec}\times Z_{bifun}
=\sum_{\alpha^{i}_{a}}\varphi_{i}^{|\alpha^{(i)}|}\prod_{i=1}^{M}\prod_{a,b=1}^{N}\frac{\vartheta_{\alpha^{(i+1)}_{a}\alpha^{(i)}_{b}}(Q_{ab}Q_{m})}
{\vartheta_{\alpha^{(i)}_{a}\alpha^{(i)}_{b}}(Q_{ab})}\\\nn
&=&\sum_{\alpha^{i}_{a}}\varphi_{i}^{|\alpha^{(i)}|}\Big(\prod_{i=1}^{M}\prod_{a=1}^{N}
\frac{\vartheta_{\alpha^{(i+1)}_{a}\alpha^{(i)}_{a}}(Q_{m})}{\vartheta_{\alpha^{(i)}_{a}\alpha^{(i)}_{a}}(1)}\Big)\Big(\prod_{1\leq a<b\leq N}\prod_{i=1}^{M}
\frac{\vartheta_{\alpha^{(i+1)}_{a}\alpha^{(i)}_{b}}(Q_{ab}Q_{m})\vartheta_{\alpha^{(i)}_{a}\alpha^{(i+1)}_{b}}(Q^{-1}_{ab}Q_{m})}
{\vartheta_{\alpha^{(i)}_{a}\alpha^{(i)}_{b}}(Q_{ab})\vartheta_{\alpha^{(i)}_{a}\alpha^{(i)}_{b}}(Q^{-1}_{ab})}\Big)\,,
\eea
which is precisely the partition function $\widehat{Z}(M,N)$ given in Eq.(\ref{zmnpf}).

\section{Free Energy for $M=N=1$}\label{Sect:TopString}
%{Topological Strings and ${\cal N}=2^{*}$ Gauge Theory}\label{Sect:TopString}

In this section we provide another approach to computing the mass-deformed gauge theory partition functions, which make particularly the modular properties in \emph{e.g.}~(\ref{ModularProperties}) (see also \cite{mstrings}) somewhat more tangible. We will focus on the partition function for the simplest case (\emph{i.e.} $M=N=1$) as written in equ.~(\ref{u1pf}).
%%%%%%%%%%%%%%%%%%%%%%%%%%
\subsection{Genus Expansion}
\begin{figure}[h!]
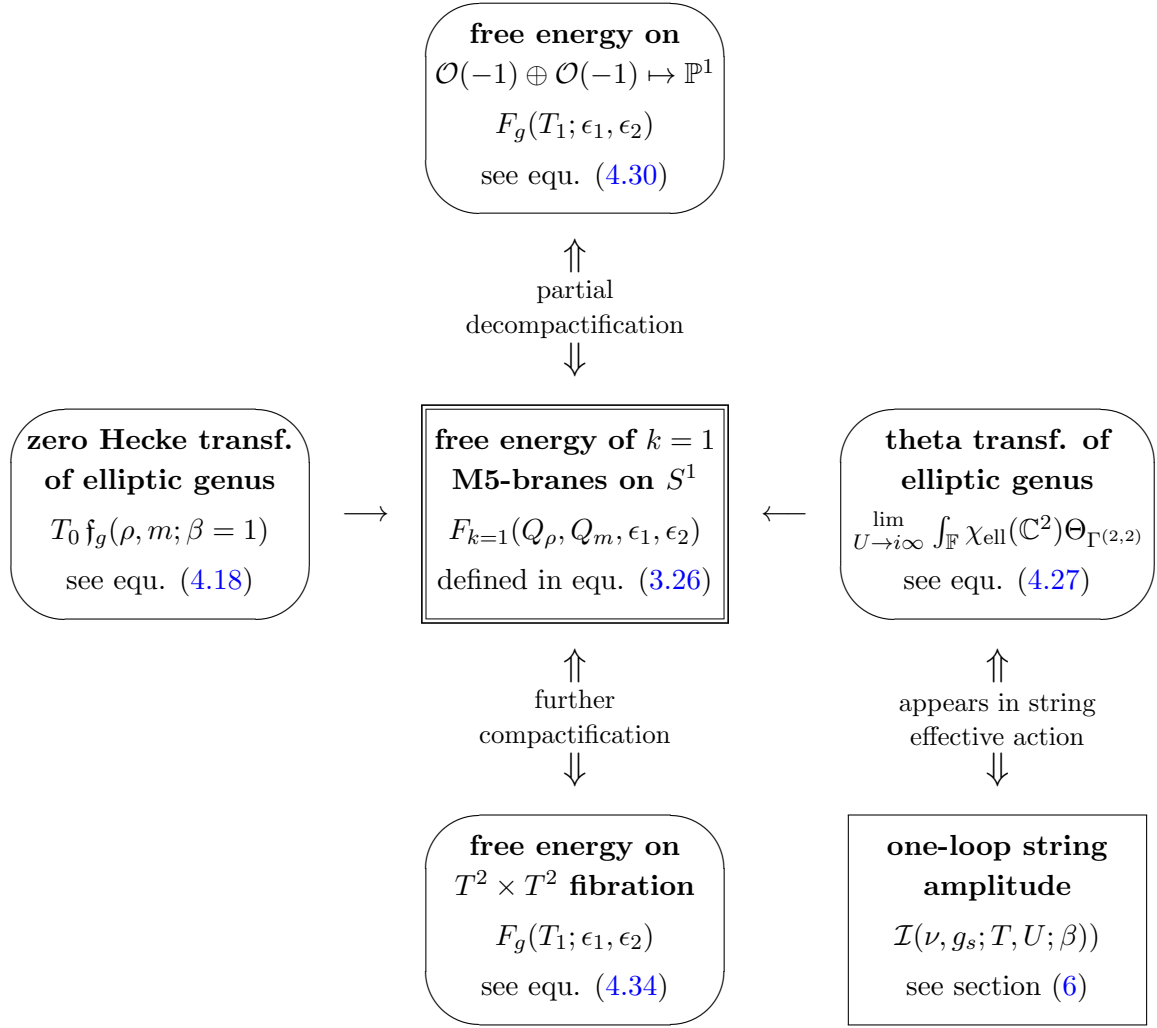

\begin{small}
\begin{center}
\begin{tabular}{ccccc}
&&\ovalbox{\parbox{3.7cm}{\begin{center}${}$\\[-16pt]\textbf{free energy on ${\cal O}(-1) \oplus {\cal O}(-1) \mapsto \mathbb{P}^{1}$} \\[4pt] $F_g(T_1;\epsilon_1,\epsilon_2)$  \\[4pt] see equ. (\ref{DecompactifiedFreeEnergy})\\[-20pt]${}$\end{center}}} &&\\
&&&&\\[-6pt]
&& {\large$\Uparrow$} &&\\[-12pt]
&& \parbox{3cm}{\footnotesize\begin{center}partial decompactification\end{center}} &&\\[-12pt]
&& {\large $\Downarrow$} &&\\
&&&&\\[-6pt]
\ovalbox{\parbox{3.7cm}{\begin{center}${}$\\[-16pt]\textbf{zero Hecke transf. of elliptic genus} \\[4pt] $T_0\, \mathfrak{f}_g(\rho,m;\beta=1)$  \\[4pt] see equ. (\ref{FinAnswerHecke})\\[-20pt]${}$\end{center}}}    &      $\longrightarrow$      &  \doublebox{\parbox{3.7cm}{\begin{center}${}$\\[-16pt]\textbf{free energy of $k=1$ M5-branes on $S^1$} \\[4pt] $F_{k=1}(Q_\rho,Q_m,\epsilon_1,\epsilon_2)$  \\[4pt] defined in equ. (\ref{u1pf})\\[-20pt]${}$\end{center}}}    &     $\longleftarrow$        &\ovalbox{\parbox{3.9cm}{\begin{center}${}$\\[-16pt]\textbf{theta transf. of elliptic genus} \\[4pt] ${\lim\atop{U\to i\infty}}\int_{\mathbb{F}}\chi_{\text{ell}}(\mathbb{C}^2)\Theta_{\Gamma^{(2,2)}}$  \\[4pt] see equ. (\ref{TorusIntegralFin})\\[-20pt]${}$\end{center}}}\\
&&&&\\[-6pt]
&& {\large$\Uparrow$} && {\large$\Uparrow$}\\[-12pt]
&& \parbox{3cm}{\footnotesize\begin{center}further compactification\end{center}} && \parbox{3cm}{\footnotesize\begin{center}appears in string effective action\end{center}}\\[-12pt]
&& {\large $\Downarrow$} &&{\large $\Downarrow$}\\
&&&&\\[-6pt]
&&\ovalbox{\parbox{3.7cm}{\begin{center}${}$\\[-16pt]\textbf{free energy on $T^2\times T^2$ fibration} \\[4pt] $F_g(T_1;\epsilon_1,\epsilon_2)$  \\[4pt] see equ. (\ref{FreeEnergyCompact})\\[-20pt]${}$\end{center}}} &&\fbox{\parbox{3.7cm}{\begin{center}${}$\\[-16pt]\textbf{one-loop string amplitude} \\[4pt] $\mathcal{I}(\nu,g_s;T,U;\beta))$  \\[4pt] see section (\ref{Sect:OneLoopStringAmp})\\[-20pt]${}$\end{center}}}
\end{tabular}
\end{center}
\end{small}
\caption{\emph{Connections between the free energy of a single M5 brane and objects discussed in subsequent sections:} Simple arrows denote relations which we checked for $\beta=1$, while double arrows denote relations holding in a general $\Omega$ background.}
\label{Fig:RefsFreeEngk1}
\end{figure}

Let us consider the partition function $W_1(\emptyset)$ in equ.~(\ref{u1pf}) and study in more detail the corresponding free energy
\begin{align}
F_{k=1}&(Q_\rho,Q_m,\epsilon_{1},\epsilon_{2})=\ln \,W_{1}(\emptyset )\nonumber\\
&=-\sum_{n,m=1}^{\infty}\,\frac{Q_\rho^{nm}}{n(1-t_1^n)(1-t_2^{-n})}\left[(e^{nT_1}+e^{nT_2})(t_1/t_2)^{n/2}-(1+(t_1/t_2)^n)\right]\,,\nonumber
\end{align}
where we have introduced the K\"ahler parameters $(T_1,T_2)$, by writing
\begin{align}
&Q_\rho=e^{-T}=e^{-(T_1+T_2)}\,,&&\text{and} &&Q_m=e^{-T_2}\,.
\end{align}
Furthermore, we are using the notation
\begin{align}
&t_1=e^{ig_s \sqrt{\beta}}\,,&&t_2=e^{-ig_s/\sqrt{\beta}}\,,\label{EquivariantParameters}
\end{align}
instead of $(t,q)$ which is closer to the literature on equivariant elliptic genera. Furthermore, we write for the deformation parameters
\begin{align}
&\epsilon_1=g_s\sqrt{\beta}\,,&&\text{and} &&\epsilon_2=-g_s/\sqrt{\beta}\,,
\end{align}
which allows us to expand the free energy in powers of $g_s$
\begin{align}
F_{k=1}&(Q_\rho,Q_m,\beta,g_s)=\sum_{g=0}^\infty g_s^{2g-2} F_g^{(k=1)}(Q_\rho,Q_m,\beta)\,,\label{GenExpansion}
\end{align}
with the coefficients $F_g^{(k=1)}$ given by
\bea
F_g^{(k=1)}(Q_\rho,Q_m,\beta)=-\sum_{n,m=1}^{\infty}\,\frac{Q_\rho^{nm}}{n^{3-2g}}\Big[(e^{nT_1}+e^{nT_2})f_{g}(\beta)-a_{g}(\beta)\Big]\,.\label{GenusgFree}
\eea
From the perspective of the (refined) topological string, $g_s$ has the interpretation of the topological string coupling, such that (\ref{GenExpansion}) corresponds to a genus expansion. The functions $f_{g}(\beta)$ and $a_g(\beta)$ appear as coefficients in the following series expansion
\begin{align}
\sum_{g=0}^{\infty}\,f_{g}(\beta)\,g_{s}^{2g-2}=\frac{\sqrt{t_1/t_2}}{(1-t_1)(1-t_2^{-1})}\,,&&\sum_{g\geq 0}a_{g}(\beta)\,g_{s}^{2g-2}=\frac{1+t_1/t_2}{(1-t_1)(1-t_2^{-1})}\,.
\end{align}
and can explicitly be written in the form ($\mathfrak{c}_g=(-1)^{g-1}\frac{B_{2g}(2g-1)}{(2g)!}$)
\begin{align}
f_{g}(\beta)&=\mathfrak{c}_g+\sum_{m=1}^{g}\Big(\sqrt{\beta}-\frac{1}{\sqrt{\beta}}\Big)^{2m}\,b_{g,m}\,,\label{CoeffsBeta}\\
a_{g}(\beta)&=(2\mathfrak{c}_g+\delta_{g,1})+\sum_{m=1}^{g}(\sqrt{\beta}-\frac{1}{\sqrt{\beta}})^{2m}\,a_{g,m}\,.\label{agbCoeffs}
\end{align}
The coefficients $a_{g,m}$ and $b_{g,m}$ can equally be expressed in terms of Bernoulli numbers $B_{2g}$, which, however, will not be important in the following. Indeed, from now on we consider $\beta=1$, such that the sums in (\ref{CoeffsBeta}) and (\ref{agbCoeffs}) do not contribute. In the following we discuss the connection of $F_g^{(k=1)}(Q_\rho,Q_m,\beta=1)$ to the equivariant elliptic genus of $\mathbb{C}^2$, which is further discussed in section~\ref{Sect:EllGen}:
\begin{enumerate}
\item The coefficient $F_g^{(k=1)}$ of order $g$ can be written as a particular type of Hecke transform of the elliptic genus, introduced in~\cite{hep-th/0002169}
\item The genus $g$ free energy can be obtained from the elliptic genus through a particular type of $SL(2,\mathbb{Z})$ invariant integration. This relation is known in the literature as theta-transform~\cite{Borcherds2,Kontsevich,Prasad}. The torus integral in turn appears as a one-loop scattering amplitude computing a particular class of higher derivative terms in type II string theory on $T^2$, as we discuss in section~\ref{Sect:OneLoopStringAmp}
\end{enumerate}
For the reader's convenience, we have summarised these connections in \figref{Fig:RefsFreeEngk1}.
%%%%%%%%%%%%%%%%%%%%%%%%%%%%
%%%%%%%%%%
\subsection{Hecke Transformation and elliptic genus of $\mathbb{C}^2$}\label{Sect:Hecke}
Using the definition of the polylogarithm $\text{Li}_n(z)=\sum_{m=1}^\infty\frac{z^m}{m^n}$,
we can rewrite expansion coefficients (\ref{GenusgFree}) in the following form
\begin{align}
&F_{g}^{(k=1)}(Q_\rho,Q_m,\beta=1)\nonumber\\
&=-\mathfrak{c}_g\sum_{n=1}^{\infty}\Big(Li_{3-2g}(Q_\rho^{n}\,Q_m^{-1})+Li_{3-2g}(Q_\rho^{n-1}\,Q_m)-2Li_{3-2g}(Q_\rho^{n})\Big)+\delta_{g,1}\sum_{n=1}^{\infty}Li_{1}(Q_\rho^{n})\nn\\
&=-D_{g}(0,1;\beta=1)\,Li_{3-2g}(Q_m)-\sum_{\ell\in\mathbb{Z}}\sum_{n=1}^{\infty}\Big[D_{g}(0,\ell;\beta=1)\,Li_{3-2g}(Q_\rho^{n}Q_m^\ell)\Big]\label{Hecke}
%&=D_{g}(0,1)\,Li_{3-2g}(y)+\sum_{k=1}^{\infty}\Big[D_{g}(0,1)\,Li_{3-2g}(z^{k}y)+D_{g}(0,-1)Li_{3-2g}(z^{k}y^{-1})+D_{g}(0,0)Li_{3-2g}(z^k)\Big]\nn
\end{align}
where we have introduced the coefficients
\bea
D_{g}(0,1;\beta=1)&=&D_{g}(0,-1;1)=\mathfrak{c}_g=\frac{(-1)^{g-1}B_{2g}(2g-1)}{(2g)!}\,,\\
D_{g}(0,0;\beta=1)&=&-2\mathfrak{c}_g+\delta_{g,1}=-2\frac{(-1)^{g-1}B_{2g}(2g-1)}{(2g)!}+\delta_{g,1}\,,\\
D_g(0,\ell;\beta=1)&=&0\,\hspace{1cm}\forall |\ell|>1\,.\label{Vanishl0}
\eea
These are in fact the Fourier coefficients of the equivariant elliptic genus of $\mathbb{C}^2$ evaluated at $\beta=1$\footnote{Here we denote the modular parameter by $\tau$ and introduce $\nome=e^{2\pi i\tau}$ to avoid confusion with previous sections while still remaining close to the literature on modular forms.}
\begin{align}
\chi_{\text{ell}}&(\mathbb{C}^{2};\tau,\nu,g_s;\beta)=\frac{\theta_1(\tau;\,y^{-1}\sqrt{\frac{t_{2}}{t_{1}}})\,\theta_1(\tau;\,y^{-1}\sqrt{\frac{t_{1}}{t_{2}}})}{\theta_1(\tau;\,t_{1}^{-1})\,\theta_1(\tau;\,t_{2}^{-1})}\nonumber\\
&=\sum_{g=0}^\infty g_s^{2g-2}\,\mathfrak{f}_g(\tau,\nu,\beta)=\sum_{g,n=0}^{\infty}\sum_{\ell}\,g_{s}^{2g-2}e^{2\pi i(n\tau+\ell \nu)} D_g(n,\ell;\beta)\,,\label{BetaExpansionEllk0firstapp}
\end{align}
which we discuss in great deal in the following section~\ref{Sect:EllGen}. Note that (\ref{Vanishl0}) reflects the asymptotic behaviour of the equivariant elliptic genus at the cusp at $\tau=i\infty$
\begin{align}
\lim_{\tau\to i\infty}\chi_{\text{ell}}(\mathbb{C}^2;\tau,\nu,g_s;\beta)&=\sum_{g=0}^\infty\sum_{\ell}  g_s^{2g-2}y^\ell D_g(0,l;\beta)\nonumber\\
&=-\frac{e^{\frac{ig_s}{2}\left(\sqrt{\beta}+\tfrac{1}{\sqrt{\beta}}\right)}\left[y+\frac{1}{y}-e^{\frac{ig_s}{2}\left(\sqrt{\beta}+\tfrac{1}{\sqrt{\beta}}\right)}-e^{-\frac{ig_s}{2}\left(\sqrt{\beta}+\tfrac{1}{\sqrt{\beta}}\right)}\right]}{\left(e^{ig_s/\sqrt{\beta}}-1\right)\left(e^{ig_s\sqrt{\beta}}-1\right)}\,.\label{AsymCusp}
\end{align}
Since, as we discuss in section~\ref{Sect:EllGen}, $\mathfrak{f}_g(\tau,\nu,\beta=1)$ is a weak Jacobi form of weight $2g-2$ and index $1$, we observe that the structure of (\ref{Hecke}) is very reminiscent of a Hecke transformation. Indeed, for $\ell\in\mathbb{N}$ define the Hecke transform of a weight $k$ Jacobi form $\Phi_k(\tau,z)=\sum_{m,n}h(m,n)\nome^m z^n$ in the following manner \cite{EichlerZagier}
\begin{align}
T_\ell \Phi_k(\tau,z)=\ell^{k-1}\sum_{{ad=\ell}\atop{a>0}}\sum_{b=0}^{d-1}d^{-k}\Phi\left(\frac{a\tau+b}{d},az\right)\,,
\end{align}
while for $\ell=0$ we use the definition \cite{hep-th/0002169}
\begin{align}
T_0 \Phi_k(\tau,z)=\frac{h(0,0)}{2} \zeta(1-k) +\sum_{(0,m,n)>0}h(0,n)\text{Li}_{1-k}(\nome^m z^n)\,,\label{Hecke0}
\end{align}
with the range of summation defined as
\begin{align}
(0,m,n)>0\Longleftrightarrow \left\{\begin{array}{l}m>0\,, \\ m=0,\,n>0\,.\end{array}\right.
\end{align}
Remembering furthermore $\text{Li}_n(1)=\zeta(n)$, we find indeed
\begin{align}
F_g^{(k=1)}(Q_\rho,Q_m,\beta=1)=T_0 \,\mathfrak{f}_g(\rho,m,\beta=1)\,.\label{FinAnswerHecke}
\end{align}
Notice, however, that (\ref{Hecke0}) differs from the Hecke transformations introduced in \cite{EichlerZagier}, explaining the modular properties of the partition functions discussed in the previous sections.
%%%%%%%%%%%%%%%%%%%%%%%%%%%%
\subsection{Theta Correspondence and Torus Integration}\label{Sect:TorusIntegration}
Another connection between the equivariant elliptic genus and the free energy is via so-called (singular) theta-correspondences \cite{Borcherds2,Kontsevich,Prasad}. To explain this, consider two Lie groups $(G_1,G_2)$ such that $G_1\times G_2$ is a subgroup (of the universal cover) of a larger symplectic group $Sp(W)$, for $W$ a symplectic vector space. We further assume that $G_1$ is the centraliser of $G_2$ in $Sp(W)$ and vice versa. In this case, $(G_1,G_2)$ are called a \emph{dual reductive pair} \cite{Howe}. Defining automorphic forms for both Lie groups as irreducible components in the decomposition of the spaces $L^2(G_i/G_i(\mathbb{Z}))$ (for some discrete subgroup $G_i(\mathbb{Z})$), theta correspondences are defined as integral transforms of automorphic representations of $G_1$ into automorphic representations of $G_2$.

In the case at hand $(G_1,G_2)=(SL(2,\mathbb{Z}),SO(2,2;\mathbb{Z}))$ with $SL(2,\mathbb{Z})\times SO(2,2;\mathbb{Z})\subset Sp(8)$. The kernel of the transformation is a  suitable Siegel-Narain-theta function which can be written as a sum over the Narain lattice $\Gamma^{(2,2)}$ of signature $(2,2)$ (see (\ref{NarainPartitionFunction}) for the specific denfition\footnote{Our notation follows the string conventions in view of the string amplitude computation in section~\ref{Sect:OneLoopStringAmp}.}) with possible insertions of the Narain momenta $(P_L,P_R)$. We consider a theta transform of $\chi_{\text{ell}}$, with automorphic transformations of $SL(2,\mathbb{Z})$ acting on the parameters $\tau$ and $\nu$, while $SO(2,2;\mathbb{Z})$ acts on the complex structure and K\"aler moduli $(U,T)$ of a $T^2$ torus respectively. To be more precise, let us consider
\begin{align}
\mathcal{I}(\nu,g_s;T,U;\beta)=\int_{\mathbb{F}} \frac{d^2\tau}{\tau_2}\sum_{P_L,P_R\in\Gamma^{(2,2)}}\chi_{\text{ell}}(\mathbb{C}^2;\tau,\nu\tau_2 P_R,g_s\tau_2 P_R,\beta)\nome^{\frac{1}{2}|P_L|^2}\bar{\nome}^{\frac{1}{2}|P_R|^2}\,.\label{LoopIntegral}
\end{align}
Despite the fact that (\ref{LoopIntegral}) can be defined in full generality, we will from now on consider the simpler case $\beta=1$ (and drop it as an argument from now on). To perform the integration over the fundamental domain $\mathbb{F}$ of $SL(2,\mathbb{Z})$, there are several methods available in the literature. To make contact with (\ref{Hecke}) we can make use of the method of orbits, first introduce in \cite{Dixon:1990pc,Harvey:1995fq} and further developed in \cite{Borcherds2,Lerche:1999ju,Foerger:1998kw}.\footnote{More recent developments can be found in \cite{Angelantonj:2011br}.} We essentially follow the same steps as in \cite{HS} and start out by performing a Poisson resummation
\begin{align}
\mathcal{I}(\nu,g_s;T,U)=T_2\int\frac{d^2\tau}{\tau_2^2}\sum_{k_{1,2},n_{1,2}\atop n,g,\ell}D_g(n,\ell;1)\,\nome^ne^{-\frac{\pi T_2}{U_2\tau_2}|\mathcal{A}|^2-2\pi iT\text{det}A+\frac{\pi i(\lambda g+\nu \ell)}{U_2}\mathcal{A}}\,,\nonumber
\end{align}
where we have introduced the $2\times 2$ matrix
\begin{align}
&A=\left(\begin{array}{cc}n_1 & -k_2 \\ n_2 & k_1\end{array}\right)\,,&&\text{and} &&\mathcal{A}=(1,U) A\big(^\tau_1\big)\,.
\end{align}
Moreover, we have used a slightly different expansion of the elliptic genus than in equation (\ref{BetaExpansionEllk0firstapp})
\begin{align}
\chi_{\text{ell}}(\mathbb{C}^2;\tau,\nu,g_s;\beta)=\sum_{n=0}^\infty\sum_{\ell,m\in\mathbb{Z}}e^{2\pi i(n\tau+g_s m+\ell\nu)} c(n,\ell,m;\beta)\,.\label{OtherFourierExpand}
\end{align}
The coefficients $c(n,\ell,m;\beta)$ can be related to $D_g(n,\ell;\beta)$ introduced in (\ref{BetaExpansionEllk0firstapp}) through
\begin{align}
D_g(n,\ell;\beta)=\sum_{m\in\mathbb{Z}}\frac{(2\pi i m)^{2g-2}}{(2g-2)!}\,c(n,\ell,m;\beta)\,.\label{IdentityFourierCoeffs}
\end{align}

After these preliminary remarks, we can proceed to compute the integral in (\ref{LoopIntegral}). Using the method of orbits, there are three different pieces one needs to consider
\begin{align}
\mathcal{I}(\nu,g_s;T,U)=\mathcal{I}_0(\nu,g_s;T,U)+\mathcal{I}_{\text{DG}}(\nu,g_s;T,U)+\mathcal{I}_{\text{NDG}}(\nu,g_s;T,U)\,,
\end{align}
corresponding to the zero-, degenerate and non-degenerate orbits respectively. All three terms can be computed explicitly, using standard methods. The results can be summarised as follows
\begin{align}
&\mathcal{I}_0(\nu,g_s;T,U)=T_2\int\frac{d^2\tau}{\tau_2^2}\sum_{n}c(n,0,0;1)\,\nome^n=\frac{\pi T_2}{3},,\\
&\mathcal{I}_{\text{DG}}(\nu,g_s;T,U)=\sum_{n=0}^\infty\sum_{\ell,m\in\mathbb{Z}} c(0,\ell,m;1)\text{Li}_1\left(e^{2\pi i (nU+mg_s+\nu\ell)}\right)
\\
&\mathcal{I}_{\text{NDG}}(\nu,g_s;T,U)=\sum_{\ell,m\in\mathbb{Z}}\sum_{r,n=0}^\infty c(nr,\ell,m;1)\text{Li}_1\left|e^{2\pi i (nU+rT+mg_s+\nu\ell)}\right|
\end{align}
In order to make contact with (\ref{Hecke}) we consider the limit $U\to i\infty$
\begin{align}
&\lim_{U\to i\infty}\mathcal{I}_0(\nu,g_s;T,U)=\frac{T_2\pi}{3}\,,\nonumber\\
&\lim_{U\to i\infty}\mathcal{I}_{\text{DG}}(\nu,g_s;T,U)=\sum_{(\ell,m)\neq (0,0)} c(0,\ell,m;1)\text{Li}_1\left(e^{2\pi i (mg_s+\nu\ell)}\right)+\gamma_E-1-\ln\frac{8\pi }{3\sqrt{3}}\,,\nonumber\\
&\lim_{U\to i\infty}\mathcal{I}_{\text{NDG}}(\nu,g_s;T,U)=\sum_{\ell,m\in\mathbb{Z}}\sum_{r=1}^\infty c(0,\ell,m;1)\text{Li}_1\left|e^{2\pi i (rT+mg_s+\ell\nu))}\right|\nonumber
\,.
\end{align}
where we have used the usual regularisation procedure to remove possible divergences as we take the limit. Thus, we can indeed write
\begin{align}
&\lim_{U\to i\infty}\mathcal{I}(\nu,g_s;T,U)\nonumber\\
&=\frac{T_2\pi }{3}+\sum_{(\ell,m)\neq(0,0)} c(0,\ell,m;1)\text{Li}_1\left(e^{2\pi i (mg_s+\nu\ell)}\right)+\sum_{\ell,m\in\mathbb{Z}}\sum_{r=1}^\infty c(0,\ell,m;1)\text{Li}_1\left|e^{2\pi i (rT+mg_s+\nu\ell))}\right|\nonumber\\
&=\frac{T_2\pi }{3}+\sum_{(\ell,m)\neq(0,0)}\sum_{g=0}^\infty\frac{(2\pi i m g_s)^{2g-2}}{(2g-2)!}\,c(0,\ell,m;1)\frac{\partial^{2g-2}}{\partial x^{2g-2}}\text{Li}_1\left(e^{x+2\pi i \ell\nu)}\right)\big|_{x=0}+\nonumber\\
&\hspace{0.5cm}+\sum_{\ell,m\in\mathbb{Z}}\sum_{g,r=1}^\infty\frac{(2\pi i m g_s)^{2g-2}}{(2g-2)!}\,c(0,\ell,m;1)\frac{\partial^{2g-2}}{\partial x^{2g-2}}\text{Li}_1\left(e^{x+2\pi i (rT+\ell\nu))}\right)\big|_{x=0}+\text{c.c.}\nonumber
\end{align}
Using the identity $\partial_x \text{Li}_n(x)=\text{Li}_{n-1}(x)/x$, as well as equation (\ref{IdentityFourierCoeffs}) we can write
\begin{align}
\lim_{U\to i\infty}\mathcal{I}(\nu,g_s;T,U)=&\sum_{\ell\in\mathbb{Z}}\sum_{g,r=1}^\infty g_s^{2g-2}\,D_g(0,\ell;1)\text{Li}_{3-2g}\left(e^{2\pi i (rT+\ell\nu))}\right)\nonumber\\
&+\sum_{g}^\infty g_s^{2g-2}\,D_g(0,1;1)\text{Li}_{3-2g}\left(e^{2\pi i \nu}\right)+\text{c.c.}\label{TorusIntegralFin}
\end{align}
where we have neglected moduli independent terms, as well as the zero orbit. Comparing the sum over $g$ with (\ref{Hecke}), we indeed find agreement upon identifying
\begin{align}
&Q_\rho=e^{2\pi i T}\,,&&\text{and} &&Q_m=e^{2\pi i\nu}\,.
\end{align}
for $g\geq 1$. Thus, indeed, the free energy $F_g^{(k=1)}$ can be understood as the theta-transform of the equivariant elliptic genus of $\mathbb{C}^2$.
%%%%%%%%%%%%%%%%%%%%%%%%%%%%
\subsection{Decompactification}
Consider the resolved conifold, a toric Calabi-Yau threefold which is the total space
\bea\nn
X={\cal O}(-1) \oplus {\cal O}(-1) \mapsto \mathbb{P}^{1}\,.
\eea
It has one K\"ahler parameter (the size of the $\mathbb{P}^{1}$) which we will denote by $T_{1}$. The refined topological string partition function for this space can be calculated using the refined topological vertex formalism \cite{Iqbal:2007ii} and is given by
\bea
Z(T_{1},\epsilon_{1},\epsilon_{2})&=&\prod_{i,j=1}^{\infty}\Big(1-e^{-T_{1}}\,t_1^{i-\frac{1}{2}}\,t_2^{-j+\frac{1}{2}}\Big)\,,\nn
%(t_{1},t_{2})&=&(e^{i\epsilon_{1}},e^{i\epsilon_{2}})\,.
\eea
where $t_{1,2}$ are defined as in (\ref{qtDef}). The genus $g$ free energy is given by
\bea
Z(T_{1},\epsilon_{1},\epsilon_{2})&=&\sum_{g=0}^{\infty}g_{2}^{2g-2}\,F_{g}(T_{1},\beta)\\\nn
F_{g}&=&-f_{g}(\beta)\sum_{n=1}^{\infty}\,e^{-nT_{1}}\,n^{2g-3}=-f_{g}(\beta)\,\mbox{Li}_{3-2g}(e^{-T_{1}})\,.\label{DecompactifiedFreeEnergy}
\eea
%%%%%%%%%%%%%%%%%%%%%%%%%%%%
\subsection{Compactification}\label{Sect:Compact}

As discussed in section 2, by further compactification of the setup we have a geometry which is a $T^2\times T^2$ fibration and whose mirror is a genus two curve \cite{Hollowood:2003cv}.  This Calabi-Yau threefold engineers an ${\cal N}=2$ theory which is the compactification of an M5-brane on a $T^{2}$ with a mass deformation (corresponding to turning on $c_{1}$ on the torus) \cite{Hollowood:2003cv}. The same theory also has a description as an M5-brane wrapped on a genus two curve. The complex structure parameter $T$, the K\"ahler structure parameter $U$ and the mass deformation $\nu$ combine to give the period matrix of the genus two curve
\bea
\begin{pmatrix}
T&\nu\\\nu&U
\end{pmatrix}
\eea
The corresponding Calabi-Yau threefold has three $\mathbb{P}^{1}$'s, whose K\"ahler parameters we denote as $T_{1}, T_{2}$ and $T_{3}$. Taking the limit of any of these parameters going to infinity, gives the Calabi-Yau discussed in the last subsection. We are using the following variables which will also be useful later:
\begin{align}
&U=\frac{i}{2\pi}(T_{3}+T_{2})\,,&&\nu=\frac{i}{2\pi}\,T_{2}\,,&&Q=e^{2\pi i\,U}\,,
\end{align}
A different identification of the Kahler parameters with $(T,U,\nu)$ corresponds to an $Sp(4,\mathbb{Z})$ transformation of the period matrix. The partition function of the gauge theory is given by
\bea
Z&=&Z_{\text{pert}}\,Z_{\text{instanton}}\nonumber\\
Z_{\text{pert}}&=&Z(z,y;t_1,t_2)\nonumber\\
Z_{\text{instanton}}&=&\sum_{k=0}^{\infty}Q^{k}\,\chi_{\text{ell}}\Big(\text{Hilb}^{k}[\mathbb{C}^{2}];T,\nu,t_{1},t_{2}\Big)\,,
\eea
where $\chi_{\text{ell}}(X;T,\nu, t_{1},t_{2})$ is the equivariant elliptic genus of $X$ which we discuss in more detail later. In this case since $\text{Hilb}^{k}[\mathbb{C}^2]$ is non-compact the elliptic genus is calculated equivariantly with respect to the action $(z_{1},z_{2})\mapsto (t_{1}\,z_{1},t_{2}\,z_{2})$ which lifts to a $\mathbb{C}^{\times}\times \mathbb{C}^{\times}$ action on $\text{Hilb}^{k}[\mathbb{C}^{2}]$.

Denoting the class of the three $\mathbb{P}^{1}$'s by $C_{1},C_{2}$ and $C_{3}$ then $\chi_{ell}(\mathbb{C}^{2})$ (see equ.~(\ref{BetaExpansionEllk0firstapp})) is counting curves in the class
$C_{1}+k\,C_{2}+m\,C_{3}$ for $k\geq 1,m\geq 0$,
\begin{align}
\chi_{\text{ell}}(\mathbb{C}^{2})&=\frac{F(Q_\rho,Q_m;t_{1},t_{2})}{(t_{1}^{\frac{1}{2}}-t_{2}^{-\frac{1}{2}})(t_{2}^{-\frac{1}{2}}-t_{2}^{\frac{1}{2}})}
\end{align}
where we have introduced
\begin{align}
&F(Q_\rho,Q_m;t_{1},t_{2})\nonumber\\
&=\sum_{n=0}^{\infty}\sum_{k=-n}^{\infty}Q_\rho^{n}\,Q_m^{k}\,(-1)^{2j_{L}+2j_{R}}\,N^{j_{L},j_{R}}_{(n,k)}\Big(\sum_{m_{L}=-j_{L}}^{+j_{L}}\Big(\frac{t_{1}}{t_{2}}\Big)^{m_{L}}\Big)\Big(\sum_{m_{R}=-j_{R}}^{+j_{R}}\Big(t_{1}\,t_{2}\Big)^{m_{R}}\Big)\label{FreeEnergyCompact}
\end{align}
$N^{j_{L},J_{R}}_{(n,k)}$ is the number of BPS states with spin content $(j_{L},j_{R})$ and charge $C_{1}+(n+k+1)C_{2}+nC_{3}$.

%{\bf $y\mapsto y^{-1}$ symmetry and flop invariance}

The above curve counting function $F$ is actually invariant under the transformation $y\mapsto y^{-1}$. This is clear from its definition in terms of the equivariant elliptic genus of $\mathbb{C}^{2}$. But it also follows from the geometry of this Calabi-Yau threefold. Recall that $Q_m=e^{-T_{2}}$ and therefore $Q_m\mapsto Q_m^{-1}$ corresponds to the flop transition $T_{2}\mapsto -T_{2}$. Under the flop of $C_{2}$ the new curve classes and K\"ahler parameters are $C_{i}'=C_{i}+2(C_{i}\cdot C_{2})C_{2}$
\begin{align}
&C_{1}':=C_{1}+2C_{2}\,,&&C_{2}'=-C_{2}\,,&&C_{3}'=C_{3}+2C_{2}
T_{1}'=T_{1}+2T_{2}\,,\nonumber\\
&T_{2}'=-T_{2}\,,&&T_{3}'=T_{3}+2T_{2}\,,
\end{align}
which particularly implies the following simple transformation properties
\begin{align}
&T_{1}'+T_{2}'=T_{1}+T_{2}\,,&&T_{2}'+T_{3}'=T_{2}+T_{3}\,,&&T_{2}'=-T_{2}\\\nn
&Q_\rho'=Q_\rho\,,&&Q_m'=Q_m^{-1}\,.
\end{align}
From this we may conclude
\bea
F(Q_\rho,Q_m;t_{1},t_{2})=F(Q_\rho,Q_m^{-1};t_{1},t_{2})
\eea
which implies $N^{(j_{L},j_{R})}_{(n,m)}=0$ for $m>n$. Therefore
\bea\nn
\chi_{\text{ell}}(\mathbb{C}^{2})=\frac{\sum_{n=0}^{\infty}\sum_{k=-n}^{n}Q_\rho^{n}\,Q_m^{k}\,(-1)^{2(j_{L}+j_{R})}\,N^{j_{L},j_{R}}_{(n,k)}\Big(\sum_{m_{L}=-j_{L}}^{+j_{L}}\Big(\frac{t_{1}}{t_{2}}\Big)^{m_{L}}\Big)\Big(\sum_{m_{R}=-j_{R}}^{+j_{R}}\Big(t_{1}\,t_{2}\Big)^{m_{R}}\Big)}{(t_{1}^{\frac{1}{2}}-t_{2}^{-\frac{1}{2}})(t_{2}^{-\frac{1}{2}}-t_{2}^{\frac{1}{2}})}
\eea
The full instanton partition function can be expressed in terms of the multiplicities of $\chi_{ell}(\mathbb{C}^{2})$. Define $c(n,k,r,s)$ as Fourier coefficients of $\chi_{\text{ell}}$ (similar as in (\ref{OtherFourierExpand}))
\bea\label{FourierEllGen}
\chi_{\text{ell}}(\mathbb{C}^{2})&=&\sum_{n,k,r,s}\,c(n,k,r,s)\,Q_\rho^{n}\,Q_m^{k}\,t_{1}^{r}\,t_{2}^{-s}\\\nn
c(n,k,r+k_{1},s+k_{2})&=&\begin{cases}
(-1)^{2j_{L}+2j_{R}}\,N^{j_{L},j_{R}}_{(n,k)} & \mbox{for}\,\, \left\{\begin{array}{l}-j_{R}\leq \frac{s+r}{2}\leq j_{R}\\-j_{R}\leq\frac{s-r}{2}\leq j_{R}\,,\\k_{1,2}\geq 0 \end{array}\right.\\[10pt]
0 &\mbox{otherwise}
\end{cases}
\eea
then we find explicitly
\begin{align}
Z_{\text{instanton}}&=&\sum_{k=0}^{\infty}Q^{k}\,\chi_{ell}\Big(\text{Hilb}^{k}[\mathbb{C}^{2}];T,\nu,t_{1},t_{2}\Big)=\prod_{\ell,n,k,r,s}\Big(1-Q^{\ell}\,Q_\rho^{n}\,Q_m^{k}\,t_{1}^{r}\,t_{2}^{-s}\Big)^{-c(n\ell,k,r,s)}\label{ProductRep}
\end{align}

\section{Elliptic Genera}\label{Sect:EllGen}
In the previous section we have seen the appearance of the elliptic genus in the computation of the partition function for $M=N=1$. In this section we study $\chi_{\text{ell}}$ in more detail.
%%%%%%%%%%%%%%%%%%%%%%%%%%%%
\subsection{Equivariant Elliptic Genera}
We first introduce a slight generalisation of the elliptic genus, which is non-trivial also for non-compact manifolds but reduces to the standard definition in the case of compact manifolds with vanishing first Chern class. We explicitly work out the example of $\mathbb{C}^2$.

\subsubsection{General Definition}
Elliptic genera have recently attracted a lot of interest from various perspectives, owing mostly to their interesting modular properties. Indeed, for a compact complex manifold $\mathcal{M}$ with complex dimension $d=\text{dim}_{\mathbb{C}}(\mathcal{M})$ and vanishing first Chern class $c_1(\mathcal{M})=0$, the elliptic genus $\phi(\mathcal{M};\tau,z)$ is a weak Jacobi form of weight 0 and index $d/2$ (see \cite{Witten:1993jg,Eguchi:1988vra}). If one relaxes the second assumption, \emph{i.e.} $c_1(\mathcal{M})\neq 0$, the automorphic properties of $\phi(\mathcal{M};\tau,z)$ are lost. However, as has been shown in \cite{Gritsenko:1999nm}, one can define an \emph{automorphic correction of the elliptic genus} (also referred to as the modified Witten genus)
\bea
\chi_{\text{ell}}(\mathcal{M};\tau,\nu)&:=&\Big(\frac{i\theta_1(\tau;\nu)}{\eta^{3}(\tau)}\Big)^{d}\,\int_{\mathcal{M}}P(\mathcal{M};\tau,\nu,\vec{x})\,W(\mathcal{M};\tau,\vec{x})\,,\label{DefEquEllGen}
\eea
where $x_{i}$ are the Chern roots of the tangent bundle of $\mathcal{M}$ and we have defined the quantities
\begin{align}
P(\mathcal{M};\tau,\nu,\vec{x})&=\mbox{exp}\Big(-\sum_{n\geq 2}\frac{{\cal P}^{(n-2)}(\tau,\nu)}{(2\pi i)^{n}n!}\sum_{i=1}^{d}x_{i}^{n}\Big)\nn\\
W(\mathcal{M};\tau,\vec{x})&=\mbox{exp}\Big(2\sum_{k\geq 2}\frac{G_{2k}(\tau)}{(2k)!}\sum_{i=1}^{d}x_{i}^{2k}\Big)\,.
\end{align}
Here ${\cal P}^{(n)}(\tau,\nu)$ is defined as the $n$-th derivative of the Weierstrass function (see \cite{Gritsenko:1999nm} as well as appendix \ref{App:Defs} for the notation). The definition of $\chi_{\text{ell}}(\mathcal{M};\tau,\nu)$ can further be extended to non-compact manifolds $\mathcal{M}$ which have some $U(1)$ action with non-degenerate fixed points $\{p_{1},\cdots,p_{N}\}$. In this case, the integration in (\ref{DefEquEllGen}) is calculated equivariantly with respect to this action and is given by
\begin{align}
\chi_{\text{ell}}(\mathcal{M};\tau,\nu)=\sum_{a=1}^{N}\Big(\frac{i\theta_1(\tau;\nu)}{\eta^{3}(\tau)}\Big)^{d}\,(w_{1,a}w_{2,a}\cdots w_{d,a})^{-1}\,P(\mathcal{M};\tau,\nu,\vec{w}_{a})\,W(\mathcal{M};\tau,\vec{w}_{a})
\end{align}
where $\vec{w}_{a}$ are the weights associated with the $U(1)$ action at the fixed point $p_{a}$.
%%%%%%%%%%%%%%%%%%%%%%%%%%%%%%%%%%%%%%%%%%%%%%%%%%%%%%%%%%%%%%%%%%%%%%%%%%
\subsubsection{Equivariant Elliptic Genus for $\mathbb{C}^2$}
In the following we are mostly interested in the case $\mathcal{M}=\mathbb{C}^{2}$ for which we can define an equivariant $U(1)$ action as\footnote{This corresponds to $\beta=1$ compared to (\ref{EquivariantParameters}).}
\begin{align}
&\mathbb{C}^2\ni(z_{1},z_{2})\longmapsto (e^{ig_s}z_{1},e^{-ig_s}z_{2})\,,&&\text{with} &&g_s\in\mathbb{R}_+\,.\label{C2U1Action}
\end{align}
This action has a single fixed point at the origin and the corresponding weights are given by $w_{1,1}=-w_{2,1}=g_{s}$. We then obtain
\begin{align}
\chi_{\text{ell}}(\mathbb{C}^2;\tau,\nu,g_s)=\Big(\frac{\theta_1(\tau,\nu)}{g_s\eta^{3}(\tau)}\Big)^{2}\,\mbox{exp}
\Big(-2\sum_{k\geq 1}\frac{{\cal P}^{(2k-2)}(\tau,\nu)}{(2\pi i)^{2k}(2k)!}\,g_{s}^{2k}\Big)\mbox{exp}\Big(4\sum_{k\geq 2}\frac{G_{2k}(\tau)}{(2k)!}g_{s}^{2k}\Big)\,,\label{EquivEllGenGeomRed}
\end{align}
which can also be written in the following manner
\begin{align}
&\chi_{\text{ell}}(\mathbb{C}^2;\tau,\nu,g_s)\nonumber\\
&=-\frac{\phi_{-2,1}(\tau,\nu)}{g_{s}^2}\mbox{exp}\Big(-\frac{g_{s}^2\phi_{0,1}(\tau;\nu)}{12\phi_{-2,1}(\tau;\nu)}-\sum_{k\geq 2}\frac{2g_{s}^{2k}}{(2k)!}\Big(\frac{{\cal P}^{(2k-2)}(\tau;\nu)}{(2\pi i)^{2k}}-G_{2k}(\tau)\Big)\Big)\nonumber
\end{align}
In fact, for most of the remainder of this work we prefer yet a different form. To this end we use equation (\ref{ExplicitWeierstrass}) and write
\begin{align}
-2\sum_{k=1}^\infty\frac{\mathcal{P}^{(2k-2)}(\tau,\nu)}{(2\pi i)^{2k}(2k)!}\,g_s^{2k}&=-2\sum_{2\leq n\in 2\mathbb{Z}}\frac{\mathcal{P}^{(n-2)}(\tau,\nu)}{(2\pi i)^{n}n!}\,g_s^{n}\nonumber\\
&=-\sum_{n=1}^\infty\frac{\mathcal{P}^{(n-2)}(\tau,\nu)}{(2\pi i)^{n}n!}\,g_s^{n}-\sum_{n=1}^\infty\frac{\mathcal{P}^{(n-2)}(\tau,\nu)}{(2\pi i)^{n}n!}\,(-g_s)^{n}\,.
\end{align}
Notice that all odd powers of $g_s$ in the last line cancel between the two terms. Using again (\ref{ExplicitWeierstrass}) we can write
\begin{align}
-&2\sum_{k=1}^\infty\frac{\mathcal{P}^{(2k-2)}(\tau,\nu)}{(2\pi i)^{2k}(2k)!}\,g_s^{2k}=-\sum_{n=1}^\infty\frac{\partial^{n-2}}{\partial\nu^{n-2}}\left[8\pi^2G_2-\frac{\partial^2}{\partial\nu^2}\log(i\theta_1(\tau;\nu))\right]\frac{g_s^n+(-g_s)^n}{(2\pi i)^nn!}\nonumber\\
%&\hspace{0.5cm}=2g_s^2G_2-2\log(i\theta_1(\tau;\nu))+\sum_{n=0}^\infty\frac{\partial^n}{\partial\nu^n}\log(i\theta_1(\tau;\nu))\frac{g_s^n+(-g_s)^n}{(2\pi i)^nn!}\nonumber\\
&\hspace{0.5cm}=2g_s^2G_2-2\log(i\theta_1(\tau;\nu))+\log\left(i\theta_1\left(\tau;\nu+\frac{g_s}{2\pi }\right)\right)+\log\left(i\theta_1\left(\tau;\nu-\frac{g_s}{2\pi}\right)\right)\,.\nonumber
\end{align}
Inserting this into (\ref{EquivEllGenGeomRed}) we obtain
\begin{align}
\chi_{\text{ell}}(\mathbb{C}^2;\tau,\nu,g_s)%&=-\frac{\theta_1^2(\tau;\nu)}{g_s^2\eta^6(\tau)}\,\frac{\theta_1\left(\tau;\nu+\frac{g_s}{2\pi }\right)\theta_1\left(\tau;\nu-\frac{g_s}{2\pi}\right)}{\theta_1^2(\tau;\nu)}\,e^{2g_s^2G_2}\,\text{exp}\left[4\sum_{k=2}^\infty\frac{G_{2k}}{(2k)!}\,g_s^{2k}\right]\nonumber\\
%&=-\frac{\theta_1\left(\tau;\nu+\frac{g_s}{2\pi}\right)\theta_1\left(\tau;\nu-\frac{g_s}{2\pi}\right)}{g_s^2\eta^6(\tau)}\,\text{exp}\left[4\sum_{k=1}^\infty\frac{G_{2k}}{(2k)!}\,g_s^{2k}\right]\nonumber\\
&=-\frac{\theta_1\left(\tau;\nu+\frac{g_s}{2\pi}\right)\theta_1\left(\tau;\nu-\frac{g_s}{2\pi}\right)}{\theta_1^2(\tau;\frac{g_s}{2\pi})^2}\,,\label{EquEllGenb1}
\end{align}
which is the expression we will mostly be concerned with in this work. For completeness we also note that upon expanding $\chi_{\text{ell}}$ in powers of $g_s$
\begin{align}
\chi_{\text{ell}}(\mathbb{C}^{2};\tau,\nu,g_s)=\sum_{g=0}^{\infty}\,g_{s}^{2g-2}\,\,\fgel^{(0)}_{g}(\tau,\nu)\,,\label{BetaExpansionEllk0}
\end{align}
the coefficient functions have a very simple structure
\begin{align}
\fgel^{(k=0)}_{g}(\tau,\nu)=\left\{\begin{array}{lcl}
-\phi_{-2,1}(\tau;\nu) & &g=0\\[7pt]
\frac{1}{12}\phi_{0,1}(\tau,\nu) & &g=1\\[7pt]
\frac{B_{2g}(2g-1)}{(2g)!}\,E_{2g}(\tau)\,\phi_{-2,1}(\tau;\nu) & &g\geq 2\,\end{array}\right.\label{ExplicitFg}
\end{align}
which is proven in appendix~\ref{App:Expansion}.
%%%%%%%%%%%%%%%%%%%%%%%%%%%%%%%%%%%%%%%%%%%%%%

%%%%%%%%%%%%%%%%%%%%%%%%%%%%%%%%%%%%%%%%%%%%%%%%%%%%%%%%%%%
\subsection{Equivariant Elliptic Genus of $\beta$-deformed $\mathbb{C}^{2}$}
There exists an interesting generalisation of (\ref{EquEllGenb1}), namely the elliptic genus of a $\beta$-deformed version of $\mathbb{C}^2$. For this $\Omega$-background \cite{Moore:1997dj,Lossev:1997bz} (see also appendix~\ref{App:OmegaBackground}) the $U(1)$ action (\ref{C2U1Action}) is generalised to
\begin{align}
&\mathbb{C}_{\beta}^2\ni(z_{1},z_{2})\longmapsto (t_1z_{1},t_2z_{2})\,,\label{C2U1ActionBeta}
\end{align}
with two new parameters $(t_1,t_2)$ (introduced in (\ref{EquivariantParameters})) which we remind, can be expressed in terms of $g_s$ of equation (\ref{C2U1Action}) and the deformation parameter $\beta$
\begin{align}
&t_{1}:=\mbox{exp}(i\,g_{s}\sqrt{\beta})\,,&&\text{and} &&t_{2}:=\mbox{exp}(-i\,g_{s}/\sqrt{\beta})\,.
\end{align}
For $\beta=1$ the deformation vanishes and we are dealing again with $\mathbb{C}^2$. Repeating the computation of the previous section for generic $\beta$, we find for the elliptic genus
\begin{align}
&\chi_{\text{ell}}(\mathbb{C}^{2};\tau,\nu,g_s;\beta)\nonumber\\
&=\frac{\sqrt{t_{1}t_{2}}}{y}\,\prod_{k\geq 1}\frac{(1-y\sqrt{\frac{t_{1}}{t_{2}}}\,z^{k-1})(1-y^{-1}\sqrt{\frac{t_{2}}{t_{1}}}\,z^{k})
(1-y\sqrt{\frac{t_{2}}{t_{1}}}\,z^{k-1})(1-y^{-1}\sqrt{\frac{t_{1}}{t_{2}}}\,z^{k})}{(1-t_{1}\,z^{k-1})(1-t_{1}^{-1}z^{k})(1-t_{2}\,z^{k-1})(1-t_{2}^{-1}\,z^{k})}\nonumber\\
&=\frac{\theta_1(z;\,y^{-1}\sqrt{\frac{t_{2}}{t_{1}}})\,\theta_1(z;\,y^{-1}\sqrt{\frac{t_{1}}{t_{2}}})}{\theta_1(z;\,t_{1}^{-1})\,\theta_1(z;\,t_{2}^{-1})}\label{eq1}
\end{align}
The expansion in $g_s$ can be organized in the following manner
\bea
\chi_{\text{ell}}(\mathbb{C}^{2};\tau,\nu,g_s;\beta)=\sum_{g=0}^{\infty}\,g_{s}^{2g-2}\,\sum_{m=0}^{g}\,\Big(\sqrt{\beta}-\frac{1}{\sqrt{\beta}}\Big)^{2m}\,\fgel^{(m)}_{g}(\tau,\nu)\,.\label{BetaExpansionEll}
\eea
Here all the $\fgel^{(m)}_{g}(\tau,\nu)$ are weak Jacobi forms of $SL(2,\mathbb{Z})$ with weight $(2g-2)$ and index $1$. For the reader's convenience we have compiled the first few of them explicitly in table~\ref{Tab:FgTab}.
\begin{table}[ht]\centerline{\scalebox{.90}{
\rotatebox{90}{\footnotesize$\begin{array}{|c||l|l|l|l|l|}\hline
&&&&&\\[-8pt]
g\diagdown m & m=0 & m=1 & m=2 & m=3 & m=4 \\[4pt]\hline\hline
&&&&&\\[-8pt]
0 & -\phi_{-2,1} &&&&\\[4pt]\hline
&&&&&\\[-8pt]
1 & \frac{\phi_{0,1}}{12} & \frac{E_{2}\,\phi_{-2,1} + \phi_{0,1}}{48} &&&\\[4pt]\hline
&&&&&\\[-8pt]
2 & -\frac{E_{4}\phi_{-2,1}}{240} & -\frac{5E_{2}\phi_{0,1}+9E_{4}\phi_{-2,1}}{2880} & -\frac{10E_{2}\phi_{0,1}+\left(5E_{2}^2 +13E_{4}\right)\phi_{-2,1}}{23040} &&\\[4pt]\hline
&&&&&\\[-8pt]
3 & \frac{E_{6}\phi_{-2,1}}{6048} & \text{\parbox{2.8cm}{$\frac{21E_{4}\,\phi_{0,1}}{241920}+$\\[4pt]$\frac{\left(21E_{4}E_{2}+40E_{6}\right)\phi_{-2,1}}{241920}$\\[-4pt]}} &\text{\parbox{2.9cm}{$\frac{\left(35E_{2}^2+91E_{4}\right)\phi_{0,1}}{1935360}+$\\[4pt]$\frac{6\left(21E_{4}E_{2}+20E_{6}\right)\phi_{-2,1}}{1935360}$}} & \text{\parbox{3.8cm}{$\frac{\left(184E_{6}+273E_{4}E_{2}+35E_{2}^3\right)\phi_{-2,1}}{23224320}+$\\[4pt]$\frac{\left(147E_{4}+105E_{2}^2\right)\phi_{0,1}}{23224320}$\\[-4pt]}} &\\[4pt]\hline
&&&&&\\[-8pt]
4 & -\frac{E_{4}^2\phi_{-2,1}}{172800} & \text{\parbox{2.8cm}{$-\frac{E_6\phi_{0,1}}
{290304}-$\\[4pt]$\frac{\left(25E_6E_2+84E_4^2\right)\phi_{-2,1}}{7257600}$\\[-4pt]}} & \text{\parbox{4.1cm}{$-\frac{\left(21E_4E_2+40E_6\right)\phi_{0,1}}{11612160}-$\\[4pt]$\frac{\left(21E_4(E_2^2+9E_4)+80E_6E_2\right)\phi_{-2,1}}{23224320}$\\[-4pt]}} & \text{\parbox{4.4cm}{$-\frac{\left(35E_2^3 + 273E_4E_2+304E_6\right)\phi_{0,1}}{278691840}-$\\[4pt]$\frac{\left(7E_4(15E_2^2 +53E_4) + 200E_6E_2\right)\phi_{-2,1}}{154828800}$}} & \text{\parbox{5.8cm}{$-\frac{\left(35E_2^3 + 147E_4E_2 + 124E_6\right)}{1114767360}-$\\[4pt]$\frac{\left(175E_2^4 + 2730E_4E_2^2+5583E_4^2+3680E_6E_2\right)\phi_{-2,1}}{22295347200}$\\[-4pt]}}\\[4pt]\hline
\end{array}$}}}
\caption{Explicit expressions for the first few $\fgel^{(m)}_g$, to save writing we have not put explicit arguments for all modular forms.}
\label{Tab:FgTab}
\end{table}
Upon close inspection of the latter we observe a number of interesting relations between $\fgel^{(m)}_{g}$ of different $g$ and $m$. In particular, we find
\bea
\frac{\partial \fgel_{g}^{(m)}}{\partial E_{2}}=-\frac{1}{48}\,\fgel_{g-1}^{(m-1)}\,
\eea
For the complete genus $g$ contribution
\begin{align}
\fgel_{g}(\tau;\nu):=\sum_{m=0}^{g}\left(\sqrt{\beta}-\frac{1}{\sqrt{\beta}}\right)^{2m}\,\fgel_{g}^{(m)}(\tau;\nu)
\end{align}
this entails the following holomorphicity type of relation
\begin{align}
\frac{\partial \fgel_{g}(\tau;\nu)}{\partial E_2(\tau)}=-\frac{1}{48}\Big(\sqrt{\beta}-\frac{1}{\sqrt{\beta}}\Big)^2\, \fgel_{g-1}(\tau;\nu)\,.
\end{align}
Notice in particular that the 'anomalous' contribution on the right hand side vanishes in the case of $\beta=1$.
%%%%%%%%%%%%%%%%%%%%%%%%%%%%%%%%%%%%%%%%%%%%%%%%%%%%%%%%%%
\section{String Amplitudes and the Elliptic Genus}\label{Sect:OneLoopStringAmp}
In this section we try to find the structures that appeared in the previous sections within the setting of perturbative string theory. More precisely, we exhibit a series of one-loop string scattering amplitudes, which provides a generating functional for the equivariant elliptic genus of $\mathbb{C}^2$ in its $\beta$-deformed version. Here we only sketch this computation skipping most of the details. A brief review of our notation as well as some more details can be found in appendix~\ref{Sect:LoopAmps}.
%%%%%%%%%%%%%%%%%%%%%%%%%%%%%%%%%%%%%%%%%%%%%%%%%%%%%%%%%

Following a similar logic as in \cite{HS} we consider one-loop amplitudes with massive external states. The main difference with respect to the latter work is that the internal $K3$ compactification has been decompactified such that we consider type~II amplitudes on $T^2$. Moreover, in order to account for the modified Witten genus of $\mathbb{C}^2$ (see section~\ref{Sect:EllGen}) rather than the elliptic genus (which would vanish for $\mathbb{C}^2$), we also require a number of additional vertex insertions compared to \cite{HS}. Finally, in order to obtain the $\beta$-deformation of the modified Witten genus, we will follow a similar logic as in \cite{Antoniadis:2010iq,Antoniadis:2013bja}: we implement the $\beta$-deformation through the insertion of additional vertex operators related to the $T^2$ compactification.
%%%%%%%%%%%%%%%%%%%%%%%%%%%%%%%%%%%%%%%%%%%%%%%%%%%%%%%%%

To be precise, here we achieve this by including a number of massive vertices of the type $V^{(-1,0)}(\gamma,p)$, which have been introduced in appendix~\ref{App:Vertices}. Indeed, the string correlation functions we have in mind are given by
{\allowdisplaybreaks
\begin{align}
&\mathcal{G}_{M,N,k}(p^{(1)},p^{(1)},p^{(I)})\nonumber\\
&=\bigg\langle V_R^{(0,0)}\left(h,p^{(1)}\right)V_R^{(0,0)}\left(h,p^{(2)}\right)\prod_{I=1}^NV_F^{(0,0)}\left(p^{(I)}\right)V_F^{(0,0)}\left(p^{(I)}\right)\left(V_0^{(-1,-1)}(p^{(0)})\right)^2\nonumber\\
&\times \prod_{J=1}^{2M}V^{(-1,0)}(\gamma^{(M)},p^{(M)})\,\prod_{\ell=1}^kV_{++}^{(-1,0)}(p^{(\ell)})V_{--}^{(-1,0)}(\bar{p}^{(\ell)})\left(V_{PCO}\,\bar{V}_{PCO}\right)^{2M+2k+2}\bigg\rangle\,,\nonumber
\end{align}}
where in addition $V_R$ are insertions of graviton vertex operators and $V_F$ correspond to KK-gauge fields coming from the reduction on $T^2$. Finally, the $V_{\pm\pm}$ are the analogue of a particular class of states considered in \cite{HS}. Here we have inserted a total number of $2M+2k+2$ picture changing operators at fixed position which drop out at the end of the computation due to BRST invariance of the amplitude. Our conventions are summarised in appendix~\ref{App:Vertices}.

We are interested in the contribution to $\mathcal{G}_{M,N,k}$ proportional to $\left(p^{(1)}\right)^2\left(p^{(2)}\right)^2$ $\prod_{I=1}^Np^{(I)}p^{(I)}$. Following similar steps as in \cite{HS} and working in a fixed kinematical configuration, we indeed find
\begin{align}
\mathcal{G}_{M,N,k}\big|_{\left(p^{(1)}\right)^2\left(p^{(2)}\right)^2\prod_{I=1}^Np^{(I)}p^{(I)}}=\mathcal{A}_{M,N,k}\sim \int _{\mathbb{F}}\frac{d^2\tau}{\tau_2^2}\int d^2x \sum_{m=0}^M\sum_{n=0}^N\left(^{2M}_{2m}\right)\left(^{2N}_{2n}\right)\mathcal{A}_{M,N,k}^{(n,m)}
\end{align}
where $x$ collectively denotes integration over all world-sheet positions of the vertex insertions and $\tau=\tau_1+i\tau_2$ denotes the integral over the world-sheet torus, with $\mathbb{F}$ the fundamental domain of $SL(2,\mathbb{Z})$. $\mathcal{A}_{M,N,k}^{(n,m)}$ is finally given in terms of correlation functions of the world-sheet CFT (for our notation, see appendix~\ref{Sect:LoopAmps}.)
\begin{align}
&\mathcal{A}_{M,N,k}^{(n,m)}=\left\langle\prod_{i=0}^n\psi_1\psi_2(x_i)\bar{\psi}_1\bar{\psi}_2(y_i)\prod_{c=1}^k\psi_1\bar{\psi}_2(q_c)\bar{\psi}_1\psi_2(p_c)\right\rangle(\partial X_3)^{2M+2k+2}(\bar{\partial} X_3)^{2N+4M+4k+2}\nonumber\\
&\times \bigg\langle\prod_{j=1}^{N-n}X_1\partial X_2(u_j)\bar{X}_2\partial\bar{X}_1(v_j)\prod_{a=1}^m\bar{X}_1\partial X_2(w_a)\bar{X}_2\partial X_1(z_a)\prod_{b=1}^{M-m}\bar{X}_1\partial X_2(s_b)\bar{X}_2\partial X_1(t_b)\bigg\rangle\,,\nonumber
\end{align}
where angular brackets denote free correlation functions of the world-sheet fields. Using path integral methods we can determine each correlator through a generating functional, using similar methods as in \cite{Polchinski} (see also \cite{Antoniadis:2013bja}), focusing on the anomaly-free contributions. Furthermore, the bosonic world-sheet fields of the internal $T^2$ ($X_3$) cannot contract among each other and only yield bosonic zero modes in the form of explicit insertions of the Narain momenta $P_L$ and $P_R$. Therefore, after summing over $m$ and $n$ respectively, we find
\begin{align}
\mathcal{A}_{M,N,k}=\int _{\mathbb{F}}\frac{d^2\tau}{\tau_2}&\left[\frac{\partial^{2N}}{\partial\lambda_1^{2N}}\frac{\partial^{2k}}{\partial\lambda_2^{2k}}\frac{\partial^{2M}}{\partial\lambda_3^{2M}}\frac{\theta_1\left(\tau;\lambda_1+\lambda_2\right)\theta_1\left(\tau;\lambda_1-\lambda_2\right)}{\theta_1\left(\tau;\lambda_1+\lambda_3\right)\theta_1\left(\tau;\lambda_1-\lambda_3\right)}\right]_{\lambda_{1,2,3}=0}\nonumber\\
&\times  \sum_{\Gamma^{(2,2)}}(P_L)^{2M+2k+2}(P_R)^{2N+4M+4k+2}\nome^{\frac{1}{2}|P_L|^2}\bar{\nome}^{\frac{1}{2}|P_R|^2}\,,
\end{align}
where the summation is over the $\Gamma^{(2,2)}$ Narain lattice which encodes the dependence on the physical moduli $(T,U)$ of the internal $T^2$. As in \cite{HS}, we introduce the covariant derivatives $\mathcal{D}_{\bar{U}}$ with respect to the $\bar{U}$ modulus, which acts on a function $f^{(w)}$ with weight $w$ as
\begin{align}
\mathcal{D}_{\bar{U}}:=\frac{U-\bar{U}}{2}\left(\frac{\partial}{\partial\bar{U}}-\frac{w}{2(U-\bar{U})}\right)f^{(w)}\,.
\end{align}
With this definition we can write
\begin{align}
\mathcal{A}_{M,N,k}=\mathcal{D}_{\bar{U}}^{2M+2k+2}\mathcal{I}^{\beta\neq 1}_{M,N,k}\,,
\end{align}
where we have introduced the expression
\begin{align}
\mathcal{I}^{\beta\neq 1}_{M,N,k}=\int\frac{d^2\tau}{\tau_2}& \left[\frac{\partial^{2N}}{\partial\lambda_1^{2N}}\frac{\partial^{2k}}{\partial\lambda_2^{2k}}\frac{\partial^{2M}}{\partial\lambda_3^{2M}}\frac{\theta_1\left(\tau;\lambda_1+\lambda_2\right)\theta_1\left(\tau;\lambda_1-\lambda_2\right)}{\theta_1\left(\tau;\lambda_1+\lambda_3\right)\theta_1\left(\tau;\lambda_1-\lambda_3\right)}\right]\bigg|_{\lambda_1=\lambda_2=\lambda_3=0}\nonumber\\
&\times \sum_{\Gamma^{2,2}}(\tau_2P_R)^{2N+2M+2k}\nome^{\frac{1}{2}|P_L|^2}\bar{\nome}^{\frac{1}{2}|P_R|^2}\,.
\end{align}
Instead of performing the world-sheet integration separately for each $(M,N,k)$, we define the following 'master integral'
\begin{align}
\mathcal{I}&(T,U;\lambda_1,\lambda_2,\lambda_3)=\sum_{N=0}^{\infty}\sum_{k=0}^{\infty}\sum_{M=0}^{\infty}\frac{\lambda_1^{2N}}{(2N)!}\frac{\lambda_2^{2k}}{(2k)!}\frac{\lambda_3^{2M}}{(2M)!}\,\mathcal{I}^{\beta\neq 1}_{M,N,k}(T,U)\nonumber\\
&=\int\frac{d^2\tau}{\tau_2}\sum_{\Gamma^{2,2}}\frac{\theta_1\left(\tau;(\lambda_1+\lambda_2)\tau_2P_R\right)\theta_1\left(\tau;(\lambda_1-\lambda_2)\tau_2P_R\right)}{\theta_1\left(\tau;(\lambda_1+\lambda_3)\tau_2P_R\right)\theta_1\left(\tau;(\lambda_1-\lambda_3)\tau_2P_R\right)}\,\nome^{\frac{1}{2}|P_L|^2}\bar{\nome}^{\frac{1}{2}|P_R|^2}
\end{align}
To write the answer in a form which makes contact with the torus integral we discussed in section~\ref{Sect:TorusIntegration},  we identify
\begin{align}
&\lambda_1=\frac{g_s}{2}\left(\sqrt{\beta}+\frac{1}{\sqrt{\beta}}\right)\,,&&\text{and}&&\lambda_2=-2\pi \nu\,,&&\text{and}&&\lambda_3=-\frac{g_s}{2}\left(\sqrt{\beta}-\frac{1}{\sqrt{\beta}}\right)\,,
\end{align}
such that we find
\begin{align}
\mathcal{I}(T,U;g_s,\nu,\beta)&=\int\frac{d^2\tau}{\tau_2}\sum_{\Gamma^{2,2}}\chi_{\text{ell}}(\mathbb{C}^2;\tau,\nu\tau_2P_R,g_s\tau_2P_R,\beta)\,\nome^{\frac{1}{2}|P_L|^2}\bar{\nome}^{\frac{1}{2}|P_R|^2}\label{MasterBeta}
\end{align}
which for $\beta=1$ is exactly the integral considered in (\ref{LoopIntegral}).
%%%%%%%%%%%%%%%%%%%%%%%%%%
\section{Conclusions}\label{Sec:Conclusions}
In this paper we have computed partition functions for mass deformed $\mathcal{N}=2$ supersymmetric gauge theories. There are various different points of view how to think about these theories. In the first part of the paper we have used the fact that three different dual representations of these theories exist:
\begin{enumerate}
\item through $(p,q)$-brane webs in type II string theory
\item through the (refined) topological string on elliptically fibered CY3folds
\item through particular configurations of M5/M2-branes (M-strings)
\end{enumerate}
Each of these descriptions offers powerful computational techniques to compute the gauge theory partition functions with interesting dualities in each case.

In the second part of the paper, we have explored a fourth approach, namely linking the gauge theory partition function to the equivariant elliptic genus of $\mathbb{C}^2$ through a particular Hecke transformation, as well as through a (singular) theta-transform. The latter takes the form of an integral over the fundamental domain of $SL(2,\mathbb{Z})$ (torus integral), with the role of the integral kernel being played by a specific Siegel-Narain theta function. We have recovered precisely this kind of integral in a series of particular one-loop scattering amplitudes in type II string theory compactified on $T^2$ with massive external vertices.

There are several directions in which our work can be extended. For example, the connection between the partition functions and the equivariant elliptic genus of $\mathbb{C}^2$ so far has only been worked out for $k=1$. It would be very interesting to extend these computations also to several M5 branes and understand which type of scattering amplitudes this corresponds to. Similarly, it would be interesting to consider M-strings on yet different internal manifolds. We have already considered further compactifications to $T^2\times T^2$ fibrations in section~\ref{Sect:Compact}. However, it would be interesting to perform a more systematic analysis of additional string backgrounds. For example, it would be interesting to make contact with certain realisations of the $\Omega$-background in string theory, \emph{e.g.} \cite{Orlando:2013yea}. Besides, it would be interesting to analyse the couplings in the $\mathcal{N}=4$ string effective action corresponding to the class of scattering amplitudes discussed in section~\ref{Sect:OneLoopStringAmp}. They are expected to compute certain physical $1/4$-BPS saturated terms, generalising the ones studied in \cite{HS}.

We believe that our results will also be interesting for different problems, related to the physics of BPS states in M-theory. For once it would be interesting to see whether our results can give any further hints on the existence of a so-called algebra of BPS states, \emph{i.e.} the idea that the BPS states in string models are organised to form an algebra. This idea was pioneered in \cite{Harvey:1995fq,Harvey:1996gc} and further studied in \cite{BKM}. Further hints for its existence in $T^2$ compactifications (relevant for the setting which we studied in section~\ref{Sect:OneLoopStringAmp}) were presented in \cite{Gaberdiel:2011qu}. The product representations (see \emph{e.g.}~(\ref{ProductRep})) look very similar to expressions found for denominator formulas of BKM algebras, which have been speculated to be candidates for algebras of BPS states. It would be interesting to see whether (\ref{ProductRep}) indeed allows a more algebraic interpretation.

A possible further connection might be made to a phenomenon called Mathieu moonshine. Indeed, in recent years, convincing evidence has been compiled for an action of the Mathieu group $\mathbb{M}_{24}$ on the BPS states contributing to the elliptic genus of $K3$
\cite{Eguchi:2010ej} and several recent developments also point to a connection to more physical problems \cite{Harvey:2013mda}. It would be curious to see whether there is a similar moonshine phenomenon connected to the equivariant elliptic genus we have been studying. In particular, note that the $g=0$ contribution in the expansion of the equivariant elliptic genus precisely corresponds to the elliptic genus of $K3$. It will be interesting in the future to further exploit this expression.

%%%%%%%%%%%%%%%%%%%%%%%%%%%%
\section*{Acknowledgements}
We would like to thank C. Kozcaz, S. Meinhardt and C. Vafa for useful discussions. Furthermore we thank the Simons Center for Geometry and Physics in Stony Brook for kind hospitality during the initial stages of this work.

%%%%%%%%%%%%%%%%%%%%%%%%%%
\appendix
%%%%%%%%%%%%%%%%%%%%%%%%%%%%%%%%%%%%%%%%%%%%%%%%%%%%%
\section{Mathematical Definitions and Conventions}

\subsection{Identity}\label{Sec:Identity}

Consider the following sum which is the $N=2$ case of Eq.(\ref{schursum}),
\bea
G({\bf x},{\bf y},{\bf Q})&=&
\sum_{\lambda_{1,2,3,4},\eta_{1,2,3,4}}Q_{1}^{|\lambda_{1}|}Q_{2}^{|\lambda_{2}|}Q_{3}^{|\lambda_{3}|}Q_{4}^{|\lambda_{4}|}\,
s_{\lambda_{1}/\eta_{1}}(x_{1})s_{\lambda_{1}^t/\eta_{2}}(y_{1})\\\nn
&&\times
s_{\lambda_{2}/\eta_{2}}(x_{2})s_{\lambda_{2}^t/\eta_{3}}(y_{2})
s_{\lambda_{3}/\eta_{3}}(x_{3})s_{\lambda_{3}^t/\eta_{4}}(y_{3})
s_{\lambda_{4}/\eta_{4}}(x_{4})s_{\lambda_{4}^t/\eta_{1}}(y_{4})
\eea
Then by summing over the partitions four times using the identities:
\bea
\sum_{\eta}s_{\eta/\nu}(x)\,s_{\eta/\mu}(y)&=&\prod_{i,j}(1-x_{i}y_{j})^{-1}\,\sum_{\tau}s_{\mu/\tau}(x)\,s_{\nu/\tau}(y)\\\nn
\sum_{\eta}s_{\eta^{t}/\nu}(x)\,s_{\eta/\mu}(y)&=&\prod_{i,j}(1+x_{i}y_{j})\,\sum_{\tau}s_{\mu^{t}/\tau}(x)\,s_{\nu^{t}/\tau^{t}}(y)\,,
\eea
we can establish the following relation
\bea
&&G(x_{1},y_{1},x_{2},y_{2},x_{3},y_{3},x_{y},y_{4};Q_{1},Q_{2},Q_{3},Q_{4})=
P({\bf x},{\bf y},{\bf Q})\label{relation}\\\nn
&&\times \,G(Q_{1}Q_{2}x_{1},Q_{1}Q_{4}y_{1},Q_{2}Q_{3}x_{2},Q_{1}Q_{2}y_{2},Q_{3}Q_{4}x_{3},Q_{2}Q_{3}y_{3},
Q_{1}Q_{4}x_{4},Q_{3}Q_{4}y_{4};Q_{3},Q_{4},Q_{1},Q_{2})\,,
\eea
where $Q_{\bullet}=Q_{1}Q_{2}Q_{3}Q_{4}$ and
\bea\nn
&&P({\bf x},{\bf y},{\bf Q}):=\prod_{i,j}\frac{(1+Q_{1}x_{1,i}y_{1,j})(1+Q_{2}x_{2,i}y_{2,j})(1+Q_{3}x_{3,i}y_{3,j})(1+Q_{4}x_{4,i}y_{4,j})}
{(1-Q_{1}Q_{4}y_{1,i}x_{4,j})(1-Q_{1}Q_{2}x_{1,i}y_{2,j})(1-Q_{2}Q_{3}x_{2,i}y_{3,j})(1-Q_{3}Q_{4}x_{3,i}y_{4,j})}\\\nn
&&\times \prod_{i,j}\frac{(1+Q_{\bullet}Q_{3}^{-1}y_{2,i}x_{4,j})(1+Q_{\bullet}Q_{4}^{-1}x_{1,i}y_{3,j})(1+Q_{\bullet}Q_{1}^{-1}x_{2,i}y_{4,j})
(1+Q_{\bullet}Q_{2}^{-1}y_{1,i}x_{3,j})}
{(1-Q_{\bullet}y_{2,i}x_{3,j})(1-Q_{\bullet}y_{3,i}x_{4,j})
(1-Q_{\bullet}x_{1,i}y_{4,j})(1-Q_{\bullet}y_{1,i}x_{2,j})}
\eea
Using Eq.(\ref{relation}) we see that
\bea
G({\bf x};Q_{1},Q_{2},Q_{3},Q_{4})=R({\bf x},{\bf y})\,
G(Q_{\bullet}{\bf x};Q_{1},Q_{2},Q_{3},Q_{4})
\eea
where
\bea
R&:=&P({\bf x},{\bf y};Q_{1},Q_{2},Q_{3},Q_{4})\\\nn
&&\times\,P(Q_{1}Q_{2}x_{1},Q_{1}Q_{4}y_{1},Q_{2}Q_{3}x_{2},Q_{1}Q_{2}y_{2},Q_{3}Q_{4}x_{3},Q_{2}Q_{3}y_{3},
Q_{1}Q_{4}x_{4},Q_{3}Q_{4}y_{4};Q_{3},Q_{4},Q_{1},Q_{2})\\\nn
&=&\prod_{i,j}\frac{(1+Q_{1}x_{1,i}y_{1,j})(1+Q_{2}x_{2,i}y_{2,j})(1+Q_{3}x_{3,i}y_{3,j})(1+Q_{4}x_{4,i}y_{4,j})}
{(1-Q_{1}Q_{4}y_{1,i}x_{4,j})(1-Q_{1}Q_{2}x_{1,i}y_{2,j})(1-Q_{2}Q_{3}x_{2,i}y_{3,j})(1-Q_{3}Q_{4}x_{3,i}y_{4,j})}\\\nn
&&\times \prod_{i,j}\frac{(1+Q_{\tau}Q_{3}^{-1}y_{2,i}x_{4,j})(1+Q_{\tau}Q_{4}^{-1}x_{1,i}y_{3,j})(1+Q_{\tau}Q_{1}^{-1}x_{2,i}y_{4,j})(1+Q_{\tau}Q_{2}^{-1}y_{1,i}x_{3,j})}
{(1-Q_{\tau}y_{2,i}x_{3,j})(1-Q_{\tau}y_{3,i}x_{4,j})(1-Q_{\tau}x_{1,i}y_{4,j})(1-Q_{\tau}y_{1,i}x_{2,j})}\\\nn
&&\times \prod_{i,j}\frac{(1+Q_{\tau}Q_{1}x_{1,i}y_{1,j})(1+Q_{\tau}Q_{2}x_{2,i}y_{2,j})(1+Q_{\tau}Q_{3}x_{3,i}y_{3,j})(1+Q_{\tau}Q_{4}x_{4,i}y_{4,j})}
{(1-Q_{\tau}Q_{1}Q_{4}y_{1,i}x_{4,j})(1-Q_{\tau}Q_{1}Q_{2}x_{1,i}y_{2,j})(1-Q_{\tau}Q_{2}Q_{3}x_{2,i}y_{3,j})(1-Q_{\tau}Q_{3}Q_{4}x_{3,i}y_{4,j})}\\\nn
&&\times \prod_{i,j}\frac{(1+Q_{\tau}^{2}Q_{3}^{-1}y_{2,i}x_{4,j})(1+Q_{\tau}^{2}Q_{4}^{-1}x_{1,i}y_{3,j})(1+Q_{\tau}^{2}Q_{1}^{-1}x_{2,i}y_{4,j})
(1+Q_{\tau}^{2}Q_{2}^{-1}y_{1,i}x_{3,j})}
{(1-Q_{\tau}^{2}y_{2,i}x_{3,j})(1-Q_{\tau}^{2}y_{3,i}x_{4,j})(1-Q_{\tau}^{2}x_{1,i}y_{4,j})(1-Q_{\tau}^{2}y_{1,i}x_{2,j})}\\\nn
\eea

Then
\bea
G({\bf x},{\bf y},{\bf Q})=\prod_{k=0}^{L-1}R(Q_{\bullet}^{k-1}{\bf x},Q_{\bullet}^{k-1}{\bf y})\, G(Q_{\bullet}^{L}{\bf x},Q_{\bullet}^{L}{\bf y},{\bf Q})\,.
\eea
Taking the limit $L\mapsto \infty$ gives
\bea
G({\bf x},{\bf y},{\bf Q})=\prod_{k=0}^{\infty}R(Q_{\bullet}^{k-1}{\bf x},Q_{\bullet}^{k-1}{\bf y})\, G(0,0,{\bf Q})\,,
\eea
where $G(0,0,{\bf Q})=\prod_{k=1}^{\infty}(1-Q_{\bullet}^{k})^{-1}$. Thus we get
\bea\nn
&&\frac{G({\bf x},{\bf y},{\bf Q})}{G(0,0,{\bf Q})}:=\\\nn
&&\prod_{n=0}^{\infty}\prod_{i,j}\frac{(1+Q_{\bullet}^{n}Q_{1}x_{1,i}y_{1,j})(1+Q_{\bullet}^{n}Q_{2}x_{2,i}y_{2,j})(1+Q_{\bullet}^{n}Q_{3}x_{3,i}y_{3,j})
(1+Q_{\bullet}^{n}Q_{4}x_{4,i}y_{4,j})}
{(1-Q_{\bullet}^{n}Q_{1}Q_{4}y_{1,i}x_{4,j})(1-Q_{\bullet}^{n}Q_{1}Q_{2}x_{1,i}y_{2,j})(1-Q_{\bullet}^{n}Q_{2}Q_{3}x_{2,i}y_{3,j})(1-Q_{\bullet}^{n}Q_{3}Q_{4}x_{3,i}y_{4,j})}\\\nn
&&\times \prod_{i,j}\frac{(1+Q_{\bullet}^{n+1}Q_{3}^{-1}y_{2,i}x_{4,j})(1+Q_{\bullet}^{n+1}Q_{4}^{-1}x_{1,i}y_{3,j})(1+Q_{\bullet}^{n+1}Q_{1}^{-1}x_{2,i}y_{4,j})
(1+Q_{\bullet}^{n+1}Q_{2}^{-1}y_{1,i}x_{3,j})}
{(1-Q_{\bullet}^{n+1}y_{2,i}x_{3,j})(1-Q_{\bullet}^{n+1}y_{3,i}x_{4,j})(1-Q_{\bullet}^{n+1}x_{1,i}y_{4,j})(1-Q_{\bullet}^{n+1}y_{1,i}x_{2,j})}\,.\eea

A generalization of the above result is given in Eq.(\ref{id}).

\subsection{Modular Forms}\label{App:Defs}
Among the many different modular objects appearing in the main body of this work, a particular role is played by the Jacobi theta functions. They can be defined in the following manner in terms of infinite products
{\allowdisplaybreaks
\begin{align}
\theta_1(\tau;\nu)&:=-i\nome^{\frac{1}{8}} y^{\frac{1}{2}}\, \prod_{n=1}^\infty(1-\nome^n)(1-y\nome^n)(1-y^{-1}\nome^{n-1})
\nonumber \\
\theta_2(\tau;\nu)&:=\nome^{\frac{1}{8}} (y^{\frac{1}{2}}+y^{-\frac{1}{2}})\, \prod_{n=1}^{\infty} (1-\nome^n)\, (1+y\nome^n)
(1+y^{-1} q^n)   \nonumber \\
\theta_3(\tau;\nu)&:=\prod_{n=1}^{\infty} (1-\nome^n) \, (1+y\nome^{n-1/2})(1+y^{-1}\nome^{n-1/2}) \nonumber\\
\theta_4(\tau;\nu)&:=\prod_{n=1}^{\infty} (1-\nome^n) \, (1-y\nome^{n-1/2}) (1-y^{-1}\nome^{n-1/2}) \ \ .
\end{align}}
where we have introduced the shorthand notation for the variables
\begin{align}
&\nome=e^{2\pi i\tau}\,,&&\text{and} &&y=e^{2\pi i\nu}\,.
\end{align}
%For later convenience we also not the following addition formulae \cite{Whittaker}
%\begin{align}
%&\theta_1(\tau;y+z)\theta_1(\tau;y-z)=\frac{\theta_3(\tau;y)^2\theta_2(\tau;z)^2-\theta_2(\tau;y)^2\theta_3(\tau;z)^2}{\theta_4(\tau;0)^2}=\frac{\theta_1(\tau;y)^2\theta_4(\tau;z)^2-\theta_4(\tau;y)^2\theta_1(\tau;z)^2}{\theta_4(\tau;0)^2}\nonumber\\
%&\theta_2(\tau;y+z)\theta_2(\tau;y-z)=\frac{\theta_4(\tau;y)^2\theta_2(\tau;z)^2-\theta_1(\tau;y)^2\theta_3(\tau;z)^2}{\theta_4(\tau;0)^2}=\frac{\theta_2(\tau;y)^2\theta_4(\tau;z)^2-\theta_3(\tau;y)^2\theta_1(\tau;z)^2}{\theta_4(\tau;0)^2}\nonumber\\
%&\theta_3(\tau;y+z)\theta_3(\tau;y-z)=\frac{\theta_4(\tau;y)^2\theta_3(\tau;z)^2-\theta_1(\tau;y)^2\theta_2(\tau;z)^2}{\theta_4(\tau;0)^2}=\frac{\theta_3(\tau;y)^2\theta_4(\tau;z)^2-\theta_2(\tau;y)^2\theta_1(\tau;z)^2}{\theta_4(\tau;0)^2}\nonumber\\
%&\theta_4(\tau;y+z)\theta_4(\tau;y-z)=\frac{\theta_3(\tau;y)^2\theta_3(\tau;z)^2-\theta_2(\tau;y)^2\theta_2(\tau;z)^2}{\theta_4(\tau;0)^2}=\frac{\theta_4(\tau;y)^2\theta_4(\tau;z)^2-\theta_1(\tau;y)^2\theta_1(\tau;z)^2}{\theta_4(\tau;0)^2}\,.\label{AdditionFormulae}
%\end{align}
Moreover, we will find the following identities useful in the main body of the text
\begin{align}
&\frac{\partial}{\partial z}\frac{\theta_1(\tau;z)}{\theta_4(\tau;z)}=\frac{\theta_2(\tau;z)\theta_3(\tau;z)\theta_4(\tau;0)^2}{\theta_4(\tau;z)^2}\,,\label{DiffIdentTheta}\\
&\frac{\partial}{\partial z}\frac{\theta_2(\tau;z)}{\theta_4(\tau;z)}=-\frac{\theta_1(\tau;z)\theta_3(\tau;z)\theta_3(\tau;0)^2}{\theta_4(\tau;z)^2}\,,\\
&\frac{\partial}{\partial z}\frac{\theta_3(\tau;z)}{\theta_4(\tau;z)}=-\frac{\theta_1(\tau;z)\theta_2(\tau;z)\theta_2(\tau;0)^2}{\theta_4(\tau;z)^2}\,,
\end{align}
as well as
\begin{align}
\frac{\theta_1(\tau;2z)}{2\theta_1(\tau;z)}=\frac{\theta_2(\tau;z)\theta_3(\tau;z)\theta_4(\tau;z)}{\theta_2(\tau;0)\theta_3(\tau;0)\theta_4(\tau;0)}\,.\label{ThreeTheta}
\end{align}
In addition to the theta-functions we also introduce the Dedekind-eta function
\begin{align}
\eta(\tau):=\nome^{\frac{1}{24}}\prod_{n=1}^\infty(1-\nome^n)\,,
\end{align}
which is a holomorphic modular form of weight $\frac{1}{2}$ of $SL(2,\mathbb{Z})$. It can be related to the Jacobi theta functions through \cite{Whittaker}
\begin{align}
\theta_2(\tau;0)\theta_3(\tau;0)\theta_4(\tau;0)=2\eta(\tau)^3\,.\label{Tripleeta}
\end{align}
With these quantities, we can finally introduce the standard weak Jacobi forms $\phi_{-2,1}(z;y)$ and $\phi_{0,1}(z;y)$ of index 1 and weight $-2$ and $0$ respectively (see \cite{EichlerZagier})
\begin{align}
&\phi_{-2,1}(\tau;\nu):=-\frac{\theta_1(\tau;\nu)^2}{\eta^6(\tau)}\,, &&\phi_{0,1}(\tau;\nu):=4\sum_{i=2}^4\frac{\theta_i(\tau;\nu)^2}{\theta_i(\tau;1)^2}\,,\label{JacobiForms}
\end{align}
$\phi_{0,1}(\tau;\nu)$ is proportional to the elliptic genus of $K3$
\begin{align}
\phi_{0,1}(\tau;\nu)=\frac{1}{2}\chi_{\text{ell}}(K3;\tau,\nu)\,.
\end{align}
Moreover, another class of modular forms which is of great importance in the main body of this work are the Eisenstein series which are defined as
\begin{align}
E_{2n}(\tau):=1-\frac{4n}{B_{2n}}\sum_{k=1}^\infty\sigma_{2n-1}(k)\,\nome^k\,,
\end{align}
where $B_{2n}$ are the Bernoulli numbers and $\sigma_{2n-1}(k)$ the divisor function. Notice that for $n\geq 2$ the Eisenstein series $E_{2n}$ are holomorphic and transform with weight $2n$ under $SL(2,\mathbb{Z})$. $E_2(\tau)$ itself is not a modular form but acquires a shift-term under $SL(2,\mathbb{Z})$ transformations. The latter can be removed by defining the quasi modular form
\begin{align}
&\hat{E}_2(\tau,\bar{\tau}):=E_2(\tau)-\frac{3}{\pi\,\tau_2}\,,&&\text{with} &&\tau_2:=\text{Im} \tau\,,
\end{align}
which, however, is manifestly not holomorphic. We should also mention that the Eisenstein series are not independent objects. Indeed, for $n\geq4$ each $E_{2n}(\tau)$ can be expressed as a finite polynomial in $(E_4,E_6)$. In some cases we will also use the un-normalized Eisenstein series which are defined as
\begin{align}
&\hat{G}_2(\tau,\bar{\tau}):=-\frac{1}{24}\,\hat{E}_2(\tau,\bar{\tau})\,,&&\text{and} &&G_{2n}(\tau):=-\frac{B_{2n}}{4n}\,E_{2n}(\tau)\,.
\end{align}
Finally, we introduce the elliptic Weierstrass function $\mathcal{P}(\tau,\nu)$
\begin{align}
\mathcal{P}(\tau,\nu):=\nu^{-2}+\sum_{{\omega\in\mathbb{Z}\tau+\mathbb{Z}}\atop {\omega\neq 0}}\left[(\nu+\omega)^{-2}-\omega^{-2}\right]\,,
\end{align}
which is a meromorphic Jacobi form of weight $2$ and index $0$. An alternative way of writing the Weierstrass function is (see \cite{Gritsenko:1999nm})
\begin{align}
\mathcal{P}(\tau,\nu)=8\pi^2 G_2-\frac{\partial^2}{\partial\nu^2}\log(i\theta_1(\tau;\nu))\,,\label{ExplicitWeierstrass}
\end{align}
which highlights the relation to the Eisenstein series. Throughout this work we denote derivatives of $\mathcal{P}(\tau,\nu)$ with respect to the second argument by
\begin{align}
&\mathcal{P}^{(n)}(\tau,\nu)=\partial_\nu^n\mathcal{P}(\tau,\nu)\,.
\end{align}
These are meromorphic Jacobi forms of weight $n+2$ and index zero and have a pole of order $n+2$ at $z=0$. Following \cite{Gritsenko:1999nm} these derivatives can be written in terms of the Eisenstein series in the following manner
\begin{align}
\mathcal{P}^{(n-2)}(\tau,\nu)=\frac{(-1)^n(n-1)!}{\nu^n}+2\sum_{k\geq 2\atop{2k\geq n}} (2\pi i)^{2k} G_{2k}(\tau)\,\frac{\nu^{2k-n}}{(2k-n)!}\,.\label{DerWeierstrass}
\end{align}
%%%%%%%%%%%%%%%%%%%%%%%%%%%%%%%%%%%%%%%%%%%%%%%%%%%%%
\subsection{Double Elliptic Gamma Function}
Following the notation of \cite{Spiridonov:2012de}, for $x\in\mathbb{C}^*$, we define the double elliptic Gamma function as
\begin{align}
\Gamma_2(x;y_1,y_2,y_3):=\prod_{i,j,k=0}^\infty\left(1-x^{-1} Q_{y_1}^{i+1}Q_{y_2}^{j+1}Q_{y_3}^{k+1}\right)\left(1-x Q_{y_1}^i Q_{y_2}^j Q_{y_3}^k\right)\,,\label{DefGamma2}
\end{align}
with the shorthand notation
\begin{align}
&Q_{y_i} =e^{2\pi i y_i}\,.
\end{align}
It satisfies the following relations
\begin{align}
&\frac{\Gamma_2(x Q_{y_1};y_1,y_2,y_3)}{\Gamma_2(x;y_1,y_2,y_3)}=\Gamma_1(x;y_2,y_3)\,,\nonumber\\
&\frac{\Gamma_2(x Q_{y_2};y_1,y_2,y_3)}{\Gamma_2(x;y_1,y_2,y_3)}=\Gamma_1(x;y_1,y_3)\,,\label{DifferenceEq}\\
&\frac{\Gamma_2(x Q_{y_3};y_1,y_2,y_3)}{\Gamma_2(x;y_1,y_2,y_3)}=\Gamma_1(x;y_1,y_2)\,,\nonumber
\end{align}
where $\Gamma_1$ is the standard elliptic Gamma function defined as
\begin{align}
\Gamma_1(x;y_1,y_2):=\prod_{i,j=0}^\infty\frac{1-x^{-1}Q_{y_1}^{i+1}Q_{y_2}^{j+1}}{1-x Q_{y_1}^iQ_{y_2}^j}\,.\label{DefGamma1}
\end{align}

The double elliptic Gamma function $\Gamma_2$ has very interesting modular properties
\begin{align}
\Gamma_{2}(z;\rho,\epsilon_1,\epsilon_2)=\Gamma_{2}(\frac{z}{\rho};-\frac{1}{\rho},\frac{\epsilon_1}{\rho},\frac{\epsilon_2}{\rho})
\Gamma_{2}(\frac{z}{\epsilon_1};\frac{\rho}{\epsilon_1},-\frac{1}{\epsilon_1},\frac{\epsilon_2}{\epsilon_1})
\Gamma_{2}(\frac{z}{\epsilon_2};\frac{\rho}{\epsilon_2},\frac{\epsilon_1}{\epsilon_2},-\frac{1}{\epsilon_2})\,\mbox{exp}\Big(\frac{i\pi}{12}B_{44}\Big)\,,\label{GammaMod}
\end{align}
where $B_{4,4}$ is given by
\bea
B_{4,4}(z;\rho,\epsilon_1,\epsilon_2)=\frac{d^{4}}{dx^4}\frac{x^4\,e^{z\,x}}{(e^{\rho\,x}-1)(e^{\epsilon_{1}\,x}-1)(e^{\epsilon_{2}\,x}-1)}|_{x=0}\,.
\eea

%%%%%%%%%%%%%%%%%%%%%%%%%%%%%%%%%%%%%%%%%%%%%%%%%
\subsection{Equivariant Cohomology and Integration}
In defining the modified elliptic genus for a non-compact manifold in section~\ref{Sect:EllGen} we have used a particular \emph{equivariant integration}. In this appendix we will give a brief review of this technique and point the reader to the relevant literature.

Our starting point is a manifold $\mathcal{M}$ with a non-trivial action of a Lie group $G$. We will denote the corresponding Lie algebra by $\mathfrak{g}$ and its positive definite invariant quadratic form by $(\cdot,\cdot)_{\mathfrak{g}}$. For $\Omega^\star(\mathcal{M})$ the de Rham complex of $\mathcal{M}$, following \cite{Atiyah,Givental,Witten:1992xu} we define the $G$-\emph{equivariant de Rham complex} $\Omega_G^\star(X)$ as the $G$-invariant part of $(\Omega^\star(X)\otimes \text{Fun}(\mathfrak{g}))$, where $\text{Fun}(\mathfrak{g})$ is the algebra of graded polynomial functions on $\mathfrak{g}$. We will call elements of $\Omega_G^\star(X)$ equivariant differential forms $\alpha$.

For $g\in\mathfrak{g}$ let $V_g\in\mathcal{M}$ be the vector field generated by the infinitesimal action of $g$. Then we define the equivariant de Rahm operator as\footnote{We are using the definition of \cite{Witten:1992xu} which includes a factor $i$, usually omitted in the mathematical literature.}
\begin{align}
D:=d-i\iota_{V_g}\,,
\end{align}
where $\iota_{V_g}$ is the contraction by the vector field $V_g$. Following \cite{Witten:1992xu}
\begin{align}
D^2=-i\mathcal{L}_{V_g}=-id\iota_{V_g}-i\iota_{V_g}\,,
\end{align}
which vanishes manifestly on $\Omega_G^\star(X)$. We denote the cohomology $H^\star_G(\mathcal{M})$ of $D$ as the G-\emph{equivariant cohomology of} $\mathcal{M}$.

For equivariant differential forms we can define the notion of an integral as the map
\begin{align}
\Omega^\star_G(\mathcal{M})&\longrightarrow \mathbb{C}\\
\alpha&\longrightarrow\frac{1}{\text{vol}(G)}\int_{\mathfrak{g}\times \mathcal{M}}\frac{d\phi_1\ldots d\phi_s}{(2\pi)^s}\cdot\alpha\,,
\end{align}
where $(\phi_1,\ldots,\phi_s)$ is a Euclidean basis of $\mathfrak{g}$. Since this integration generically does not converge, we can introduce the following regularization factor
\begin{align}
\oint \alpha=\frac{1}{\text{vol}(G)}\int_{\mathfrak{g}\times \mathcal{M}}\frac{d\phi_1\ldots d\phi_s}{(2\pi)^s}\cdot\alpha\cdot\text{exp}\left(-\frac{\epsilon}{2}(\phi,\phi)_{\mathfrak{g}}\right)\,,&&\text{with} &&\epsilon\in\mathbb{R}_+\,.
\end{align}
Integrals of this type can be computed as a sum over contributions of isolated fixed points of the $G$-action using the stationary phase approximation (Atiyah-Bott-Lefschetz fixed point theorem), which we will make use of in the main body of this text.

%%%%%%%%%%%%%%%%%%%%%%%%%%%%%%%%%%%%%%%%%%%%%%%%%%%%%
\subsection{$\beta$-deformed Backgrounds}\label{App:OmegaBackground}
Recently tremendous progress in the computation of the instanton partition functions of four-dimensional $\mathcal{N}=2$ gauge theories has been achieved \cite{Nekrasov:2002qd} by using particular localization techniques in a particular four-dimensional background. This so-called \emph{$\Omega$-background} is parameterized by two rotational parameters, which were called $\epsilon_1,\epsilon_2\in\mathbb{C}$ in the original work \cite{Moore:1997dj,Lossev:1997bz}: The starting point is six-dimensional $\mathcal{N}=1$ Super-Yang-Mills theory which yields a four-dimensional $\mathcal{N}=2$ gauge theory when compactified on a torus $T^2$. However, instead of the 'standard' product space compactification $\mathbb{R}^4\times T^2$ we will rather consider an $\mathbb{R}^4$ vector bundle over the two-torus $T^2$ with flat connection
\begin{align}
\text{Spin}(4)=SU(2)_L\times SU(2)_R\,.
\end{align}
To be more precise, the metric on this 6-dimensional space has the following structure
\begin{align}
ds^2=A\,dz\,d\bar{z}+g_{\mu\nu}\left(dx^\mu+V^\mu dz+\bar{V}^\mu d\bar{z}\right)\left(dx^\nu+V^\nu dz+\bar{V}^\nu d\bar{z}\right)\,,
\end{align}
with $A$ the volume of the $T^2$, $\{z,\bar{z}\}$ the complex coordinates on the torus and $x^\mu$ the four-dimensional coordinates. Moreover, we have introduced the quantities
\begin{align}
&V^\mu:{\Omega^\mu}_\nu x^\nu\,,&&\text{with} &&\Omega^{\mu\nu}=\left(\begin{array}{cccc} 0 & \epsilon_1 & 0 & 0 \\ -\epsilon_1 & 0 & 0 & 0 \\ 0 & 0 & 0 & \epsilon_2 \\ 0 & 0 & -\epsilon_2 & 0 \end{array}\right)\,.
\end{align}
To make contact with the computations in the main body of this paper, we have to identify
\begin{align}
&\epsilon_{1}=g_s\sqrt{\beta}\,,&&\text{and} &&\epsilon_2=-g_s/\sqrt{\beta}\,.
\end{align}
For generic $\beta\neq 1$, this background breaks supersymmetry. However, fermionic symmetries can be restored by introducing in addition Wilson lines in the $SU(2)$ R-symmetry group. This can be done by embedding an $SU(2)$ subgroup of the flat connection into the $SU(2)$ R-symmetry group. In the final step we will send the volume $A$ to zero.

%%%%%%%%%%%%%%%%%%%%%%%%%%%%%%%%%
\subsection{Expansion of $\chi_{\text{ell}}$ at $\beta=1$}\label{App:Expansion}
In this appendix we want to give a short proof of the expansion (\ref{ExplicitFg}). To this end, we first use addition formulae for the Jacobi-theta functions~\cite{Whittaker} to write
\begin{align}
&\chi_{\text{ell}}(\mathbb{C}^2;\tau,\nu,g_s)=-\frac{\theta_1(\tau;\nu)^2\theta_4(\tau;\frac{g_s}{2\pi})^2}{\theta_1(\tau;\frac{g_s}{2\pi})^2\,\theta_4(\tau;0)^2}+\frac{\theta_4(\tau;\nu)^2\theta_1(\tau;\frac{g_s}{2\pi})^2}{\theta_1(\tau;\frac{g_s}{2\pi})^2\,\theta_4(\tau;0)^2}\nonumber\\
%
%&=-\left[\frac{\theta_1(\tau;\nu)^2\theta_4(\tau;\frac{g_s}{2\pi})^2}{\theta_1(\tau;\frac{g_s}{2\pi})^2\,\theta_4(\tau;0)^2}-\frac{1}{3}\left(\frac{2\theta_4(\tau;\nu)^2}{\theta_4(\tau;0)^2}-\frac{\theta_2(\tau;\nu)^2}{\theta_2(\tau;0)^2}-\frac{\theta_3(\tau;\nu)^2}{\theta_3(\tau;0)^2}\right)\right]\nonumber\\
%&\phantom{=}\hspace{0.2cm}-\frac{1}{3}\left(\frac{2\theta_4(\tau;\nu)^2}{\theta_4(\tau;0)^2}-\frac{\theta_2(\tau;\nu)^2}{\theta_2(\tau;0)^2}-\frac{\theta_3(\tau;\nu)^2}{\theta_3(\tau;0)^2}\right)+\frac{\theta_4(\tau;\nu)^2}{\theta_4(\tau;0)^2}\nonumber\\
%
&=-\left[\frac{\theta_1(\tau;\nu)^2\theta_4(\tau;\frac{g_s}{2\pi})^2}{\theta_1(\tau;\frac{g_s}{2\pi})^2\,\theta_4(\tau;0)^2}-\frac{1}{3}\left(\frac{2\theta_4(\tau;\nu)^2}{\theta_4(\tau;0)^2}-\frac{\theta_2(\tau;\nu)^2}{\theta_2(\tau;0)^2}-\frac{\theta_3(\tau;\nu)^2}{\theta_3(\tau;0)^2}\right)\right]+\frac{\phi_{0,1}(\tau,\nu)}{12}\,,\nonumber
\end{align}
where we have used the definition (\ref{JacobiForms}). Explicitly expanding the term in the bracket into a Laurent series, one can check that it contains no term of order $g_s^0$. This not only immediately proves the middle line of equation (\ref{ExplicitFg}) but also implies that all $\fgel_g^{(k=0)}$ are proportional to $\phi_{-2,1}$. To argue for the form of the remaining terms, we will focus on the bracket and consider
\begin{align}
\tilde{\chi}_{\text{ell}}:=&\,2\pi\frac{\partial}{\partial g_s}\left[\frac{\theta_1(\tau;\nu)^2\theta_4(\tau;\frac{g_s}{2\pi})^2}{\theta_1(\tau;\frac{g_s}{2\pi})^2\,\theta_4(\tau;0)^2}-\frac{1}{3}\left(\frac{2\theta_4(\tau;\nu)^2}{\theta_4(\tau;0)^2}-\frac{\theta_2(\tau;\nu)^2}{\theta_2(\tau;0)^2}-\frac{\theta_3(\tau;\nu)^2}{\theta_3(\tau;0)^2}\right)\right]\nonumber\\
=&\phi_{-2,1}(\tau;\nu)\,\frac{\eta(\tau)^6\,\theta_2(\tau;\nu)\theta_3(\tau;\nu)\theta_4(\tau;\nu)}{\theta_1(\tau;\frac{g_s}{2\pi})^3}=\phi_{-2,1}(\tau;\nu)\,\frac{\theta_1(\tau;\frac{g_s}{\pi})\,\eta(\tau)^9}{\theta_1(\tau;\frac{g_s}{2\pi})^4}\nonumber
\end{align}
where we have used identities (\ref{DiffIdentTheta}), (\ref{ThreeTheta}) and (\ref{Tripleeta}). Using the notation of \cite{Gritsenko:1999nm}, the factor
\begin{align}
\mathfrak{p}\!\left(\tau;\frac{g_s}{2\pi}\right):=\frac{\theta_1(\tau;\frac{g_s}{\pi})\,\eta(\tau)^9}{\theta_1(\tau;\frac{g_s}{2\pi})^4}=\frac{\phi_{0,\frac{3}{2}}(\tau;\frac{g_s}{2\pi})}{\phi_{-1,\frac{1}{2}}(\tau;\frac{g_s}{2\pi})^3}\in J^{\text{mer}}_{(3,0)}\,,
\end{align}
is a meromorphic Jacobi form of weight $3$ and index $0$, which means that it can be understood as a differential
\begin{align}
&\mathfrak{p}\!\left(\tau;\frac{g_s}{2\pi}\right)=2\pi \frac{\partial}{\partial g_s}\,\mathfrak{h}\!\left(\tau;\frac{g_s}{2\pi}\right)\,,&&\text{with} &&\mathfrak{h}\in J^{\text{mer}}_{(2,0)}\,,
\end{align}
of a meromorphic Jacobi form of weight $2$ and index 0 and has a second order pole at $g_s=0$. Thus, $\mathfrak{h}$ is proportional to the Weierstrass function $\mathcal{P}(\tau;\frac{g_s}{2\pi})$. To see this, we note that $\mathfrak{h}$ can be made into a meromorphic Jacobi form of weight 0 and index 0 through multiplication by
\begin{align}
\mathfrak{g}\!\left(\tau;\frac{g_s}{2\pi}\right)=\frac{E_4(\tau)E_6(\tau)}{\eta(\tau)^{24}}\,\mathfrak{h}\!\left(\tau;\frac{g_s}{2\pi}\right)\in J^{\text{mer}}_{(0,0)}
\end{align}
As shown in \cite{Berndt} (see also \cite{Proc}), the field of $J^{\text{mer}}_{(0,0)}$ is the field $K\left(\frac{E_4(\tau)E_6(\tau)}{\eta(\tau)^{24}}\mathcal{P}(\tau,\nu),j(\tau)\right)$, where $j(\tau)$ is the Klein invariant function. Since $\mathfrak{g}$ has a second order pole in $g_s$,
\begin{align}
&\mathfrak{g}\!\left(\tau;\frac{g_s}{2\pi}\right)=\mathcal{P}\left(\tau;\frac{g_s}{2\pi}\right)\,j(\tau)^a\,,&&\text{for} &&a\in\mathbb{Z}\,.
\end{align}
Using moreover, that $\mathfrak{g}$ has only a first order pole in $q$ it follows that $a=0$.\footnote{$a=0$ can also be determined by series expanding $\mathfrak{g}$ in $g_s$ and comparing the coefficients.} Using finally the series expansion of the Weierstrass function (\ref{DerWeierstrass}) finally proves (\ref{ExplicitFg}).

%%%%%%%%%%%%%%%%%%%%%%%%%%%%%%%%%%%%%%%%%%%%%%%%%%%%%
\section{One-Loop Amplitudes in Type II Superstring}\label{Sect:LoopAmps}
In section~\ref{Sect:OneLoopStringAmp} we consider a class of one-loop amplitudes in type II string theory compactified on $T^2$. In this appendix we compile a few details about our notation and conventions
%%%%%%%%%%%%%%%%
\subsection{Narain Partition Function}
The one loop partition function for the internal part of type II string theory compactified on $T^2$ can be written as a lattice sum over the Narain lattice $\Gamma^{(2,2)}$ of signature $(2,2)$. This theta-series depends explicitly on th K\"ahler and complex structure modulus $(T,U)=(T_1+i T_2,U_1+iU_2)$ of $T^2$, as well as the parameter $\tau=\tau_1+i\tau_2$ of the world-sheet torus
\begin{align}
&\Theta^{(2,2)}(T,U;\tau)=\sum_{m_{1,2},n_{1,2}} q^{\frac{1}{2}|P_L|^2}\bar{q}^{\frac{1}{2}|P_R|^2}\,,\label{NarainPartitionFunction}
\end{align}
Here $(P_L,P_R)$ are the Narain momenta, which can explicitly be written
\begin{align}
P_L=\frac{1}{\sqrt{2T_2U_2}} (m_1+m_2 U+n_1\bar{T} +n_2\bar{T}U)\,,\\
P_R=\frac{1}{\sqrt{2T_2U_2}} (m_1+m_2 U+n_1T +n_2TU)\,,
\end{align}
which satisfy $|P_L|^2-|P_R|^2=2(m_1n_2-m_2n_1)$.
%%%%%%%%%%%%%%%%%%%%%%%%%%%%%%%%%%%%%%%%%%%%%%%%%%%%%
\subsection{Vertex Operators in Type II String Theory on $T^2$}\label{App:Vertices}
To explicitly compute string amplitudes, we need the world-sheet emission vertex operators for the string states discussed in the previous section. We consider type II string theory compactified on $T^2$ and use a similar notation as in \cite{Antoniadis:1993ze} (see also \cite{HS}): we introduce a basis of 10-dimensional bosonic complex coordinates $(X_1,\ldots,X_5)$ where $X_3$ corresponds to the torus $T^2$. The superpartners of these will be denoted $(\psi_1,\ldots,\psi_5)$ in the left moving sector and $(\tilde{\psi}_1,\ldots,\tilde{\psi}_5)$ in the right moving sector. For explicit computations we will bosonize the fermions in terms of free scalar fields $\phi_{1,\ldots,5}$ (and $\tilde{\phi}_{1,\ldots,5}$ respectively), such that
\begin{align}
\begin{array}{l}\psi_i=e^{i\phi_i}\,,\\[4pt]\tilde{\psi}_i=e^{i\tilde{\phi}}\,,\end{array}&&\text{and}&&\begin{array}{l}\bar{\psi}_i=e^{-i\phi_i}\,,\\[4pt]\bar{\tilde{\psi}}_i=e^{-i\tilde{\phi}}\,,\end{array}&&\forall\,i=1,\ldots,5\,.\label{Bososnization}
\end{align}
The relevant bosonic contractions for the non-compact directions are
\begin{align}
\langle X_I(z)\bar{X}_J(0)\rangle_{1-\text{loop}}&=-\delta_{IJ}\log\left(e^{-\frac{2\pi \text{Im}(z)}{\tau_2}^2}\left|\frac{\theta_1(\tau;z)}{2\pi\eta(\tau)^3}\right|^2\right)\,,&&\forall I,J\neq 3\,,\label{BosonicContract}\\
\left\langle\psi_I(z)\bar{\psi}_J(0)\right\rangle\!\big[^a_b\big]&=2\pi\delta_{IJ}\frac{\theta\big[^a_b\big](\tau;z)\eta(\tau)^3}{\theta\big[^a_b\big](\tau;0)\theta_1(\tau;z)}\,.
\end{align}
For the bosonic torus coordinates, we will use the correlators
\begin{align}
&\left\langle\partial X_3(z)\partial X_3(0)\right\rangle=P_L^2\,,&&
\left\langle\bar{\partial} X_3(z)\bar{\partial} X_3(0)\right\rangle=P_R^2\,,&&\left\langle\partial X_3(z)\bar{\partial} X_3(0)\right\rangle=P_LP_R\,,
\end{align}
where $(P_L,P_R)$ are the Narain-momenta as discussed in the previous section.

From the spectrum discussed in the previous section, we see that there are massless as well as massive states. Among the massless states, the ones relevant for us is the graviton, whose vertex in the $(0,0)$-ghost picture takes the form
\begin{align}
V_R^{(0,0)}(h,p)=:h^{\mu\nu}\left[\partial X_\mu+i(p\cdot\psi)\psi_\mu\right]\,\left[\bar{\partial} X_\nu+i(p\cdot\tilde{\psi})\tilde{\psi}_\nu\right]e^{ip\cdot X}:\,,
\end{align}
which is characterized by a symmetric traceless polarization tensor $h^{\mu\nu}$ and a momentum $p^{\mu}$, such that $p_{\mu}h^{\mu\nu}=0$. Besides this, among the massless states we will make use of the KK vector fields stemming from the reduction on $T^2$. Their vertices are given by in the $(0,0)$-ghost picture are given by
\begin{align}
V_F^{(0,0)}(\epsilon,p)=:\epsilon^\mu\left[\partial X_\mu+i(p\cdot\psi)\psi_\mu\right]\bar{\partial} X_3\,e^{ip\cdot X}:\,,
\end{align}
and are determined by a polarization vector $\epsilon^{\mu}$ together with a space-time momentum $p^\mu$ such that $\epsilon^{\mu}p_{\mu}=0$.

Vertices of massive string states have first been introduced in \cite{Koh:1987hm} (see also \cite{Feng:2010yx,HS} for their explicit use in string scattering amplitudes). All vertices of physical states at the first massive level have been classified and written explicitly in \cite{Schlotterer} (see also \cite{Lust:2012zv}). Among those we only need a very particular class. Our strategy is to consider a generalisation of the computations in \cite{HS} by thinking of the directions $X^{4,5}$ as arising from the large volume limit of a $K3$ compactification. In this manner, we consider the analogue of the free $SU(2)$-currents in the compact picture, which is the way how they appeared in \cite{HS}. Their vertices in the $(-1,0)$-ghost picture take the form
\begin{align}
V^{(-1,0)}_A(p)=:e^{-\varphi}\psi_3\,\mathcal{J}_A\,e^{ip\cdot X}:\,,&&\text{with} &&\mathcal{J}_{A}=\left\{\begin{array}{lcl}\psi_4\psi_5 & \text{for} & A=++\,, \\[4pt] \psi_4\bar{\psi}_4+\psi_5\bar{\psi}_5 & \text{for} & A=0\,, \\[4pt] \bar{\psi}_4\bar{\psi}_5 & \text{for} & A=--\,.\end{array}\right.
\end{align}
Finally there we use one further massive state, which can be found in \cite{Schlotterer}
\begin{align}
&V^{(-1,0)}(\gamma,p)=:i\gamma^\mu\,e^{-\varphi}\,\left[\partial X_\mu\psi_3+\partial X_3\psi_\mu\right]\bar{\partial} X_3\,e^{ip\cdot X}:\,.
\end{align}
These vertex operators will turn out to be sufficient for all computations performed in the main body of this work.
%%%%%%%%%%%%%%%%%%%%%%%%%%%%%%%%%%%%%%%%%%%%%%%%%%%%%%%%%%%%%%%%%%%%%%%%%%%%%%%%%%%%%%%%%%%%%%%%%


\begin{thebibliography}{99}
\bibitem{Witten:1995zh} E.~Witten, {\it Some comments on string dynamics,}
  In *Los Angeles 1995, Future perspectives in string theory* 501-523
  [hep-th/9507121].
  %%CITATION = HEP-TH/9507121;%%
\bibitem{Bonetti:2012st}
 M.~R.~Douglas, {\it On D=5 super Yang-Mills theory and (2,0) theory,}
  JHEP {\bf 1102} (2011) 011
  [arXiv:1012.2880 [hep-th]].
  %%CITATION = ARXIV:1012.2880;%%
    N.~Lambert, C.~Papageorgakis and M.~Schmidt-Sommerfeld,
  {\it M5-Branes, D4-Branes and Quantum 5D super-Yang-Mills,}
  JHEP {\bf 1101} (2011) 083
  [arXiv:1012.2882 [hep-th]].
  %%CITATION = ARXIV:1012.2882;%%
 P.~-M.~Ho, K.~-W.~Huang and Y.~Matsuo, {\it A Non-Abelian Self-Dual Gauge Theory in 5+1 Dimensions,}
  JHEP {\bf 1107} (2011) 021
  [arXiv:1104.4040 [hep-th]].
  %%CITATION = ARXIV:1104.4040;%%
 C.~-S.~Chu and S.~-L.~Ko, {\it Non-abelian Action for Multiple Five-Branes with Self-Dual Tensors,}
  JHEP {\bf 1205} (2012) 028
  [arXiv:1203.4224 [hep-th]].
  %%CITATION = ARXIV:1203.4224;%%
F.~Bonetti, T.~W.~Grimm and S.~Hohenegger, {\it A Kaluza-Klein inspired action for chiral p-forms and their anomalies,}
  Phys.\ Lett.\ B {\bf 720} (2013) 424
  [arXiv:1206.1600 [hep-th]].
  %%CITATION = ARXIV:1206.1600;%%
K.~-W.~Huang, {\it Non-Abelian Chiral 2-Form and M5-Branes,}
  arXiv:1206.3983 [hep-th].
  %%CITATION = ARXIV:1206.3983;%%
  F.~Bonetti, T.~W.~Grimm and S.~Hohenegger, {\it Non-Abelian Tensor Towers and (2,0) Superconformal Theories,}
  JHEP {\bf 1305} (2013) 129
  [arXiv:1209.3017 [hep-th]].
  %%CITATION = ARXIV:1209.3017;%%
\bibitem{Aharony:1997bh}
  O.~Aharony, A.~Hanany and B.~Kol,
  ``Webs of (p,q) five-branes, five-dimensional field theories and grid diagrams,''
  JHEP {\bf 9801}, 002 (1998)
  [hep-th/9710116].
\bibitem{Hollowood:2003cv}
T.~J.~Hollowood, A.~Iqbal and C.~Vafa, {\it Matrix models, geometric engineering and elliptic genera,} JHEP {\bf 0803} (2008) 069 [hep-th/0310272].
%%CITATION = HEP-TH/0310272;%%
\bibitem{Nekrasov:2002qd} N.~A.~Nekrasov, {\it Seiberg-Witten prepotential from instanton counting,} Adv.\ Theor.\ Math.\ Phys.\  {\bf 7 } (2004)  831-864. [hep-th/0206161].
\bibitem{mstrings} B.~Haghighat, A.~Iqbal, C.~Kozcaz, G.~Lockhart and C.~Vafa, {\it M-Strings,} arXiv:1305.6322 [hep-th].
%%CITATION = ARXIV:1305.6322;%%
\bibitem{Gopakumar:1998ii}
R.~Gopakumar and C.~Vafa, {\it M theory and topological strings. 1.,} hep-th/9809187.

\bibitem{Gopakumar:1998jq}
  R.~Gopakumar and C.~Vafa,
  ``M theory and topological strings. 2.,''
  hep-th/9812127.

%%CITATION = HEP-TH/9802016;%%
\bibitem{BCOV}
M. Bershadsky, S. Cecotti, H. Ooguri and C. Vafa, {\it Kodaira-Spencer theory of gravity and
exact results for quantum string amplitudes}, Commun. Math. Phys. 165 (1994) 311
[hep-th/9309140].
\bibitem{Iqbal:2007ii}
A.~Iqbal, C.~Kozcaz and C.~Vafa, {\it The Refined topological vertex,} JHEP {\bf 0910} (2009) 069 [hep-th/0701156].
%%CITATION = HEP-TH/0701156;%%
\bibitem{Katz:1996fh}
  S.~H.~Katz, A.~Klemm and C.~Vafa, {\it Geometric engineering of quantum field theories,}
  Nucl.\ Phys.\ B {\bf 497}, 173 (1997)
  [hep-th/9609239].
 \bibitem{Katz:1997eq}
  S.~Katz, P.~Mayr and C.~Vafa, {\it Mirror symmetry and exact solution of 4-D N=2 gauge theories: 1.,}
  Adv.\ Theor.\ Math.\ Phys.\  {\bf 1}, 53 (1998)  [hep-th/9706110].
\bibitem{BGKV}
B.~Haghighat, C.~Kozcaz, G.~Lockhart and C.~Vafa, {\it On orbifolds of M-Strings,}
  arXiv:1310.1185 [hep-th].
\bibitem{Witten:1993yc} E.~Witten, {\it Phases of N=2 theories in two-dimensions,} Nucl.\ Phys.\ B {\bf 403}, 159 (1993)  [hep-th/9301042].
\bibitem{Gritsenko:1999nm} V.~Gritsenko, {\it Complex vector bundles and Jacobi forms,} math/9906191.
%%CITATION = MATH/9906191;%%
\bibitem{Antoniadis:1993ze} I.~Antoniadis, E.~Gava, K.~S.~Narain and T.~R.~Taylor, {\it Topological amplitudes in string theory,} Nucl.\ Phys.\ B {\bf 413} (1994) 162 [hep-th/9307158].
%%CITATION = HEP-TH/9307158;%%
I.~Antoniadis, E.~Gava, K.~S.~Narain and T.~R.~Taylor, {\it N=2 type II heterotic duality and higher derivative F terms,} Nucl.\ Phys.\ B\ {\bf 455} (1995) 109 [hep-th/9507115].
%%CITATION = NUPHA,B455,109;%%
I.~Antoniadis, E.~Gava, K.~S.~Narain and T.~R.~Taylor, {\it Topological amplitudes in heterotic superstring theory,} Nucl.\ Phys.\ B {\bf 476} (1996) 133
[hep-th/9604077].
%%CITATION = HEP-TH/9604077;%%
I.~Antoniadis, K.~S.~Narain and T.~R.~Taylor, {\it Open string topological amplitudes and gaugino masses,} Nucl.\ Phys.\ B {\bf 729} (2005) 235
[hep-th/0507244].
%%CITATION = HEP-TH/0507244;%%
I.~Antoniadis, S.~Hohenegger, K.~S.~Narain and E.~Sokatchev, {\it A New Class of N=2 Topological Amplitudes,} Nucl.\ Phys.\ B {\bf 823} (2009) 448
[arXiv:0905.3629 [hep-th]].
%%CITATION = ARXIV:0905.3629;%%
I.~Antoniadis, S.~Hohenegger, K.~S.~Narain and E.~Sokatchev, {\it Generalised N=2 Topological Amplitudes and Holomorphic Anomaly Equation,}
Nucl.\ Phys.\ B {\bf 856} (2012) 360 [arXiv:1107.0303 [hep-th]].
%%CITATION = ARXIV:1107.0303;%%
\bibitem{Lerche:1999ju}
 N.~Berkovits and C.~Vafa, {\it N=4 topological strings,}
  Nucl.\ Phys.\ B {\bf 433} (1995) 123
  [hep-th/9407190].
  %%CITATION = HEP-TH/9407190;%%
W.~Lerche and S.~Stieberger, {\it 1/4 BPS states and nonperturbative couplings in N=4 string theories,} Adv.\ Theor.\ Math.\ Phys.\  {\bf 3} (1999) 1539 [hep-th/9907133].
%%CITATION = HEP-TH/9907133;%%
I.~Antoniadis, S.~Hohenegger and K.~S.~Narain, {\it N=4 Topological Amplitudes and String Effective Action,} Nucl.\ Phys.\ B {\bf 771} (2007) 40
[hep-th/0610258].
%%CITATION = HEP-TH/0610258;%%
I.~Antoniadis, S.~Hohenegger, K.~S.~Narain and E.~Sokatchev, {\it Harmonicity in N=4 supersymmetry and its quantum anomaly,}
Nucl.\ Phys.\ B {\bf 794} (2008) 348 [arXiv:0708.0482 [hep-th]].
%%CITATION = ARXIV:0708.0482;%%
 \bibitem{Billo:2007va}
  M.~Billo, M.~Frau, F.~Fucito and A.~Lerda,
  {\it Instanton calculus in R-R background and the topological string,}
  JHEP {\bf 0611} (2006) 012
  [hep-th/0606013].
  %%CITATION = HEP-TH/0606013;%%
  M.~Billo, {\it Instanton Calculus With R-R Background And Topological Strings,}
  Fortsch.\ Phys.\  {\bf 55} (2007) 561
  [hep-th/0701072].
  %%CITATION = HEP-TH/0701072;%%
 M.~Billo, L.~Ferro, M.~Frau, F.~Fucito, A.~Lerda and J.~F.~Morales, {\it Flux interactions on D-branes and instantons,}
  JHEP {\bf 0810} (2008) 112
  [arXiv:0807.1666 [hep-th]].
%%CITATION = ARXIV:0807.1666;%%
   M.~Billo, L.~Ferro, M.~Frau, F.~Fucito, A.~Lerda and J.~F.~Morales, {\it Non-perturbative effective interactions from fluxes,}
  JHEP {\bf 0812} (2008) 102
  [arXiv:0807.4098 [hep-th]].
  %%CITATION = ARXIV:0807.4098;%%
\bibitem{Antoniadis:2013bja}
  I.~Antoniadis, I.~Florakis, S.~Hohenegger, K.~S.~Narain and A.~Z.~Assi,
{\it Worldsheet Realization of the Refined Topological String,}
  Nucl.\ Phys.\ B {\bf 875} (2013) 101
  [arXiv:1302.6993 [hep-th]].
  %%CITATION = ARXIV:1302.6993;%%
\bibitem{Billo:2013fi}
  M.~Billo, M.~Frau, L.~Gallot, A.~Lerda and I.~Pesando, {\it Deformed N=2 theories, generalized recursion relations and S-duality,}
  JHEP {\bf 1304} (2013) 039
  [arXiv:1302.0686 [hep-th]].
  %%CITATION = ARXIV:1302.0686;%%
 M.~Bill—, M.~Frau, L.~Gallot, A.~Lerda and I.~Pesando, {\it Modular anomaly equation, heat kernel and S-duality in N=2 theories,}
  arXiv:1307.6648 [hep-th].
  %%CITATION = ARXIV:1307.6648;%%
\bibitem{AHFNZ}  I.~Antoniadis, I.~Florakis, S.~Hohenegger, K.~S.~Narain and A.~Zein Assi, {\it Non-Perturbative Nekrasov Partition Function from String Theory,} arXiv:1309.6688 [hep-th].
%%CITATION = ARXIV:1309.6688;%%
\bibitem{Antoniadis:2010iq} J.~F.~Morales and M.~Serone, {\it Higher derivative F terms in N=2 strings,} Nucl.\ Phys.\ B {\bf 481} (1996) 389 [hep-th/9607193]. I.~Antoniadis, S.~Hohenegger, K.S.~Narain and T.R.~Taylor, {\it Deformed Topological Partition Function and Nekrasov Backgrounds,}
Nucl.Phys. \textbf{B838} (2010) 253-265. arXiv:1003.2832 [hep-th].
%%CITATION = ARXIV:1003.2832;%%
Y.~Nakayama, H.~Ooguri, {\it Comments on Worldsheet Description of the Omega Background,} [arXiv:1106.5503 [hep-th]].
\bibitem{Hohenegger:2011us} S.~Hohenegger and S.~Stieberger, {\it BPS Saturated String Amplitudes: K3 Elliptic Genus and Igusa Cusp Form,} Nucl.\ Phys.\ B {\bf 856} (2012) 413
 [arXiv:1108.0323 [hep-th]].
%%CITATION = ARXIV:1108.0323;%%
\bibitem{Aganagic:1999fe} M.~Aganagic, A.~Karch, D.~L\"ust and A.~Miemiec, {\it Mirror symmetries for brane configurations and branes at singularities,}
  Nucl.\ Phys.\ B {\bf 569} (2000) 277
  [hep-th/9903093].
  %%CITATION = HEP-TH/9903093;%%
\bibitem{macdonald} I.G. Macdonal, {\it Symmetric Functions and Hall Polynomials,} second ed., Oxford Mathematical Monographs, Oxford
Science Publications, 1995.
\bibitem{Lockhart:2012vp}
G.~Lockhart and C.~Vafa, {\it Superconformal Partition Functions and Non-perturbative Topological Strings}, arXiv:1210.5909 [hep-th].
\bibitem{Iqbal:2008ra}
  A.~Iqbal, C.~Kozcaz and K.~Shabbir,
  {\it Refined Topological Vertex, Cylindric Partitions and the U(1) Adjoint Theory,}
  Nucl.\ Phys.\ B {\bf 838}, 422 (2010)
  [arXiv:0803.2260 [hep-th]].
  %%CITATION = ARXIV:0803.2260;%%
\bibitem{Kawai:1994np}
  T.~Kawai and K.~Mohri,
  {\it Geometry of (0,2) Landau-Ginzburg orbifolds,}
  Nucl.\ Phys.\ B {\bf 425}, 191 (1994)
  [hep-th/9402148].
\bibitem{Smirnov}
A. Smirnov, {\it On the instanton R-matrix}, [arXiv:1302.0799].
\bibitem{Bao:2011rc}
  L.~Bao, E.~Pomoni, M.~Taki and F.~Yagi, {\it M5-Branes, Toric Diagrams and Gauge Theory Duality,}
  JHEP {\bf 1204}, 105 (2012)
  [arXiv:1112.5228 [hep-th]].


\bibitem{hep-th/0002169} T.~Kawai and K.~Yoshioka, {\it String partition functions and infinite products,} Adv.\ Theor.\ Math.\ Phys.\ \ {\bf 4} (2000) 397 [hep-th/0002169].
%%CITATION = 00203,4,397;%%
\bibitem{Borcherds2} R.E.~Borcherds, {\it Automorphic forms with singularities on Grassmannians}, Invent.\ Math.\ {\bf 132} (1998) 491.
\bibitem{Kontsevich}
M.~Kontsevich, {\it Product formulas for modular forms on $O(2,n)$ (after R. Borcherds)},
S\'eminaire Bourbaki, Vol. 1996/1997, Ast\'erisque No. 245 (1997), \texttt{[arXiv:alg-geom/9709006]}
\bibitem{Prasad}
D.~Prasad, \emph{A brief survey on the theta correspondence},
Lectures given at Trichy in January 1996, available from
{\texttt{http://www.math.tifr.res.in/$\,  \widetilde{}\, $dprasad/dp.pdf}}
\bibitem{EichlerZagier} M.~Eichler and D.~Zagier, {\it The Theory of Jacobi Forms}, Birkh\"auser (1985).
\bibitem{Howe}
R.~Howe, \emph{$\theta$-series and invariant theory},
Proc. Symp. Pure Math. {\bf 33}, ÊPart 1, 275 (1979).
\bibitem{Dixon:1990pc} L.J.~Dixon, V.~Kaplunovsky and J.~Louis, {\it Moduli dependence of string loop corrections to gauge coupling constants}, Nucl.\ Phys.\  B {\bf 355} (1991) 649.
%%CITATION = NUPHA,B355,649;%%
\bibitem{Harvey:1995fq} J.A.~Harvey and G.W.~Moore, {\it Algebras, BPS states, and strings,} Nucl.\ Phys.\  {\bf B463}, 315 (1996) [hep-th/9510182].
\bibitem{Foerger:1998kw} K.~Foerger and S.~Stieberger, {\it Higher derivative couplings and heterotic type I duality in eight-dimensions,} Nucl.\ Phys.\  {\bf B559} (1999) 277 [hep-th/9901020]; K.~Foerger and S.~Stieberger {\it String amplitudes and N=2, d = 4 prepotential in heterotic $K3 \times T^2$ compactifications,} Nucl.\ Phys.\  {\bf B514} (1998) 135 [hep-th/9709004].
\bibitem{Angelantonj:2011br}
  C.~Angelantonj, I.~Florakis and B.~Pioline, {\it A new look at one-loop integrals in string theory,}
  Commun.\ Num.\ Theor.\ Phys.\  {\bf 6} (2012) 159
  [arXiv:1110.5318 [hep-th]].
  %%CITATION = ARXIV:1110.5318;%%
 C.~Angelantonj, I.~Florakis and B.~Pioline,
  {\it One-Loop BPS amplitudes as BPS-state sums,}
  JHEP {\bf 1206} (2012) 070
  [arXiv:1203.0566 [hep-th]].
  %%CITATION = ARXIV:1203.0566;%%
 C.~Angelantonj, I.~Florakis and B.~Pioline, {\it Rankin-Selberg methods for closed strings on orbifolds,}
  Journal of High Energy Physics {\bf 2013} (2013) 7,  181
  [arXiv:1304.4271 [hep-th]].
  %%CITATION = ARXIV:1304.4271;%%
\bibitem{HS}  S.~Hohenegger and S.~Stieberger, {\it BPS Saturated String Amplitudes: K3 Elliptic Genus and Igusa Cusp Form,}
  Nucl.\ Phys.\ B {\bf 856} (2012) 413
  [arXiv:1108.0323 [hep-th]].
  %%CITATION = ARXIV:1108.0323;%%
\bibitem{Witten:1993jg} E.~Witten, {\it On the Landau-Ginzburg description of N=2 minimal models,} Int.\ J.\ Mod.\ Phys.\ A {\bf 9} (1994) 4783 [hep-th/9304026].
%%CITATION = HEP-TH/9304026;%%
\bibitem{Eguchi:1988vra} T.~Eguchi, H.~Ooguri, A.~Taormina and S.~-K.~Yang, {\it Superconformal Algebras and String Compactification on Manifolds with SU(N) Holonomy,} Nucl.\ Phys.\ B {\bf 315} (1989) 193.
%%CITATION = NUPHA,B315,193;%%.
\bibitem{Moore:1997dj} G.~W.~Moore, N.~Nekrasov and S.~Shatashvili, {\it Integrating over Higgs branches,} Commun.\ Math.\ Phys.\  {\bf 209} (2000) 97 [hep-th/9712241].
%%CITATION = HEP-TH/9712241;%%
\bibitem{Lossev:1997bz} A.~Lossev, N.~Nekrasov and S.~L.~Shatashvili, {\it Testing Seiberg-Witten solution,} In *Cargese 1997, Strings, branes and dualities* 359-372 [hep-th/9801061].
  %%CITATION = HEP-TH/9801061;%%
%\cite{Antoniadis:2010iq}
\bibitem{Polchinski} J.~Polchinski, {\it Evaluation of the One Loop String Path Integral}, Commun.\ Math.\ Phys.\  {\bf 104} (1986) 37.
%%CITATION = CMPHA,104,37;%%
\bibitem{Harvey:1996gc}
  J.~A.~Harvey and G.~W.~Moore, {\it On the algebras of BPS states,}
  Commun.\ Math.\ Phys.\  {\bf 197} (1998) 489
  [hep-th/9609017].
  %%CITATION = HEP-TH/9609017;%%
  \bibitem{BKM} R.E.~Borcherds, {\it Automorphic forms on $O_{s+2,2}(R)$ and infinite products,} Invent. Math. 120 (1995) 161. S.~Govindarajan and K.~Gopala Krishna, {\it Generalized Kac-Moody Algebras from CHL dyons,}
  JHEP {\bf 0904} (2009) 032
  [arXiv:0807.4451 [hep-th]].
  %%CITATION = ARXIV:0807.4451.
  M.~C.~N.~Cheng and A.~Dabholkar, {\it Borcherds-Kac-Moody Symmetry of N=4 Dyons,}
  Commun.\ Num.\ Theor.\ Phys.\  {\bf 3} (2009) 59
  [arXiv:0809.4258 [hep-th]].
  %%CITATION = ARXIV:0809.4258;%%
S.~Govindarajan and K.~Gopala Krishna, {\it BKM Lie superalgebras from dyon spectra in Z(N) CHL orbifolds for composite N,} JHEP {\bf 1005} (2010) 014
  [arXiv:0907.1410 [hep-th]].
  %%CITATION = ARXIV:0907.1410;%%
   S.~Govindarajan, {\it BKM Lie superalgebras from counting twisted CHL dyons,}
  JHEP {\bf 1105} (2011) 089
  [arXiv:1006.3472 [hep-th]].
  %%CITATION = ARXIV:1006.3472;%%
\bibitem{Gaberdiel:2011qu} M.~R.~Gaberdiel, S.~Hohenegger and D.~Persson, {\it Borcherds Algebras and N=4 Topological Amplitudes,} JHEP {\bf 1106} (2011) 125 [arXiv:1102.1821 [hep-th]].
  %%CITATION = ARXIV:1102.1821;%%
S.~Hohenegger and D.~Persson, {\it Enhanced Gauge Groups in N=4 Topological Amplitudes and Lorentzian Borcherds Algebras,}  Phys.\ Rev.\ D {\bf 84} (2011) 106007
  [arXiv:1107.2301 [hep-th]].
  %%CITATION = ARXIV:1107.2301;%%
\bibitem{Eguchi:2010ej} T.~Eguchi, H.~Ooguri and Y.~Tachikawa, {\it Notes on the K3 Surface and the Mathieu group $M_{24}$,} Exper.\ Math.\  {\bf 20} (2011) 91 [arXiv:1004.0956 [hep-th]]. M.~C.~N.~Cheng, {\it K3 Surfaces, N=4 Dyons, and the Mathieu Group M24}, \emph{Commun.\ Num.\ Theor.\ Phys. }{\bf 4} (2010) 659 [arXiv:1005.5415 [hep-th]]. M.~R.~Gaberdiel, S.~Hohenegger and R.~Volpato, {\it Mathieu twining characters for K3}, \emph{JHEP}\ {\bf 1009} (2010) 058 [arXiv:1006.0221 [hep-th]].  M.~R.~Gaberdiel, S.~Hohenegger and R.~Volpato, {\it Mathieu Moonshine in the elliptic genus of K3}, \emph{JHEP}\ {\bf 1010} (2010) 062 [arXiv:1008.3778 [hep-th]]. T.~Eguchi and K.~Hikami, {\it Note on Twisted Elliptic Genus of K3 Surface,} Phys.\ Lett.\ B\ {\bf 694} (2011) 446 [arXiv:1008.4924 [hep-th]].  S.~Govindarajan, {\it Brewing Moonshine for Mathieu}, arXiv:1012.5732 [math.NT].
S.~Govindarajan, {\it Unravelling Mathieu Moonshine}, arXiv:1106.5715 [hep-th].
 M.~R.~Gaberdiel, D.~Persson, H.~Ronellenfitsch and R.~Volpato,
 {\it Generalised Mathieu Moonshine,}
  Commun.\  Num.\  Theor.\  Phys.\  {\bf 7} (2013) , 145
  [arXiv:1211.7074 [hep-th]].
  T.~Gannon, {\it Much ado about Mathieu,} arXiv:1211.5531 [math.RT].
  %%CITATION = ARXIV:1211.5531;%%
  \bibitem{Harvey:2013mda}
  J.~A.~Harvey and S.~Murthy, {\it Moonshine in Fivebrane Spacetimes,}
  arXiv:1307.7717 [hep-th].
  %%CITATION = ARXIV:1307.7717;%%
  M.~C.~N.~Cheng, X.~Dong, J.~Duncan, J.~Harvey, S.~Kachru and T.~Wrase, {\it Mathieu Moonshine and N=2 String Compactifications,}
  JHEP {\bf 1309} (2013) 030
  [arXiv:1306.4981 [hep-th]].
  %%CITATION = ARXIV:1306.4981;%%
  S.~Harrison, S.~Kachru and N.~M.~Paquette, {\it Twining Genera of (0,4) Supersymmetric Sigma Models on K3,}
  arXiv:1309.0510 [hep-th].
  %%CITATION = ARXIV:1309.0510;%%
\bibitem{Whittaker} E.T.~Whittaker and G.N.~Watson, {\it A Course of Modern Analysis,} Cambridge at the University Press (1952).
\bibitem{Spiridonov:2012de}
V.~P.~Spiridonov, {\it Modified elliptic gamma functions and 6d superconformal indices,}
 arXiv:1211.2703 [hep-th].
\bibitem{Atiyah} M.~F.~Atiyah and R.~Bott, {\it The Moment Map And Equivariant Cohomology,} Topology {\bf{23}} (1984) 1.
\bibitem{Givental} A.~B.~Givental, {\it Equivariant Gromov-Witten Invariants}, International Mathematical Research Notices (1996), No 13, [alg-geom/9603021].
\bibitem{Witten:1992xu} E.~Witten, {\it Two-dimensional gauge theories revisited,} J.\ Geom.\ Phys.\  {\bf 9} (1992) 303 [hep-th/9204083].
%%CITATION = HEP-TH/9204083;%%
\bibitem{Berndt} R.~Berndt, {\it Zur Arithmetik der elliptischen Funktionenk\"orper h\"oherer Stufe}, J.~Reine Angew.~Math.~{\bf{326}} (1981) 79-94.
\bibitem{Proc} Y.~Choie and W.~Kohnen, {\it Special Values of Elliptic Functions at Points of the Divisors of Jacobi Forms}, Proc.\,AMS, Volume 131, Number 11, 3309-3317.
\bibitem{Koh:1987hm} I.G.~Koh, W.~Troost and A.~Van Proeyen, {\it Covariant Higher Spin Vertex Operators In The Ramond Sector,} Nucl.\ Phys.\  {\bf B292 } (1987) 201;\\
Y.~Tanii and Y.~Watabiki, {\it Vertex Functions In The Path Integral Formalism Of String Theories,} Nucl.\ Phys.\  {\bf B316 } (1989)  171.
\bibitem{Feng:2010yx} W.Z.~Feng, D.~L\"ust, O.~Schlotterer, S.~Stieberger and T.R.~Taylor, {\it Direct Production of Lightest Regge Resonances,} Nucl.\ Phys.\  {\bf B843 } (2011)  570 [arXiv:1007.5254 [hep-th]].
\bibitem{Schlotterer}
  W.~-Z.~Feng, D.~L\"ust and O.~Schlotterer, {\it Massive Supermultiplets in Four-Dimensional Superstring Theory,}
  Nucl.\  Phys.\  B {\bf 861} (2012) 175
  [arXiv:1202.4466 [hep-th]].
  %%CITATION = ARXIV:1202.4466;%%
  \bibitem{Lust:2012zv}
  D.~L\"ust, N.~Mekareeya, O.~Schlotterer and A.~Thomson,
  {\it Refined Partition Functions for Open Superstrings with 4, 8 and 16 Supercharges,}
  Nucl.\ Phys.\ B {\bf 876} (2013) 55
  [arXiv:1211.1018 [hep-th]].
\bibitem{Orlando:2013yea}
  S.~Hellerman, D.~Orlando and S.~Reffert,
  {\it String theory of the Omega deformation,}
  JHEP {\bf 1201} (2012) 148
  [arXiv:1106.0279 [hep-th]].
  %%CITATION = ARXIV:1106.0279;%%
  S.~Reffert,
  {\it General Omega Deformations from Closed String Backgrounds,}
  JHEP {\bf 1204} (2012) 059
  [arXiv:1108.0644 [hep-th]].
  %%CITATION = ARXIV:1108.0644;%%
  S.~Hellerman, D.~Orlando and S.~Reffert,
  {\it The Omega Deformation From String and M-Theory,}
  JHEP {\bf 1207} (2012) 061
  [arXiv:1204.4192 [hep-th]].
  %%CITATION = ARXIV:1204.4192;%%
   S.~Hellerman, D.~Orlando and S.~Reffert,
  {\it BPS States in the Duality Web of the Omega deformation,}
  JHEP {\bf 1306} (2013) 047
  [arXiv:1210.7805 [hep-th]].
  %%CITATION = ARXIV:1210.7805;%%
  D.~Orlando and S.~Reffert, {\it Deformed supersymmetric gauge theories from the fluxtrap background,}
  arXiv:1309.7350 [hep-th].
  %%CITATION = ARXIV:1309.7350;%%

\end{thebibliography}
\end{document}